\documentclass[useAMS,usenatbib]{mn2e}

\usepackage{graphicx}
\usepackage{times}
\usepackage{natbib}
\usepackage{setspace}
\usepackage{amssymb}
\usepackage{amsmath}
\usepackage{enumitem}

\newif\ifAMStwofonts
\AMStwofontstrue

\newcommand{\be}{\begin{equation}}
\newcommand{\ee}{\end{equation}}
\newcommand{\ba}{\begin{eqnarray}}
\newcommand{\ea}{\end{eqnarray}}
\newcommand{\brr}{\begin{array}}
\newcommand{\err}{\end{array}}
\newcommand{\bc}{\begin{center}}
\newcommand{\ec}{\end{center}}

\newcommand{\gra}{{\sc grasil-3d}}
\newcommand{\magphys}{{\sc magphys}}
\newcommand{\gtre}{{\sc gadget}3}

\newcommand{\mincir}{\raise
  -2.truept\hbox{\rlap{\hbox{$\sim$}}\raise5.truept \hbox{$<$}\ }}
\newcommand{\magcir}{\raise
  -2.truept\hbox{\rlap{\hbox{$\sim$}}\raise5.truept \hbox{$>$}\ }}
\newcommand{\siml}{\raise
  -2.truept\hbox{\rlap{\hbox{$\sim$}}\raise5.truept \hbox{$<$}\ }}
\newcommand{\simg}{\raise
  -2.truept\hbox{\rlap{\hbox{$\sim$}}\raise5.truept \hbox{$>$}\ }}

  \graphicspath{{figures/}}

\title[Panchromatic SEDs of simulated galaxies]
{Panchromatic Spectral Energy Distributions of simulated galaxies:
  results at redshift $z=0$}

\author[D. Goz et al.] {
\parbox[t]{\textwidth}{
David Goz$^{2}$,
Pierluigi Monaco$^{1,2}$, 
Gian~Luigi Granato$^{2}$, 
Giuseppe Murante$^{2}$,\\
Rosa Dom\'{\i}nguez-Tenreiro$^{3}$,
Aura Obreja$^{4}$,
Marianna Annunziatella$^{1}$\\
and Edoardo Tescari$^{5,6}$}
\vspace*{6pt}\\
$^{1}$ Dipartimento di Fisica - Sezione di Astronomia, Universit\`a di Trieste, via Tiepolo 11, I- 34131 Trieste -- Italy\\
$^{2}$ INAF, Osservatorio Astronomico di Trieste, Via Tiepolo 11, I-34131 Trieste -- Italy (goz, monaco, granato, murante@oats.inaf.it)\\
$^{3}$ Depto. de F\'{i}sica Te\'orica, Universidad Aut\'onoma de Madrid, E-28049 Cantoblanco Madrid -- Spain\\
$^{4}$ New York University Abu Dhabi, PO Box 129188, Saadiyat Island, Abu Dhabi, UAE\\
$^{5}$ School of Physics, University of Melbourne, Parkville, VIC 3010 -- Australia\\
$^{6}$ ARC Centre of Excellence for All-Sky Astrophysics (CAASTRO)
}

\begin{document}

\maketitle

\label{firstpage}

\begin{abstract}

We present predictions of Spectral Energy Distributions (SEDs), from
the UV to the FIR, of simulated galaxies at $z=0$. These were obtained
by post-processing the results of an N-body+hydro simulation of a
small cosmological volume, that uses the Multi-Phase Particle
Integrator (MUPPI) for star formation and stellar feedback, with the
{\gra} radiative transfer code, that includes reprocessing of UV
light by dust. Physical properties of galaxies resemble
observed ones, though with some tension at small and large stellar masses.
Comparing predicted SEDs of simulated galaxies with different samples
of local galaxies, we find that these resemble observed ones,
when normalised at 3.6 $\mu$m. A comparison with the Herschel
Reference Survey shows that, when binning galaxies in Star Formation
Rate (SFR), average SEDs are reproduced to within a factor of $\sim2$
even in normalization, while binning in stellar mass highlights the
same tension that is present in the stellar mass -- SFR plane. We use
our sample to investigate the correlation of IR luminosity in Spitzer
and Herschel bands with several galaxy properties. SFR is the quantity
that best correlates with IR light up to $160\ \mu$m, while at longer
wavelengths better correlations are found with molecular mass and, at
$500\ \mu$m, with dust mass. However, using the position of the FIR peak 
as a proxy for cold dust temperature, we assess that heating of cold dust
is mostly determined by SFR, with stellar mass giving only a minor
contribution. We finally show how our sample of simulated galaxies can
be used as a guide to understand the physical properties and selection
biases of observed samples.

\end{abstract}

\begin{keywords}
galaxies: formation - hydrodynamics - radiative transfer - dust,
extinction - submillimetre: galaxies - infrared: galaxies - method:numerical
\end{keywords}

\section{Introduction}
\label{section:intro}

The current $\Lambda$CDM cosmological model is able to reproduce a
number of fundamental observations, such as temperature fluctuations
in the cosmic microwave background, Hubble diagram of distant
supernovae, large scale distribution of visible matter in the
Universe, evolution of galaxy cluster abundances
\citep[e.g.][]{Planck2014,Betoule_2014,Vikhlinin_2009,Springel_2006,Conroy_2006}.
While this is the context in which galaxy formation takes place, a
complete theory of galaxy formation continues to be elusive and our
understanding of the physical processes that lead to the observed
properties of galaxies is far from satisfactory
\citep[e.g.][]{Silk_2012,Primack_2015}.

On the theoretical side, the problem of galaxy formation has been
faced with the two main techniques of semi-analytic models and
hydrodynamic simulations. This last technique has progressed rapidly
thanks to the development of more sophisticated numerical algorithms
and to the increasing computational power
\citep[e.g.][]{Springel_2014}. However even the current computing power makes it
impossible to resolve the wide range of scales needed to describe the
formation of a galaxy in cosmological environment from first
principles. As a consequence, processes like star formation, accretion
of matter onto a supermassive black hole, and their energetic
feedbacks, must be included into the code through suitable
sub-resolution models. Early attempts to produce realistic galaxies,
reviewed in \cite{Murante_2015}, were not successful due to a number
of problems, the most important being the catastrophic loss of angular
momentum. The {\em Aquila comparison project} \citep{Scannapieco12}
demonstrated that at that time none of the tested codes was able to
produce a really satisfactory spiral galaxy in a Milky-Way-sized dark
matter halo with quiet merger history: most simulated galaxies tended
to be too massive and too compact, with rotation curves sharply
peaking at $\sim5$ kpc. It was suggested in that paper that the crucial
process for a successful simulation is effective stellar feedback.
That paper triggered a burst of efforts to improve sub-grid models in
order to solve the discrepancies with observations. Different authors
\citep[e.g.][]{Domenech_2012,Brook_2012,Stinson13,Obreja_2013,Aumer13,Vogelsberger_2014,Marinacci_2014,
  Hopkins_2014,Cen_2014,Schaye_2015,Agertz_2015}, using different
codes, hydro scheme and sub-resolution models, are now able to produce
late-type spiral galaxies with realistic properties: extended discs,
flat rotation curves, low bulge-to-total ratios.

In \cite{Murante_2015}, our group presented cosmological simulations
of individual disk galaxies, carried out with {\gtre}
\citep{GADGET2}, including a novel sub-resolution model MUlti-Phase
Particle Integrator (MUPPI) for star formation and stellar feedback.
Using initial conditions of two Milky-Way sized halos (one being the
Aquila Comparison halo), the paper presented late-type galaxies with
properties in broad agreement with observations (disk size, mass
surface density, rotation velocity, gas fraction) and with a bar
pattern developing at very low redshift and likely due to 
marginal Toomre instability of disks \citep{Goz_2015}. Simulations of
small cosmological volumes of side 25 Mpc, obtained using
the same code of \cite{Murante_2015}, were presented in
\cite{Barai_2015}. In that paper, the main properties of outflows
generated by the feedback scheme (velocity, mass load, geometry) were
investigated and quantified, showing broad agreement with
observational evidence.

Simulations directly provide predictions on physical quantities like
stellar masses, gas masses or Star Formation Rates (SFRs), while most
observed quantities are related to radiation emitted in given bands of
the electromagnetic spectrum. Comparisons of model galaxies with
observations can be carried out either by using physical quantities
obtained from observations or by synthesizing observables from the
simulation itself. While the first approach is easier to carry out
once quantities like stellar masses, SFRs or
gas masses are available for large samples of galaxies, it is clear
that exploiting the full extent of panchromatic observations can
produce much tighter constraints on models. Moreover, many assumptions
on the stellar Initial Mass Function (IMF), galaxy SFR history
and dust attenuation are needed to infer physical quantities from
observed ones, and these assumptions would better be made on the model
side.

To properly synthesize observables from a simulation snapshot it is
necessary to compute radiative transfer of starlight through a dusty
InterStellar Medium (ISM), thus producing both attenuation of
optical-UV light and re-emission in the FIR \citep[see, e.g., the
  recent review by][]{Jones_2014}. Observations across the
electromagnetic spectrum reveal that galaxy Spectral Energy
Distributions (SEDs) strongly depend on
the relative geometry of stars and dust, as well as on the properties
of dust grains \citep[e.g.][]{Fritz_2012}. This complexity produces
results that are not easily represented by simple approximations, as
shown e.g., by \cite{Fontanot_2009b} even in the idealized case of
semi-analytic galaxies where geometry is represented by a simple
bulge/disc system.

It is only in recent years that the full SEDs of simulated
  galaxies have been studied, using different approaches to follow the
  transfer of radiation through the ISM and to calculate a global
  radiation field, and hence dust re-emission. Examples of codes for
  3D radiative transfer are the following: SUNRISE
  \citep{Jonsson_2010}, SKIRT \citep{Camps_2015}, RADISHE
  \citep{Chakrabarti_2009}, ART$^{2}$ \citep{Li_2008}, all using Monte
  Carlo techniques and DART-RAY \citep{Natale_2014}, a ray-tracing
  code. Only few codes take into account structure on scales of
  star-forming regions, or include emission lines. SUNRISE includes
  the treatment of star-forming regions using the dust and
  photoionization code MAPPINGSIII \citep{Groves_2008}. The same
  approach is adopted by \cite{Camps_2016} for particles in
  star-forming regions. Several authors have performed broadband
  radiative transfer modeling for simulations of isolated or
  cosmological galaxies \citep[][e.g.]{Jonsson_2010,
    Narayanan_2010,Scannapieco_2010,
    Hayward_2011,Hayward_2012,Hayward_2013, Hayward_2015, Guidi_2015,
    Saftly_2015,Natale_2015}.

In this work we use the {\gra} code \citep{Dominguez_2014} to
post-process simulated galaxies obtained in the cosmological volumes
presented by \cite{Barai_2015}. This code, based on {\sc grasil} formalism
\citep{Silva_1998}, performs radiative transfer of starlight in an ISM
that is divided into diffuse component ({\em cirrus}) and molecular
clouds, that are treated separately on the ground that the molecular
component is unresolved in such simulations. 
Therefore, in order
to treat the very important dust reprocessing in molecular
clouds, {\gra} introduces a further sub-resolution modelling of these
sites of star formation, on the lines of the original {\sc GRASIL}.
Other features inherited by the original code are a
self-consistent computation of dust grain temperatures under the
effect of UV radiation, re-emission of dust in the IR and a model of
polycyclic aromatic hydrocarbons (PAHs).
{\gra} has been used to study the IR emission of simulated galaxies
\citep{Obreja_2014} and galaxy clusters \citep{Granato_2015}.

This paper is devoted to the study of panchromatic SEDs of our
simulated galaxies at $z=0$. The choice of redshift is motivated by
the high quality of observations in the local Universe
\citep{Cook_2014,Cook_2014a,Gruppioni_2013,Boselli_2010,Clark_2015,Groves_2015},
that allows a more thorough comparison. We will concentrate on the
average SED of galaxies, trying to reproduce the selection of observed
samples used for the comparison, and on the scaling relations of
several basic quantities (like stellar mass, SFR, atomic and molecular
gas masses, gas metallicities) among them and with FIR luminosity.
As a result, we will show that, despite the existence of several
points of tension with observations, the panchromatic properties of
simulated galaxies resemble those of observed ones, to the
extent that they can be used as a guide to understand the physical
properties and selection biases of observed samples.

The paper is organized as follows. Section~\ref{section:simulations}
presents the simulation code, the sub-resolution model used, the run
performed, the {\gra} code and the post-processing performed with it
to produce SEDs for our galaxies. Section~\ref{section:observations}
presents the observational samples used for the comparison with data.
Section~\ref{section:comparison} is devoted to the comparison of
simulation results to observations. Firstly, the global properties of
the simulated sample at $z=0$ are presented. Then 
average galaxy SEDs, in bins of stellar
mass or SFR are shown to closely resemble observed ones. We then
quantify the correlation of IR luminosities, from Spitzer to Herschel
bands, with physical galaxy properties.
Section~\ref{section:conclusions} gives the conclusions.

\section{Simulations}
\label{section:simulations}

\subsection{Simulation code}
\label{section:code}

Simulations were performed with the Tree-PM+SPH {\gtre} code,
non-public evolution of {\sc GADGET}2 \citep{GADGET2}. Gravity is solved
with the \cite{Barnes_1986} Tree algorithm, aided by a Particle Mesh
scheme on large scales. Hydrodynamics is integrated with an SPH solver
that uses an explicitly entropy-conserving formulation with force
symmetrization. No artificial conduction is used in this simulation
\citep[see e.g.][]{Beck_2016}.

Star formation and stellar feedback are implemented using the
sub-resolution model MUPPI,
described in its latest features in \cite{Murante_2015}.
Loosely following \cite{Monaco04a}, it is assumed that a
gas particle undergoing star formation samples a multi-phase
ISM, composed of cold and hot phases in pressure equilibrium.

A gas particle enters the multi-phase regime whenever its density is
higher than a threshold value $\rho\rm{_{thr}}$ (free parameter of the
model), and its temperature drops below 10$^{5}$ K. Then its
evolution, within each SPH time-step, is governed by a system of four
ordinary differential equations that describe the mass flows among
cold, hot, and (virtual) stellar components, and the thermal energy
flows (cooling, SN feedback and hydrodynamic term) of the hot phase. A
fraction of the cold phase is assumed to be molecular and provides the
reservoir for star formation. This fraction is computed following a
phenomenological relation by \cite{Blitz06}:

\begin{equation}
 f\rm{_{mol}} = \frac{1}{1 + P_{0}/P}
\label{eq:br} \end{equation}

\noindent
where in our model $P$ is the SPH-particle pressure (used in place of
the external pressure of \citet{Blitz06}), and $P_{0}$ is the pressure
at which $f\rm{_{mol}}$ = 0.5; we used $P_0=20,000\ {\rm K\ cm}^{-3}$.

SN feedback is distributed in both thermal and kinetic forms, 
within a cone whose axis is aligned along the direction
of the least resistance path (i.e. along the direction of minus the local
density gradient) and within the particle's SPH smoothing length 
(see \citealt{Murante_2015} for full
details). There is no active galactic nucleus (AGN) feedback in this
version of the code.

Star formation is implemented with the stochastic model of
\cite{Katz96} and \cite{Springel03}, where a new star particle
is spawned with a probability that depends on the stellar mass
accumulated in one hydro time-step. Each star particle represents a
Simple Stellar Population (SSP) with a given Initial Mass Function
(IMF) from \cite{Kroupa_1993} in the mass range (0.1-100) M$_{\odot}$.
Its chemical evolution is treated as in \cite{Tornatore07}: the
stellar mass of the particle is followed since its birth (spawning)
time, accounting for the death of stars and stellar mass losses. The
code tracks the production of different species: He, C, Ca, O, N, Ne,
Mg, S, Si, Fe and a further component of generic ``ejecta'' containing
all other metals. It takes into account yields from type Ia SNe
\citep{Thielemann_2003}, 
type II SNe 
\citep{Chieffi_2004} 
and Asymptotic Giant Branch (AGB) stars
\citep{Karakas_2010}.
Such mass losses are distributed to neighbouring gas particles
together with the newly produced metals. The yields we use in
this simulation are updated with respect to \cite{Barai_2015}.
Radiative cooling is implemented by adopting cooling rates from the
tables of \cite{Wiersma09}, which include metal-line cooling by the
above-mentioned chemical elements, heated by a spatially uniform,
time-dependent photoionizing background \citep{Haardt_2001}. Gas is
assumed to be dust-free, optically thin, and in (photo)ionization
equilibrium.

\subsection{SED synthesis with {\gra}}
\label{section:g3d}

\cite{Dominguez_2014}, presented the radiative
transfer code {\gra} that post-processes the output of a simulation to
synthetize the panchromatic SED of the galaxy. It 
is based on the widely employed {\sc GRASIL} (GRAphite
and SILicate) model \citep{Silva_1998,Silva_1999,Granato_2000}, that
self-consistently computes the radiation field of a galaxy, emitted by
stars and attenuated by dust, and the temperature of dust grains
heated by the radiation itself. While the original code was applied to
galaxies composed of idealized bulge and disk components, {\gra} is
applied to the arbitrary geometry of a simulated distribution of star
particles. We refer the reader to the \cite{Dominguez_2014} paper for
an exhaustive discussion of this code. Its main features are
summarized below, focusing on the few modifications that we introduced
to interface it with the output of our version of {\gtre}.

{\gra} treats the ISM as divided into two components: optically thick
molecular clouds (hereafter MCs), where stars are assumed to spend the
first part of their life, and a more diffuse cirrus component. This
separation, that is well founded in the observed properties of young
stars, produces an age-selective dust extinction, with the youngest
and most UV-luminous stars suffering the largest extinction. The
physical quantity that regulates this division is the escape time
$t\rm{_{esc}}$ of stars out of the MC component. Inside MCs, starlight
is subject to an optical depth $\tau \propto \delta \Sigma_{\rm{MC}}$
that is proportional to the cloud surface density $\Sigma_{\rm MC}$.
Both the original {\sc grasil} code and {\gra} require two parameters
to describe MCs (assumed to be spherical): their mass $M_{\rm MC}$ and
their radius $R_{\rm MC}$. However, all the results depends only on
their surface density, which is a combination of the two parameters:
$\Sigma_{\rm{MC}} = M_{\rm MC}/R_{\rm MC}^2$ (see \citealt{Silva_1998}
for details). In the following, we will use the MC mass as a free
parameter, while keeping $R_{\rm MC}$ fixed at 15 pc, a rather typical
value for Galactic giant molecular clouds. We stress that the
specific value of this parameter is immaterial, it only determines
the relation between MC mass and surface density.

Gas and stellar mass densities are constructed by smoothing particles
on a Cartesian grid, whose cell size is set by the force softening of
the simulation. For gas mass, we used cold (T $<$ 10$^{5}$ K) and
multi-phase gas particles. The original code uses a sub-resolution
model to estimate the molecular fraction of a gas particle.
A different strategy is applied in our case,
where the molecular gas mass is computed by MUPPI using
equation~\ref{eq:br}. 

However, the stochastic star formation algorithm
does not guarantee that molecular gas is associated with the youngest
stars, so the association of star particles with MCs, a working
assumption of {\gra}, is subject to unphysical fluctuations.

We modified {\gra} to read molecular fraction of gas particles
directly from the simulation output. Atomic gas is simply identified
with the cirrus component. For the MC component, {\gra} computes the
density of stars younger than $t_{\rm esc}$, then redistribute the
molecular mass proportionally to that density, so as to guarantee
perfect coincidence of the two components. To compute unattenuated
starlight, we adopt a \cite{Chabrier_2003} IMF.
The version of GRASIL3D used in this paper is based on
models of simple stellar populations (SSPs) by \cite{Bruzual_2003}, 
so we could not simply replicate the choice of 
\cite{Kroupa_1993} made in the simulation.
But the two IMFs are so similar that we do not expect this difference 
to have any impact on the results.

  Dust is assumed to consist of a mixture of silicates and graphites, 
  where the smaller carbonaceous grains, with size lower than 100 $\AA{}$,
  are assumed to have PAH properties (either neutral or ionized).
  The dust model used for the diffuse cirrus is that proposed by 
  \cite{Weingartner_2001, Draine_2001, Li_2001}, updated by \cite{Draine_2007b},
   a model originally calibrated to match the extinction and emissivity 
   properties of the local ISM.
   For small grains, use has been made of the MW3.1\textunderscore60 model in Table 3 of 
  \cite{Draine_2007b}, with a PAH-dust mass fraction q\textunderscore PAH = 4.58 
  (i.e., the Galactic value around the Sun). 
  This q\textunderscore PAH value is also consistent with observational data for other normal galaxies 
  (i.e., no low metallicity ones), as can be seen for example in \cite{Draine_2007},
  Figure 20, \cite{Galliano_2008}, Figures 25 and 29.
  The PAH ionization fraction has also been taken from \cite{Draine_2007b}.
  This model is similar to the BARE-GR-S model introduced by \cite{Zubko_2004}.
  For MCs the original mixture used by \cite{Silva_1998} is assumed,
  calibrated against the local ISM and later on updated to fit the MIR emissions 
  of a sample of actively star-forming galaxies by \cite{Dale_2000} 
  (see \citealt{Vega_2005,Vega_2008} for more details).
  In particular, PAH in MCs have been depleted by a factor 1000 relative to
  their abundance in the diffuse component.
  
  Dust is assigned to the grid following the standard assignation:
  \be
    \frac{D}{G} = \frac{Z}{Z_\odot \times 110} ,\ \mathrm{for} \ Z_\odot = 0.02
    \label{eq:dg}
  \ee

\noindent
where $D/G$ stands for the dust-to-gas mass ratio and $Z$ is the 
total gas metallicity.
This is similar to \cite{Zubko_2004} assignation rule,
except that a different value of $Z_\odot=0.014$ is taken by these authors.
Both dust evolution models (\citealt{Seok_2014}, Fig. 5; \citealt{Calura_2008}, Fig.16)
and observational data suggest that the global relation between gas mass and dust mass for galaxies
with 12 + $\log$(O/H) $<$ 8.1 is consistent with that found in the Galactic interstellar medium, i.e.
the standard $D/G$ assignation above
\citep[see, e.g., ][]{Draine_2007, Galliano_2008, Remy_2014}.
 The global metallicity of MUPPI galaxies is never 12 + $\log$(O/H) $<$ 8.1 
 (see Fig.~\ref{fig:relations}), so that the use of Eq.~\ref{eq:dg} is justified.

In this paper we refer to original papers for a justification of the 
  assumptions made in the dust model choice, 
  and make no attempt to test the effects of changes in dust parameters, 
  with the exception of dust emissivity index $\beta$. In fact, in
  Appendix~\ref{section:black-body} we test the results of changing
  this canonical value $\beta=2$ adopted here and the recent
  determination of $\beta = 1.62 \pm 0.10$ \citep[][]{Planck_2014}.
  This affects the sub-mm side of the SEDs but does not greatly affect
  the conclusions presented here. 

  While dust mass and molecular fraction are fixed by the simulation as explained above,
  two parameters need to be decided, namely the escape time $t_{\rm esc}$ 
  and $\tau$ (the latter being represented by variations of MC
  mass $M_{\rm MC}$ at fixed radius $R_{\rm MC}$ of 15 pc). In
  Appendix~\ref{appendix:calibration} we show the robustness of
  resulting SEDs to variations of these parameters around their
  reference values. Finally, in Appendix~\ref{appendix:resolution} we
test the stability of results to the exact choice of grid size, and on
the choice of aperture to define which particles belong to the galaxy.

\subsection{Simulation run}
\label{section:simulation}

The simulation, called MUPPIBOX in this paper, represents a
cosmological periodic volume of $L_{\rm box}$ = 25 comoving Mpc
($H_0=72$ km/s/Mpc), sampled by $N_{\rm{part}}= 256^{3}$ DM particles
and as many gas particles in the initial conditions. Initial gas
particle mass is $m_{\rm{gas}} = 5.36 \times 10^{6}$ M$_{\odot}$. The
Plummer-equivalent softening length for gravitational forces is set to
$L_{\rm{soft}}= 2.08$ comoving kpc up to $z= 2$, then it is held fixed
at $L_{\rm soft}= 0.69$ kpc in physical units from $z= 2$ up to $z=
0$. The simulation is very similar to the \emph{M25std} presented in
\cite{Barai_2015}, but with improved yields and slightly different
parameters for the MUPPI sector, to absorb the differences induced in
the cooling rate by the different metal compositions: f$\rm{_{b,out}}$
= 0.2, f$\rm{_{b,kin}}$ = 0.5, P$\rm{_{kin}}$ = 0.02.

To select galaxies, we proceed as in \cite{Murante_2015} and
\cite{Barai_2015}. The simulation is post-processed with a standard
friends-of-friends algorithm and with the substructure-finding code
SubFind \citep{Springel01} to select main halos and their
substructures. Galaxies are associated to substructures (including the
main halos), their centre is defined using the position of the most
bound particle. Star and cold ($T<10^5$ K) or multi-phase gas
particles that stay within a fiducial distance $R_{\rm gal}=0.1 R_{\rm
  vir}$ (a tenth of the virial radius of the host main halo)
constitute the visible part of the galaxy; they will be named {\it
  galaxy particles} in the following. The reference frame of each
object is aligned with the inertia tensor of all galaxy particles
within a fiducial distance of 8 kpc from the galaxy centre, to avoid
misalignments due to close galaxies. The $Z$-axis is aligned with the
eigenvector corresponding to the largest eigenvalue of the inertia
tensor and on the direction of the star angular momentum, while the
other axes are aligned with the other eigenvectors so as to preserve
the property $\hat{X} \times \hat{Y} = \hat{Z}$. Whenever a disc
component is present, aligning with the inertia tensor is equivalent
to aligning the $Z$-axis with the angular momentum of stars.

Galaxy morphologies are simply quantified through the $B/T$,
bulge-over-total ratio. Star particle circularities are defined as
$\epsilon = J_{z}/J\rm_{circ}$, where $J_{z}$ is the specific angular
momentum along $Z$-axis, and $J_{circ} = r\sqrt{(GM(<r))/r}$
angular momentum of a circular orbit at the same distance. The bulge mass
is estimated as twice the mass of counter-rotating particles, the disc
mass is assumed to be the remaining stellar mass. This simple procedure
is known to over-estimate $B/T$ with respect to fitting a synthetic
image of the galaxy, reproducing the procedure performed by observers
\citep{Scannapieco_2010}.
Other bulge definitions and mass measurements are possible
\citep[see, e.g.,][]{Domenech_2012,Obreja_2013}.

We consider galaxies with stellar mass $ >2 \cdot10^{9}$ M$_{\odot}$
within $R\rm_{gal}$. This way each object is resolved with at least
$\simeq$ 2 $\cdot$ 10$^{3}$ star particles. Below this mass, the
stellar mass function is observed to drop, so this is a conservative
estimate of the completeness limit mass of the simulation. All
quantities (masses, luminosities etc.) are computed on galaxy
particles within $R_{\rm gal}$; in Appendix~\ref{section:petrosian}
we discuss how galaxy SEDs change when we adopt another definition of
aperture radius, inspired by the Petrosian radius
\citep{Petrosian_1976}. SEDs do depend on galaxy inclination, so they
were not computed using the reference frame aligned with the galaxy
inertia tensor, but in a frame aligned with the box reference frame.
This way the line of sight is along the $Z$-axis of our box, yielding
an essentially random orientation for each galaxy.

\section{Observational samples used for the comparison}
\label{section:observations}

We want to compare predicted SEDs of simulated galaxies to
pan-chromatic observations of local galaxies. To this aim we have
selected observational samples that provide SEDs as well as recovered
physical properties of observed galaxies (stellar masses, SFRs, gas
and dust masses etc.).

A first step in the analysis consists in the comparison of overall 
SED shapes, once a common normalization is adopted. 
This is done by grouping
galaxies in bins of stellar mass and comparing the median of their predicted
near-UV to far-IR SEDs with the results of two local samples, namely
Local Volume Legacy (LVL) \citep{Cook_2014,Cook_2014a} and PACS
Evolutionary Probe (PEP) \citep{Gruppioni_2013}. In this comparison,
all SEDs will be normalized to the 3.6 $\mu$m flux, so the test will
only be sensitive to the overall SED shape and not to its
normalization.
The IR part of SEDs will then be compared to the
results of the Herschel Reference Survey (HRS) of \cite{Boselli_2010},
by grouping galaxies according to several physical quantities.
Finally, tests of scaling relations of physical properties (gas and
dust mass) and observable properties (FIR luminosity) will be
performed using the Herschel-ATLAS Phase-1 Limited-Extent Spatial
Survey (HAPLESS) sample \citep{Clark_2015} and the sample by
\cite{Groves_2015}. These samples are described below.

To be consistent with the adopted stellar mass threshold of simulated galaxies, 
we will always select observed galaxies above a stellar mass limit of
$2 \cdot10^{9}$ M$_{\odot}$.

\subsection{LVL sample}
\label{section: cook}

LVL \citep{Cook_2014a,Cook_2014} consists of 258 nearby galaxies ($D
\leq 11$ Mpc), considered as a volume-limited, statistically complete
and representative sample of the local Universe.
Galaxies are selected with $m_{\rm B} < 15.5$ mag \citep{Dale_2009}
and they span a wide range in morphological type, 
but clearly the sample is dominated
by dwarf galaxies due to its volume-limited nature. LVL SEDs cover the
range from GALEX FUV band to Herschel MIPS160 band. Physical
properties like stellar mass and SFRs have been obtained through SED
fitting techniques, we refer to the original papers for all details.

\subsection{PEP sample}
\label{section: gruppioni}

PEP \citep{Gruppioni_2013} is a Herschel survey that covers the most
popular and widely studied extragalactic deep fields: COSMOS, Lockman
Hole, EGS and ECDFS, GOODS-N and GOODS-S \citep[see][for a full
description]{Lutz_2011,Berta_2010}.
Galaxies are selected with flux densities $\geq 5.0$ and $\geq 10.2$
mJy at 100 and 160 $\mu$m respectively.
SEDs cover the range from U-band to SPIRE500 Herschel band. 
Being interested in the local Universe, we selected galaxies with $z<0.1$.
Physical properties of PEP galaxies are obtained with SED fitting
technique using {\magphys} \citep{daCunha_2008}. This code uses a
Bayesian approach to find, at a given redshift, the template that best
reproduces the observed galaxy fluxes. We use spectroscopic redshifts
provided by PEP for all galaxies in our sample. SED templates are
obtained from libraries of stellar population synthesis models of
\cite{Bruzual_2003} with a \cite{Chabrier_2003} IMF and a metallicity
value which can be in the range 0.02--2 ${Z_{\odot}}$.
As for the SFR history, {\magphys} assumes an exponentially declining
(${\rm SFR}\propto \exp(-\gamma t)$) to which random bursts are
superimposed. The probability distribution function of the star
formation time-scale is ${p(\gamma) =1\, -\, \tanh(8\gamma \, -
  \, 6)}$, which is uniform between 0 and 0.6 Gyr$^{-1}$ and
then drops exponentially to zero at 1 Gyr$^{-1}$. The age of
the galaxy is another free parameter, with an upper limit provided by
the age of the Universe at the chosen redshift. The code outputs
probability distribution functions of physical parameter values
(stellar mass, SFR) that give best-fitting SEDs. We adopt median
values, that are a more robust estimator, while the 1$\sigma$
uncertainty is given by half of the difference between the 16th and
84th percentiles.

\subsection{HRS sample}
\label{section:HRS}

HRS \citep{Boselli_2010} is a volume-limited sample selected between
15 and 25 Mpc ($H_0=70$ km/s/Mpc), complete for $K\le12$ (total Vega
magnitude) for late types and $K<8.7$ for early types. 
For the late types, this corresponds to a stellar mass limit of $\sim10^9$
M$_\odot$, smaller than our limit by a factor of $\sim2$. It contains
322 galaxies, 62 early-types and 260 late-types. IR SEDs of the HRS
galaxies range from 8 to 500 $\mu$m, where MIR photometry is obtained
by Spitzer/IRAC and WISE images plus archival data from Spitzer/MIPS,
and FIR photometries are obtained from Herschel/PACS, Herschel/SPIRE,
and IRAS \citep{Ciesla_2014}. We used HRS infrared SED templates
available to the community via the
HEDAM\footnote{http://http://hedam.lam.fr/HRS/} website.

\subsection{HAPLESS sample}
\label{section:clark}

HAPLESS \citep{Clark_2015} consists of 42 nearby galaxies (15 $<$ $D$
$<$ 46 Mpc), taken from the \emph{H}-ATLAS Survey. The sample spans a
range of stellar mass from 7.4 to 11.3 in ${\rm Log} M_\star$, with a
peak at $10^{8.5}$ M$_{\odot}$, and specific star formation rate from
$-11.8$ to $-8.9$ in Log yr$^-1$. Most galaxies have late-type,
irregular morphology. HAPLESS provides masses and temperatures of cold
and warm dust for its galaxies, obtained by fitting two modified
black-bodies to the FIR SEDs.

\subsection{Groves sample}
\label{section:groves}

\cite{Groves_2015} presented a sample collected from KINGFISH
(Herschel IR), HERACLES (IRAM CO), and THINGS (VLA HI) surveys
\citep[][]{Walter_2008,Leroy_2009,Kennicutt_2011}.
The matched
datasets give 36 nearby galaxies with average distance of 10 Mpc,
ranging from dwarf galaxies to massive spirals, that are dominated by
late-type and irregular galaxies. Stellar masses span the range from
$10^{6.5}$ to $10^{10.5}$ M$_{\odot}$, with a median ${\rm Log} M_\star/{\rm
  M}_\odot$ of 9.67. SFRs cover approximately 4 orders of magnitudes
from $10^{-3}$ to $8$ M$_{\odot}$ yr$^{-1}$.

\section{Comparison of simulated and observed galaxy samples}
\label{section:comparison}

\subsection{Global properties of the simulated sample}
\label{section:global}

\begin{figure*}
\centering{
\includegraphics[angle=0,width=0.33\linewidth]{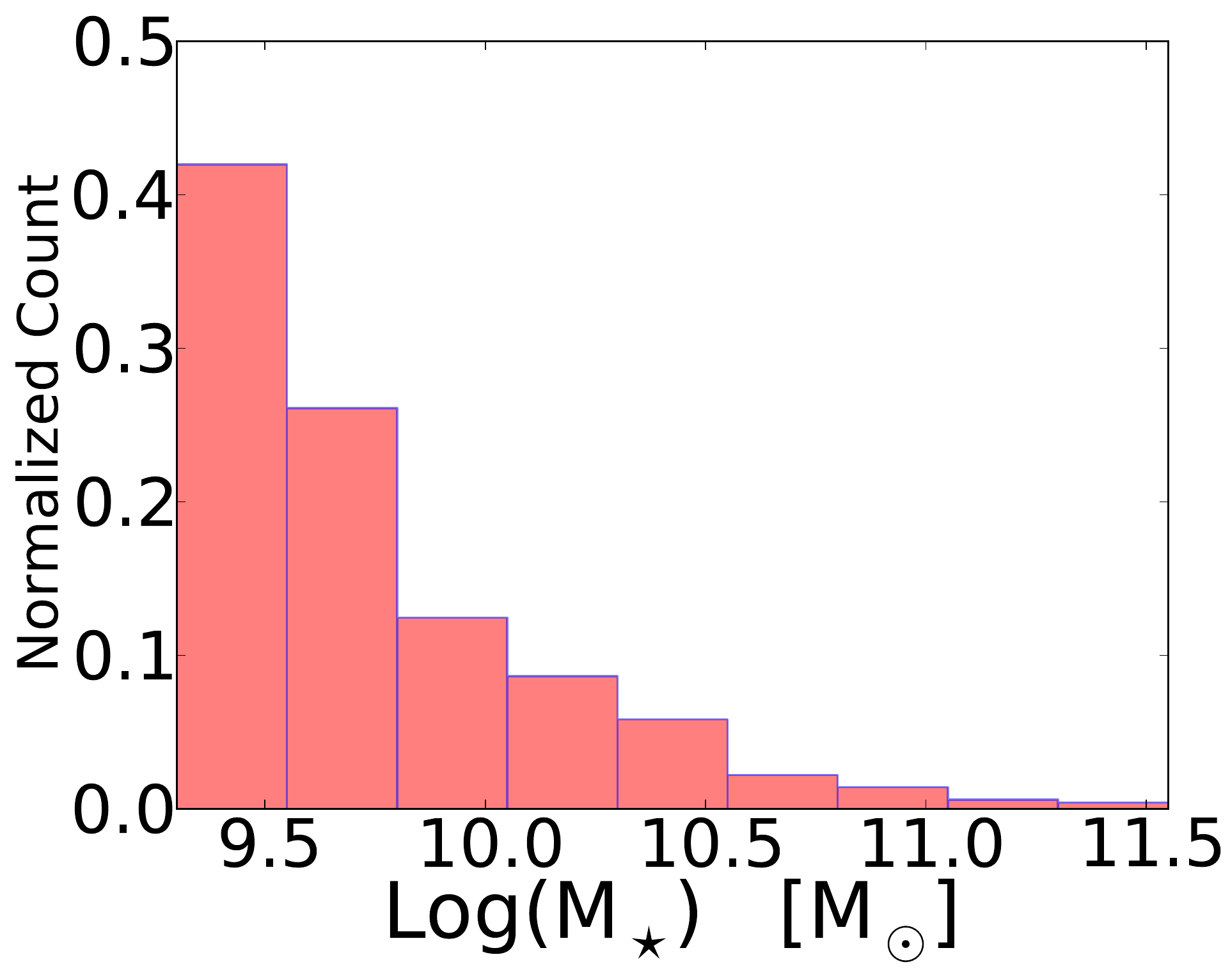}
\includegraphics[angle=0,width=0.32\linewidth]{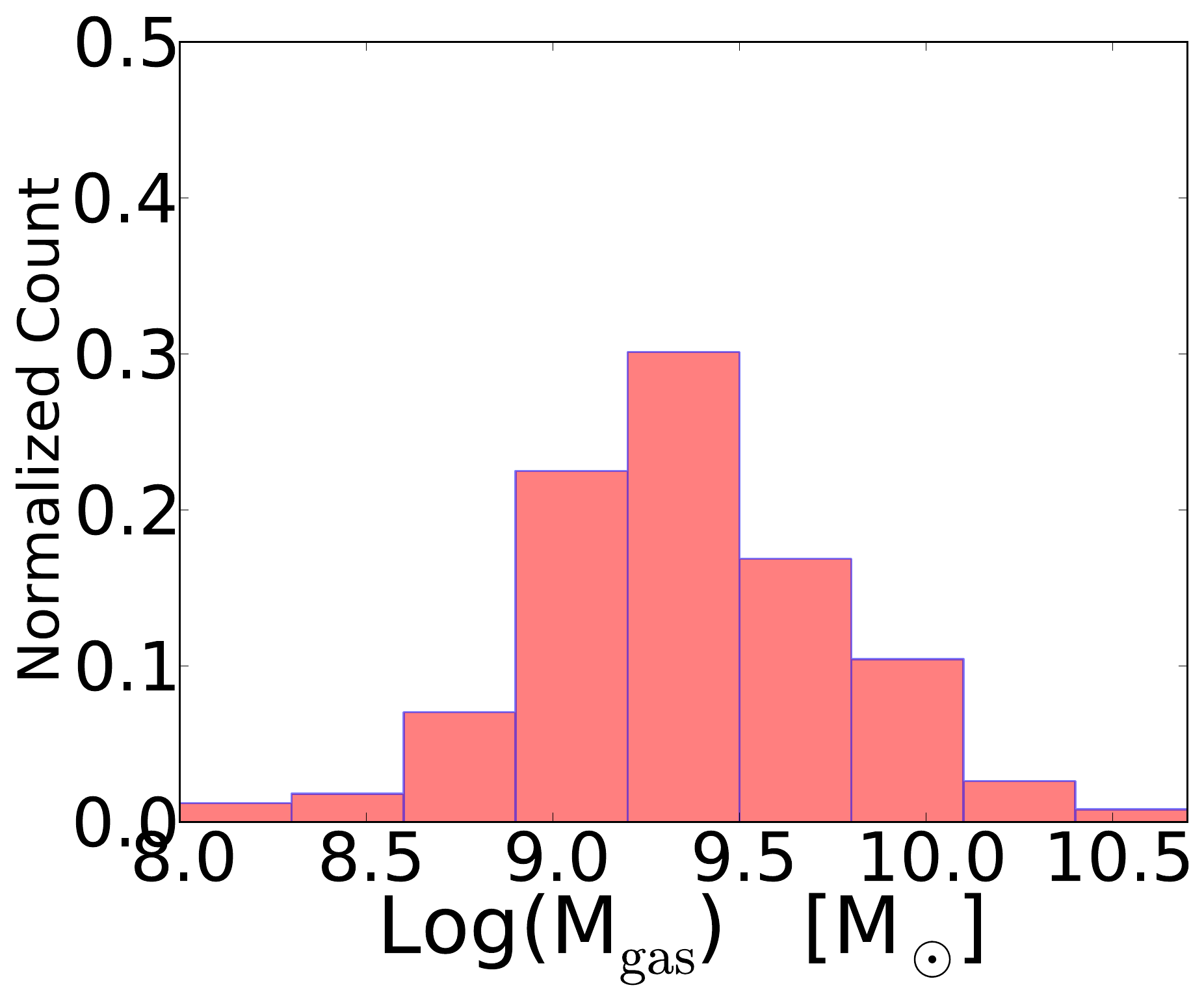}
\includegraphics[angle=0,width=0.33\linewidth]{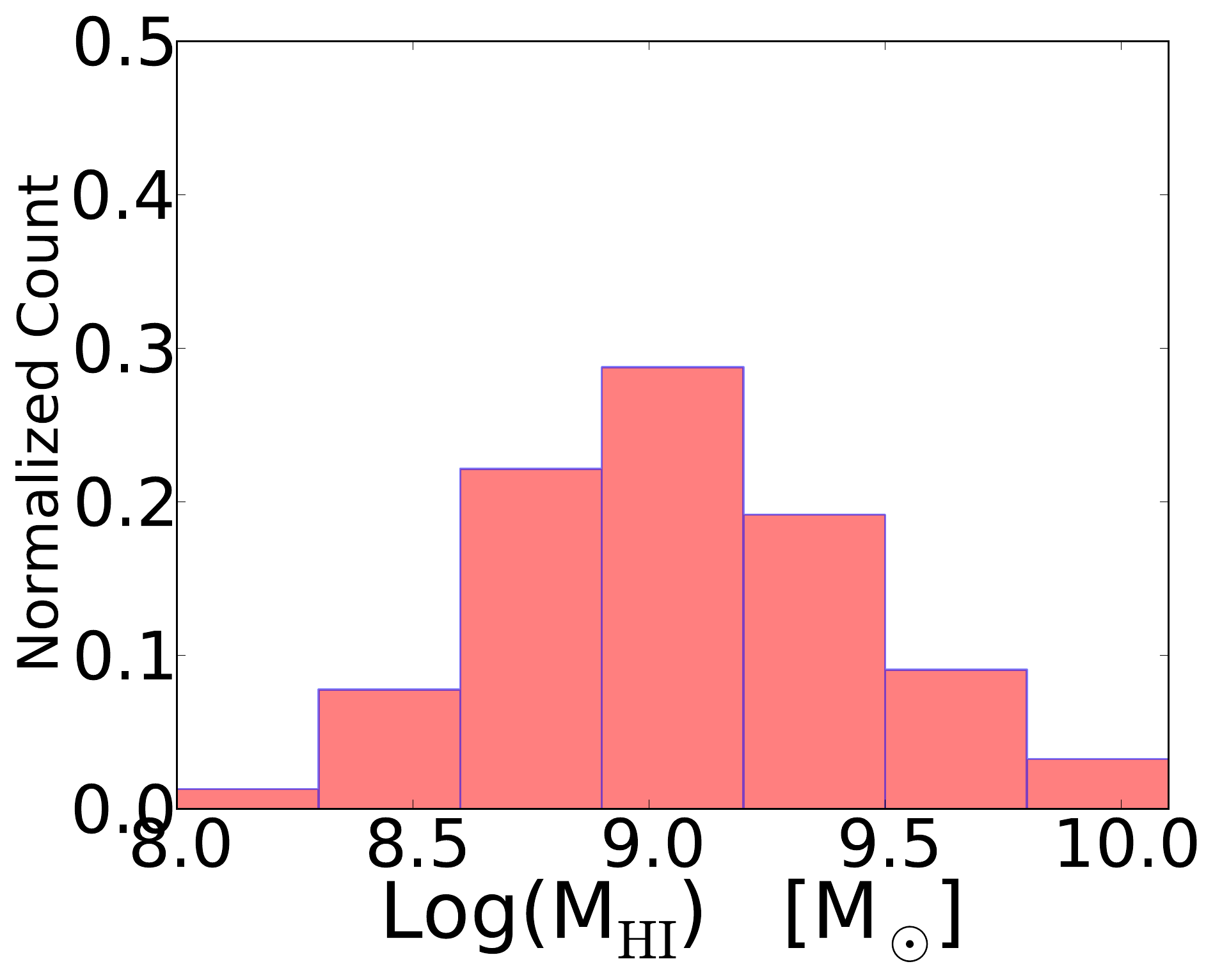}}
\centering{
\includegraphics[angle=0,width=0.32\linewidth]{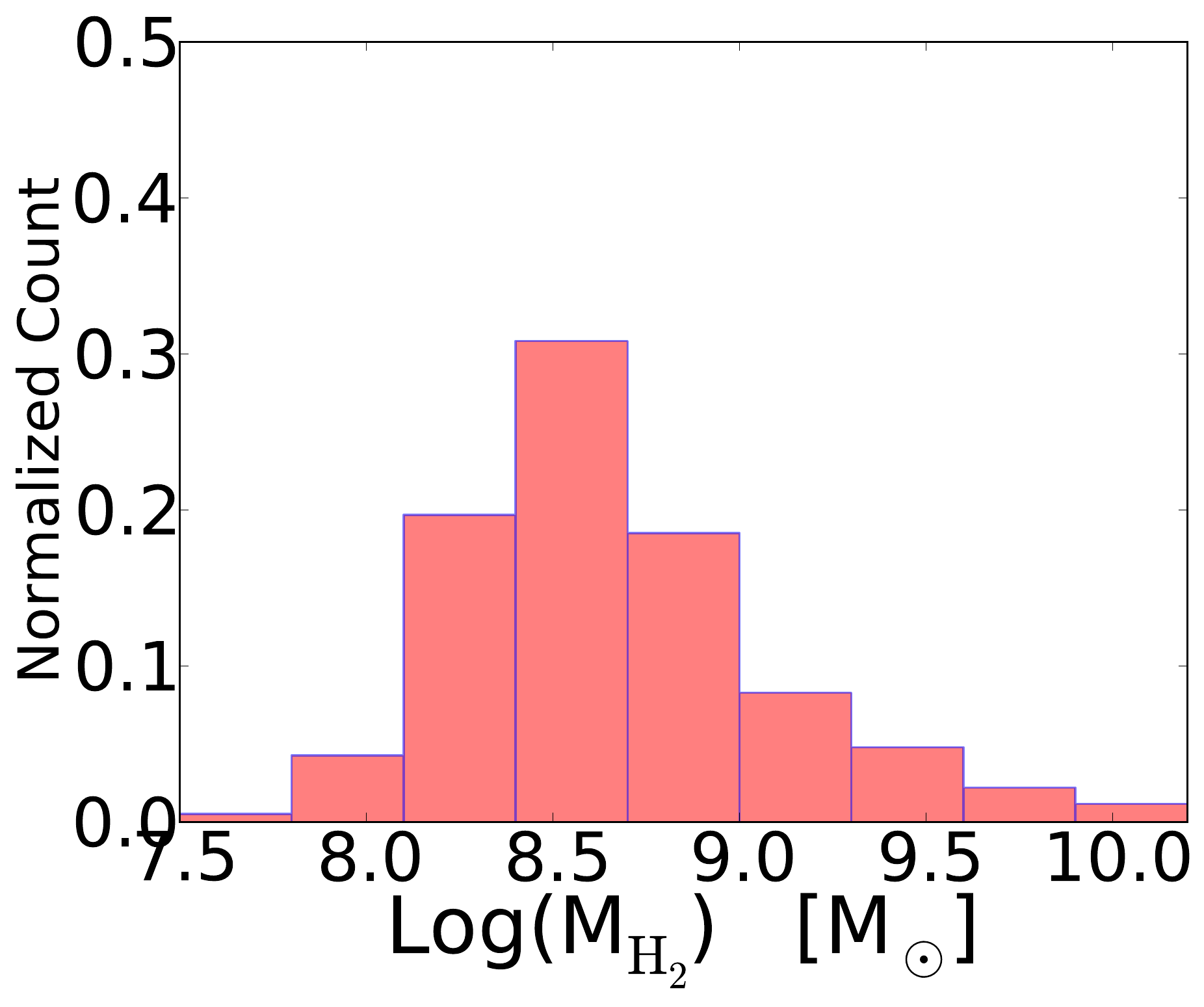}
\includegraphics[angle=0,width=0.32\linewidth]{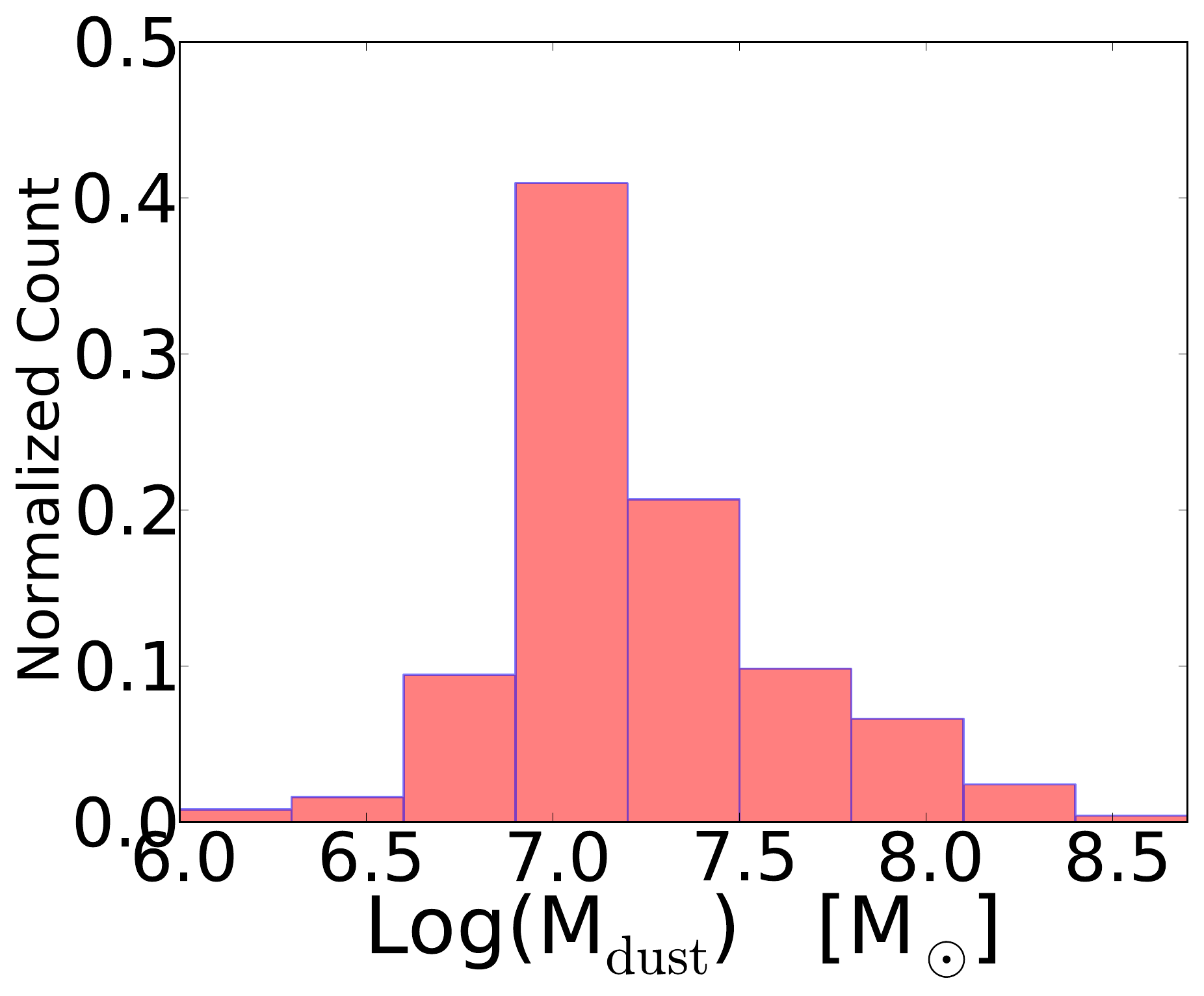}
\includegraphics[angle=0,width=0.33\linewidth]{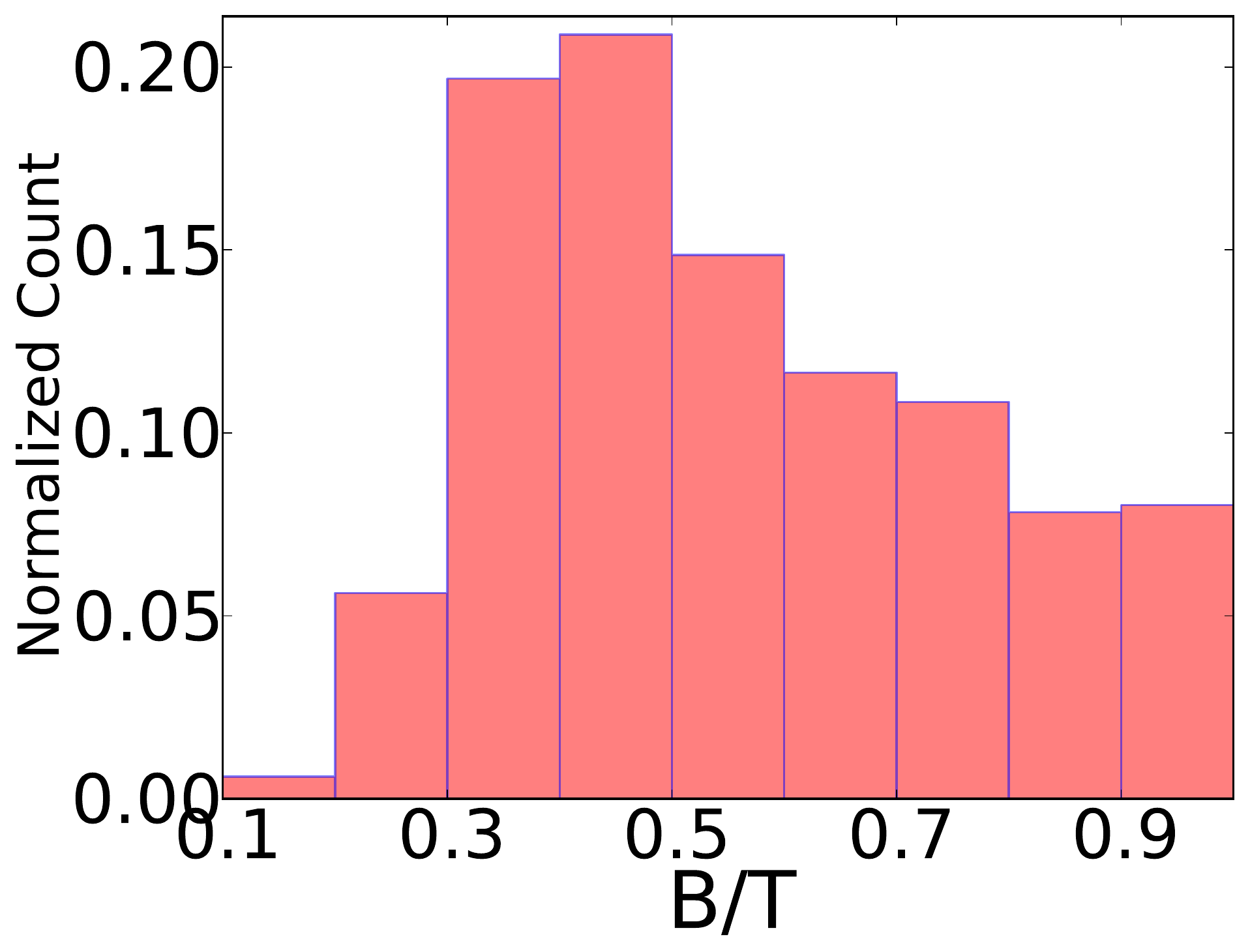}
}
\caption
{Distribution of the main physical properties of simulated
  galaxies. From left to right, and from top to bottom: stellar mass $M_{\star}$,
  total (cold and multi-phase) gas mass $M_{\rm gas}$, atomic gas mass
  $M_{HI}$, molecular gas mass $M_{H_2}$, dust mass $M_{\rm dust}$ and
  bulge-to-total ratio $B/T$.}
\label{fig:histograms}
\end{figure*}

We start from showing the range of variation of the main galaxy
physical properties, that is necessary to assess the statistical
properties of the sample.
Fig~\ref{fig:histograms} shows the stellar, atomic and molecular, dust
masses and $B/T$ distributions of the simulated galaxies at $z=0$. 
From the histograms we find that $\simeq 80$ per cent of the sample is composed of
galaxies with total stellar mass lower than $10^{10}$ M$_{\odot}$, and
$\simeq 90$ per cent of the atomic and molecular hydrogen is locked in those
galaxies. The median value for dust mass is ${\rm Log} (M\rm{_{dust}}/{\rm
  M}_\odot) \simeq 7$ and for $B/T \sim 0.5$.  

The range of values for these physical properties is realistic, but
there are a few points of tension with observations: the stellar mass function
is steeper than the observed one, though the statistics is poor to
assess the significativity of this disagreement; the few massive
galaxies ($M_* > 3\cdot10^{10}$ M$_\odot$) present in this volume
tend to be star-forming and with low $B/T$, and this can be ascribed to the
lack of AGN feedback in this code implementation.

\begin{figure*}
\centering{
\includegraphics[angle=0,width=0.49\linewidth]{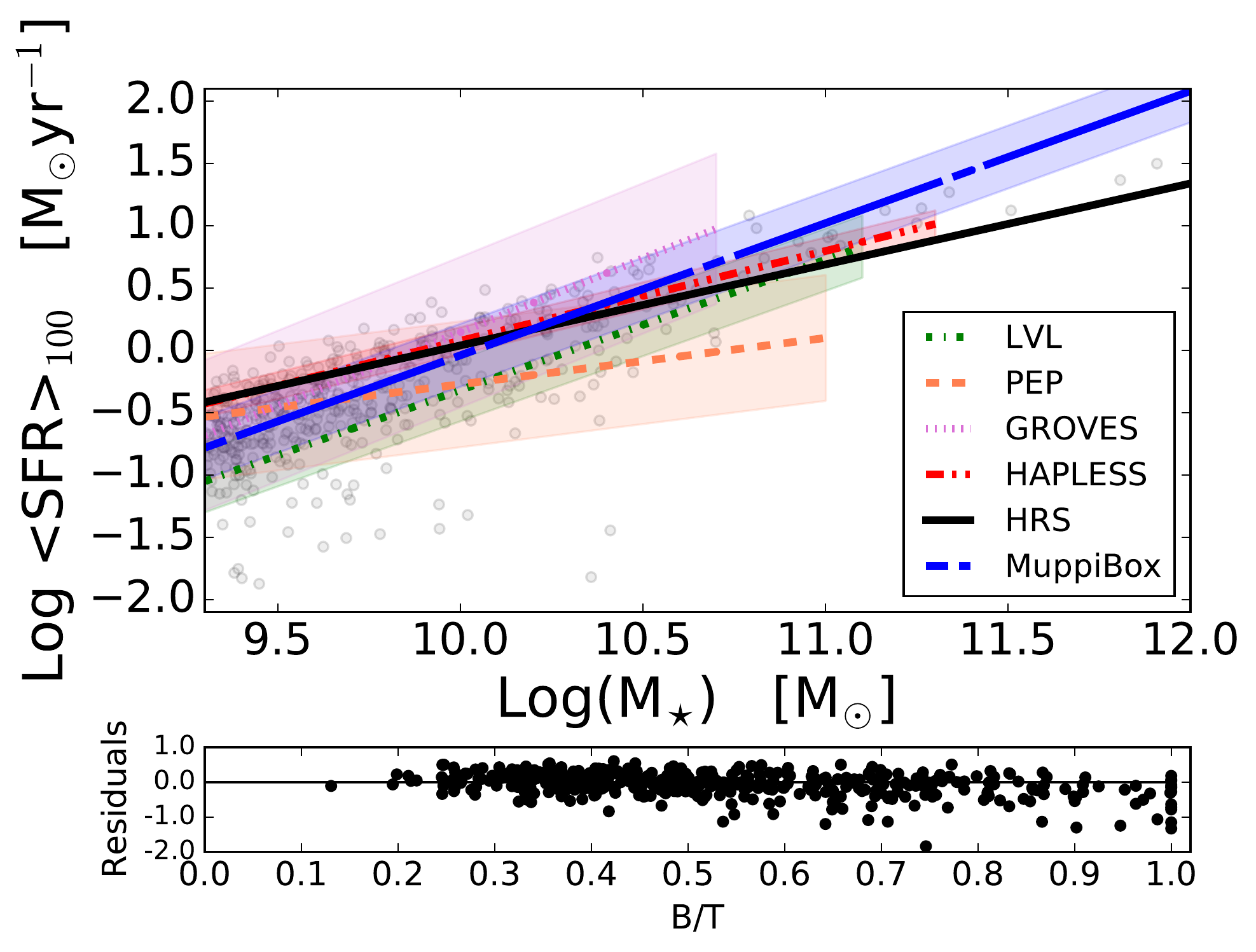}
\includegraphics[angle=0,width=0.49\linewidth]{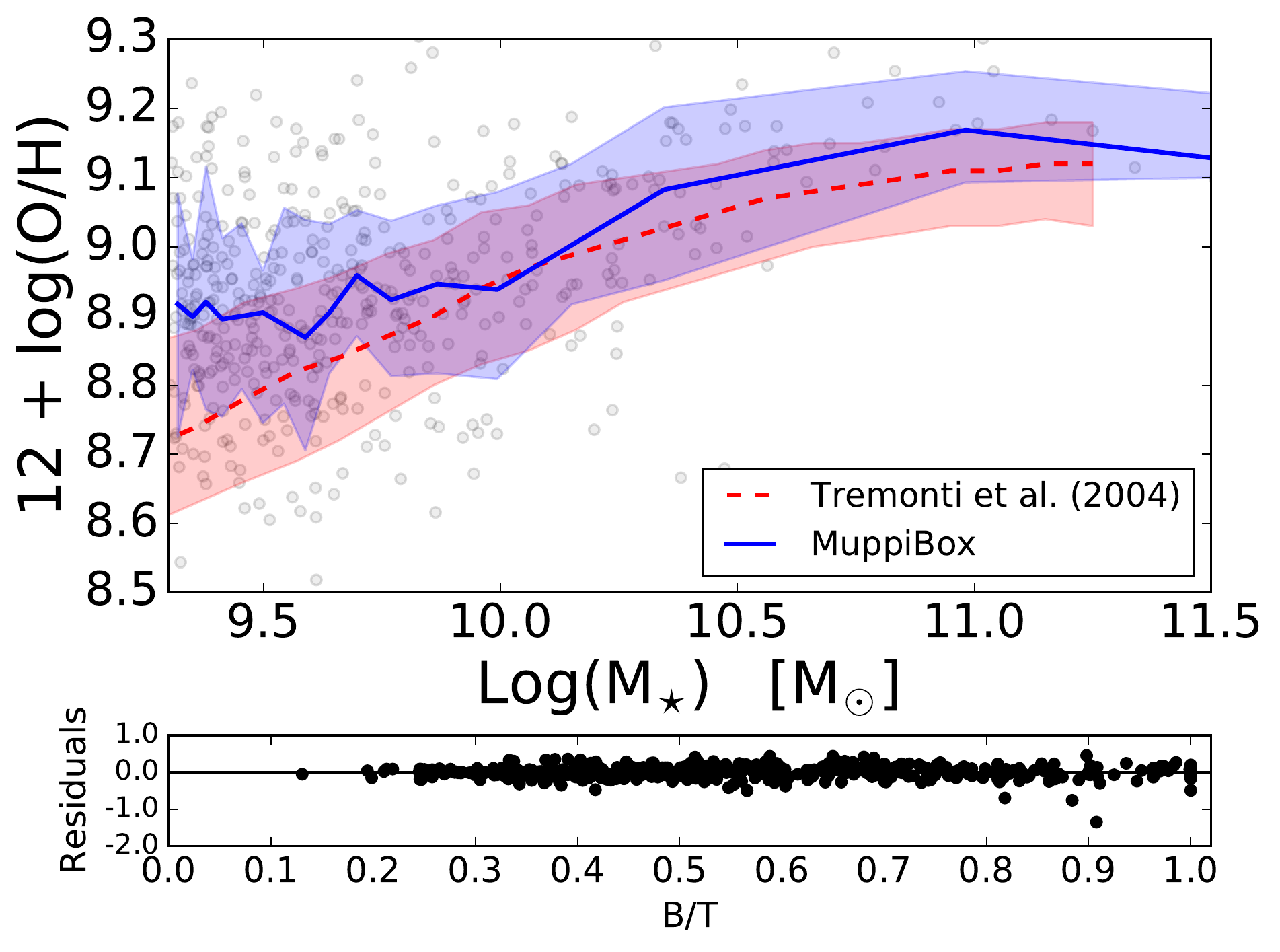}
}
\centering{
\includegraphics[angle=0,width=0.49\linewidth]{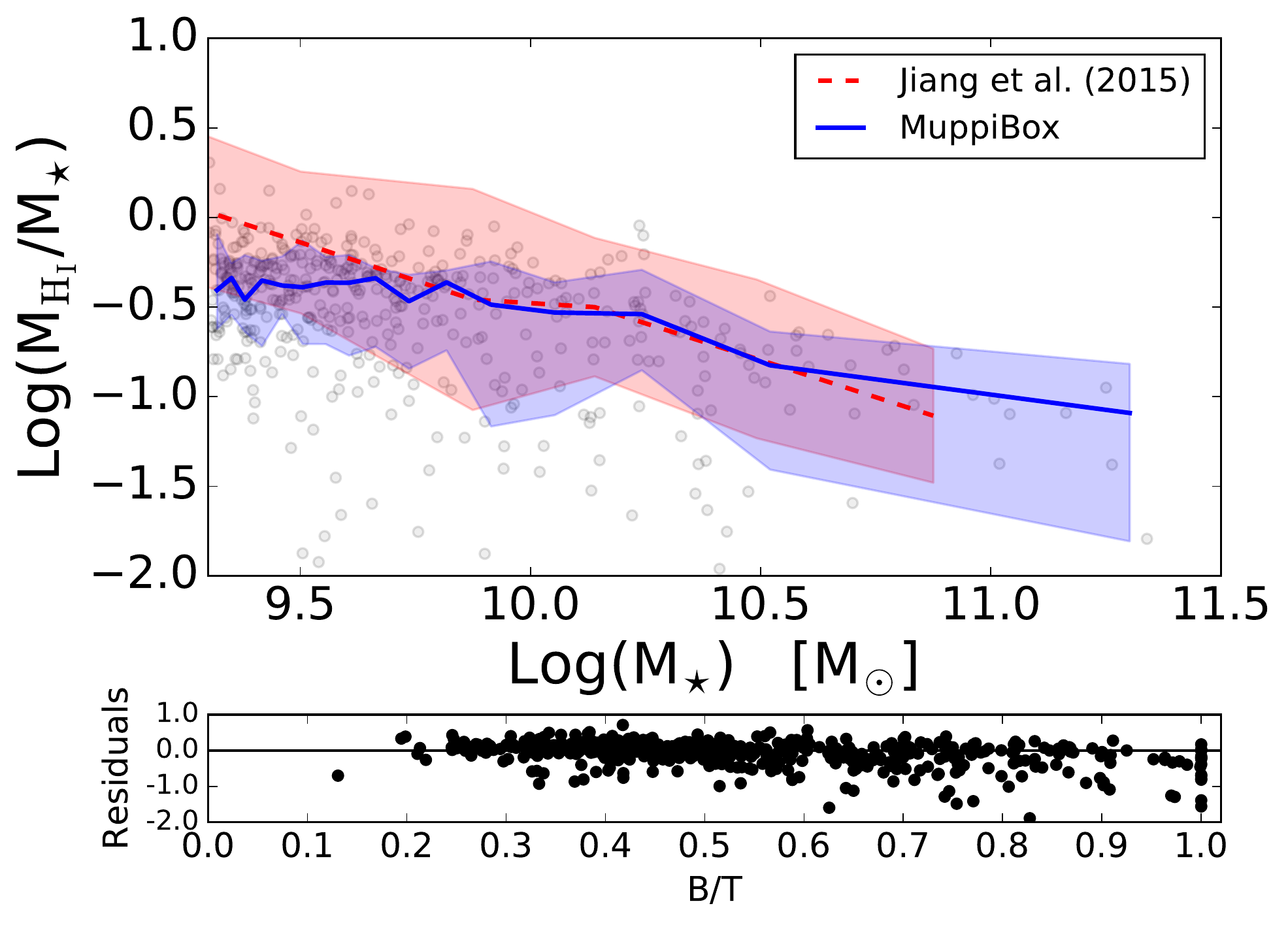}
\includegraphics[angle=0,width=0.49\linewidth]{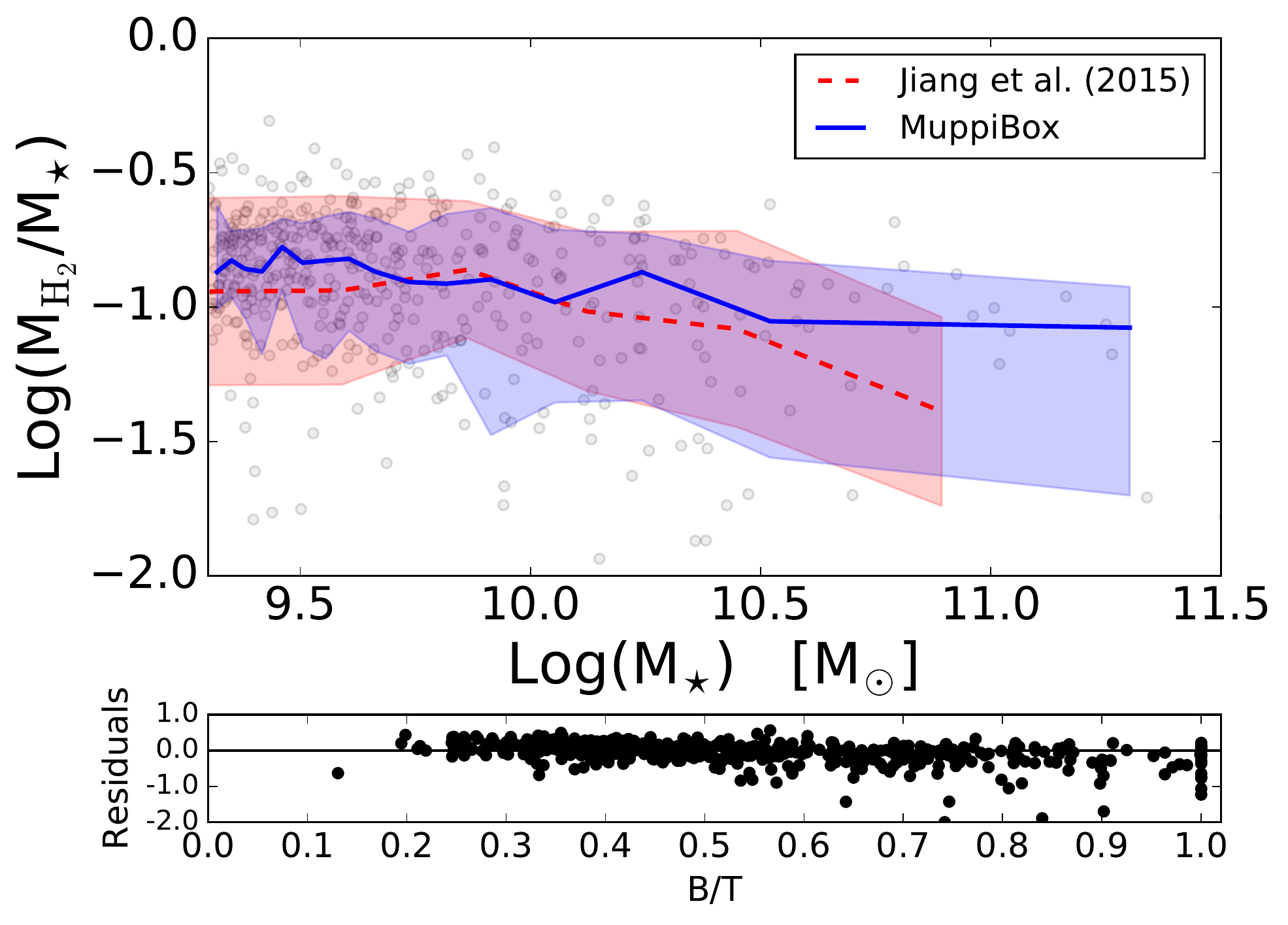}
}
\caption
{ Upper Left panels: main sequences, in the $M_\star-{\rm SFR}$
  plane, of star-forming galaxies. For LVL, PEP, Groves and HAPLESS
  samples, and for the MUPPIBOX simulated sample, main sequences were
  obtained by a linear regression (in the log-log plane) using the
  robust estimator of Cappellari et al. (2013). The plot reports the
  obtained regression as a line and the error on its normalization as
  a shaded area. The black line marks the HRS fit, that has been
  provided by Ciesla et al. (2014). Simulated galaxies are shown in
  the background as grey points. The lower sub-panel, in this and all other
  figures, gives for the MUPPIBOX sample the residual of each galaxy
  with respect to the main relation as a function of $B/T$; this is
  given to highlight possible dependencies on galaxy morphology. Upper
  right panels: distribution of galaxies in the $M_\star-12+\log(O/H)$
  plane. Grey points give simulated galaxies, dashed red and solid blue lines
  show the median value of Tremonti et al. (2004) and of the MUPPIBOX
  sample, filled regions contain 68 per cent of the data (16th and
  84th percentiles). Lower panels: distribution of galaxies in the
  $M_{HI}/M_\star-M_\star$ (left) or $M_{H_2}/M_\star-M_\star$ (right)
  plane. Grey points give simulated galaxies, dashed red and solid blue lines
  show the median value of Jiang et al. (2015), obtained using SMT,
  AMIGA-CO and COLD GASS samples, and of the MUPPIBOX sample, filled
  regions contain 68 per cent of the data (16th and 84th
  percentiles).}
\label{fig:relations}
\end{figure*}

To compare SFRs with observations, we should take into account the
range of stellar ages to which the SFR indicators used are sensitive,
with values going from a few to $\approx100$ Myr
\citep[e.g.][]{Kennicutt98,Calzetti_2005,Calzetti12}. There are two
ways to recover SFRs of simulated galaxies. One can take the
instantaneous SFR produced by the sub-resolution model at the output
time, or one can consider the number of star particles spawned by the
stochastic star formation algorithm in a given time interval. To give
an example, at the resolution of our simulation a SFR of 1 M$_\odot$
yr$^{-1}$ will produce one star particle of $1.34\cdot10^6$ M$_\odot$
each 1.34 Myr. As a consequence, a proper average of SFR would require
a time of order $\sim100$ Myr to properly suppress shot noise, the
problem being more severe at lower SFRs. We have verified that the
instantaneous and averaged SFRs show a very good correlation, with a
scatter that increases with decreasing age intervals used for the
average SFR. For this reason we use the largest age interval of 100
Myr to compute average SFRs, with the caveat that some indicators may
be sensitive to smaller age intervals.

The upper left panel of Figure~\ref{fig:relations} shows the {\em
  main sequence} of star-forming simulated galaxies
\citep[e.g.][]{Elbaz_2007,Elbaz_2011,Noeske_2007,Peng_2010} for the
MUPPIBOX sample. This has been obtained by a linear regression,
performed with the robust method of \cite{Cappellari_2013}, of the
simulated galaxies in the (log-log) $M_\star-{\rm SFR}$ plane,
shown as grey points. The fit is
reported as a blue line, the shaded area gives the 1-$\sigma$ error on
the normalization. 
The scatter of simulated galaxies around the main sequence is of order of 0.3 dex,
compatible with observations. The same fitting procedure has been applied to
the LVL, PEP,  Groves and HAPLESS samples, that are reported in the
same figure (in this case we do not show the single galaxies). In the
Figure, we also report the fit to the main sequence measured in the
HRS sample as a black line; in this case we do not show an uncertainty
on the relation. The lower sub-panel shows the residuals of simulated
galaxies with respect to the mean relation as a function of the galaxy $B/T$. This is shown to
highlight possible dependencies on galaxy morphology; in this case we
obtain that early-type galaxies tend to stay below the relation, as expected. The figure shows
an overall consistency of the main sequence of star-forming galaxies
between simulations and observations. The simulated relation is
however steeper than that found in some of the samples, especially
the HRS one, and this leads to some disagreement above
$10^{10}$ M$_\odot$. Indeed, all simulated galaxies in the sample have
significant SFR, no passive galaxy is found even at high masses. This excess of star
formation is a likely effect of the lack of AGN
feedback in this simulation. At the same time, the least massive
galaxies tend to be below the main sequence of observations (with the
exception of LVL survey) by  
$\sim0.5$ dex, a typical behaviour of dwarf galaxies in a LCDM model
\citep{Fontanot_2009a,Weinmann_2012}.

The upper right panel of Figure~\ref{fig:relations} shows the gas
metallicity of simulated galaxies as a function of $M_\star$. We also
report as a continuous line the median values in bins of stellar mass
chosen so as contain 30 galaxies, while the shaded area gives the 16th
and 84th percentiles. A similar style is used to report the
observational determination of \cite{Tremonti_2004}, relative to SDSS
galaxies at $z \sim 0.1$. For simulated galaxies, to derive the oxygen
abundance we compute the mass-weighted average of oxygen mass of
galaxy gas particles, and recast it in terms of $O/H$, assuming a
solar value of 8.69 \citep{Allende_2001}. 
Simulated galaxies show the
same trend as observations of increasing metallicity at larger stellar
mass, but we find a global offset of $\sim0.1$ dex. At large
stellar masses ($M_\star\ga10^{11}\ {\rm M}_\odot$), this is likely due
to the lack of quenching of cooling and star formation by AGN feedback,
while the offset is much less evident at the $10^{10}\ {\rm M}_\odot$ mass
scale (apart from a few high-metallicity outliers). Metallicities are
offset by up to 0.2 dex in the low-mass end of this relation.
The lower sub-panel shows again the residuals of simulated
galaxies with respect to the mean relation
as a function of the galaxy $B/T$; no trend is visible
in this case.

Next we consider the abundance of atomic and molecular gas, as a
function of stellar mass. We compare this to the determination of
\cite{Jiang_2015}, who combined the STM \citep{Groves_2015}, AMIGA-CO
\citep{Lisenfeld_2011} and COLD GASS
\citep[][]{Saintonge_2011,Catinella_2012} surveys.
The lower panels of Figure~\ref{fig:relations}
show the location of simulated galaxies in the
$M_{HI}/M_\star-M_\star$ (left panel) and $M_{H_2}/M_\star-M_\star$
(right panel) planes. As for the other panels of the same figure, 
we show simulated galaxies as grey dots, and the average of the
simulated (same binning as for the stellar mass--gas metallicity relation) and observed relations with 16th and
68th percentiles denoted as a shaded area. Lower panels show again
residuals of simulated galaxies versus $B/T$.
Besides a broadly good agreement in the stellar mass range from
$10^{10}$ to $10^{10.5}$ M$_\odot$, atomic
gas tends to be low in simulated galaxies, especially at small stellar
masses, where the relation flattens and the mass ratio saturates at
unity values, at variance with the observed relation. The relation
with molecular gas is reproduced in a much better way, with some
possible average overestimate.
Looking at residuals, we notice that many outliers with low gas masses have early type
morphologies; these objects may be absent in the observed sample 
due to selection.
Following \cite{Jiang_2015}, we computed Spearman's
correlation coefficients for these correlations, obtaining values of
$-0.40$ for $HI$ and $-0.23$ for $H_2$, to be compared with
$-0.72$ and $-0.48$ given in that paper.

This analysis shows that the simulated galaxy sample provides objects with
properties that are broadly consistent with observations.
This is in line with the findings of other simulation projects, like the
Illustris \citep{Vogelsberger_2014} and EAGLE simulation
\citep{Schaye_2015}, where good agreement of simulated galaxies with
observations at $z=0$ extends to molecular and atomic gas, obtained in
post-processing \citep{Genel_2014,Lagos_2015,Bahe_2016}.
However our massive galaxies show a lack of proper quenching and 
small galaxies are too massive, passive, 
metallic and with low atomic gas content.

\subsection{Average SEDs from UV to FIR}
\label{section:calibrating}

\begin{figure*}
\centering{
\includegraphics[angle=0,width=0.49\linewidth]{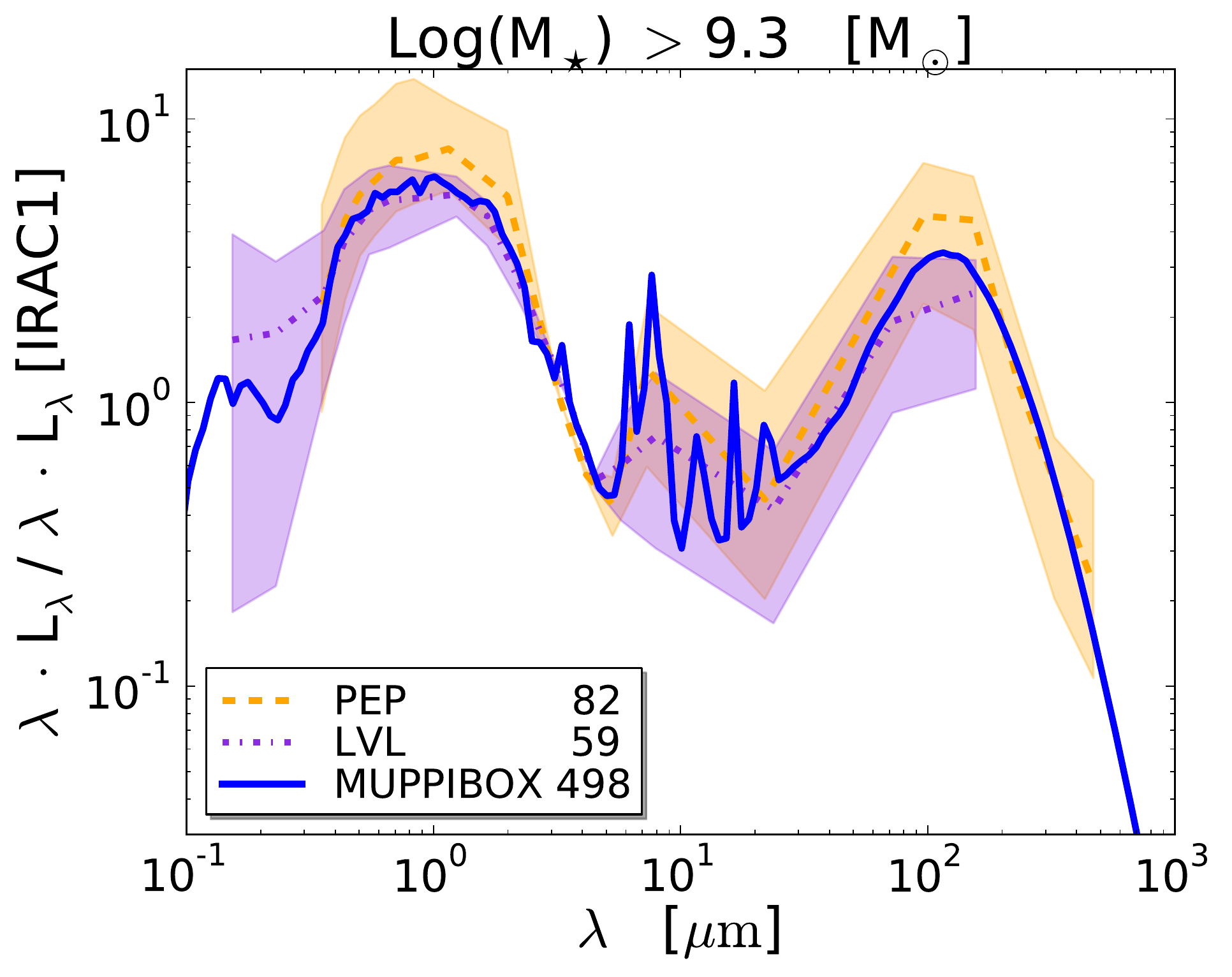}
\includegraphics[angle=0,width=0.49\linewidth]{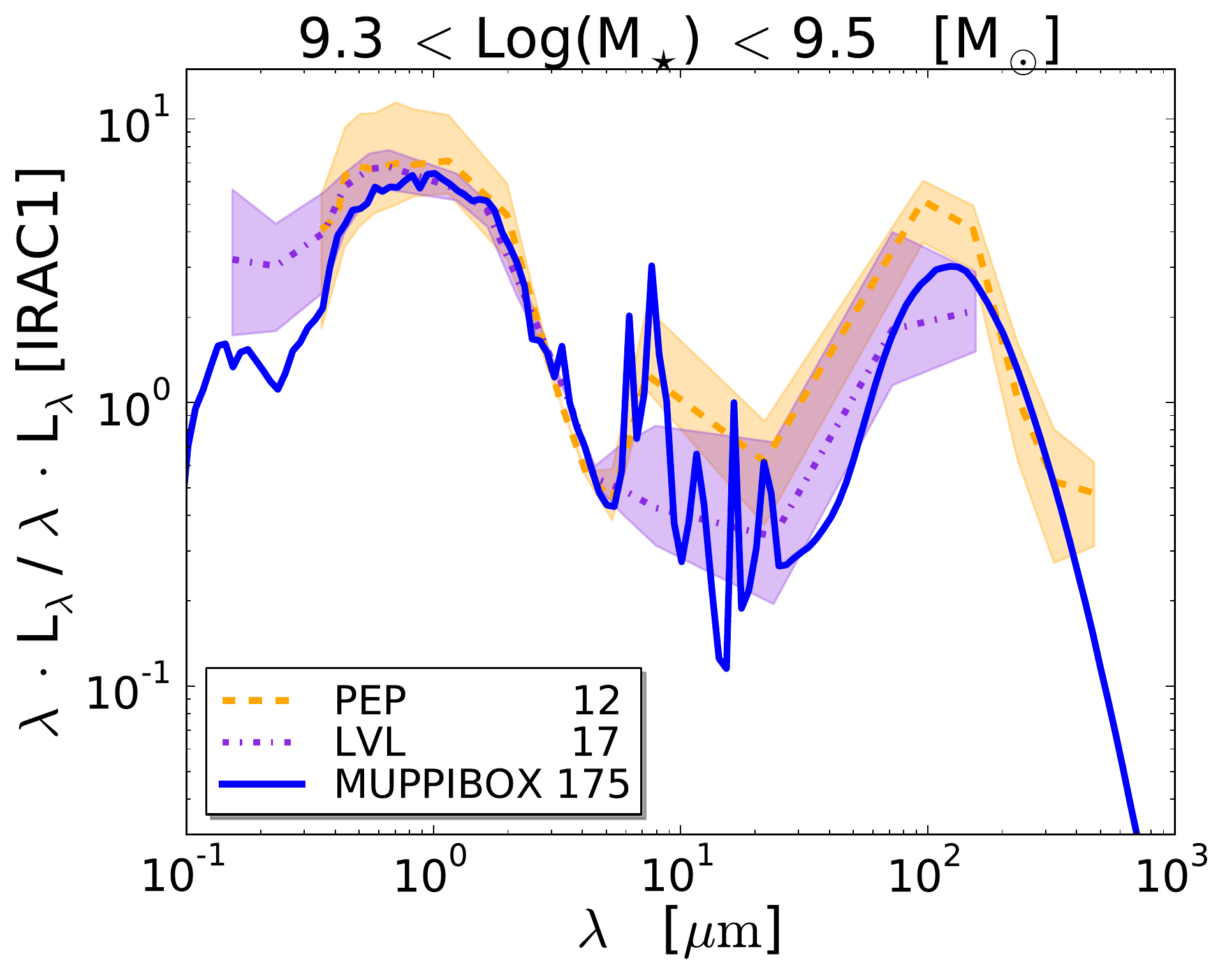}
}
\centering{
\includegraphics[angle=0,width=0.49\linewidth]{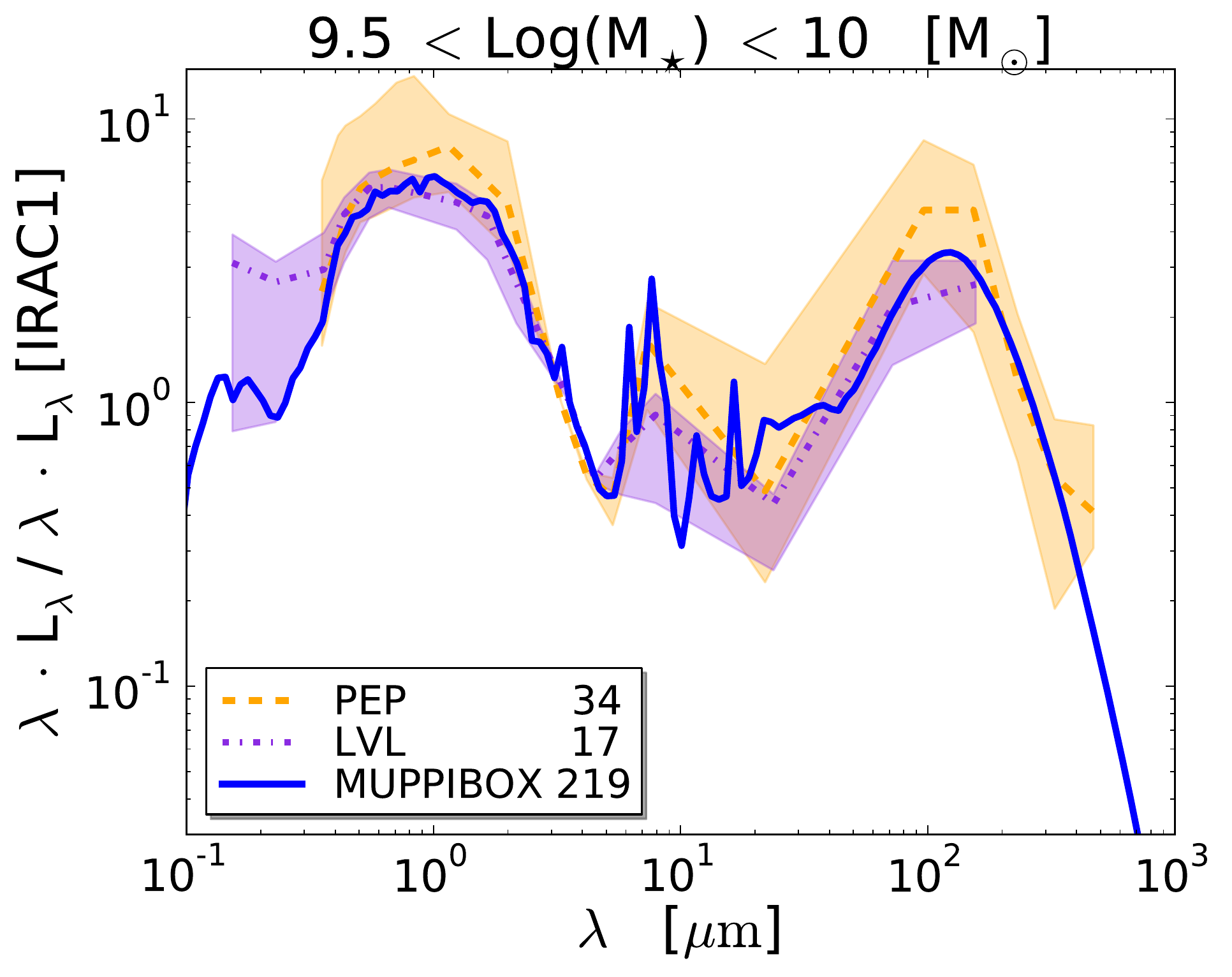}
\includegraphics[angle=0,width=0.49\linewidth]{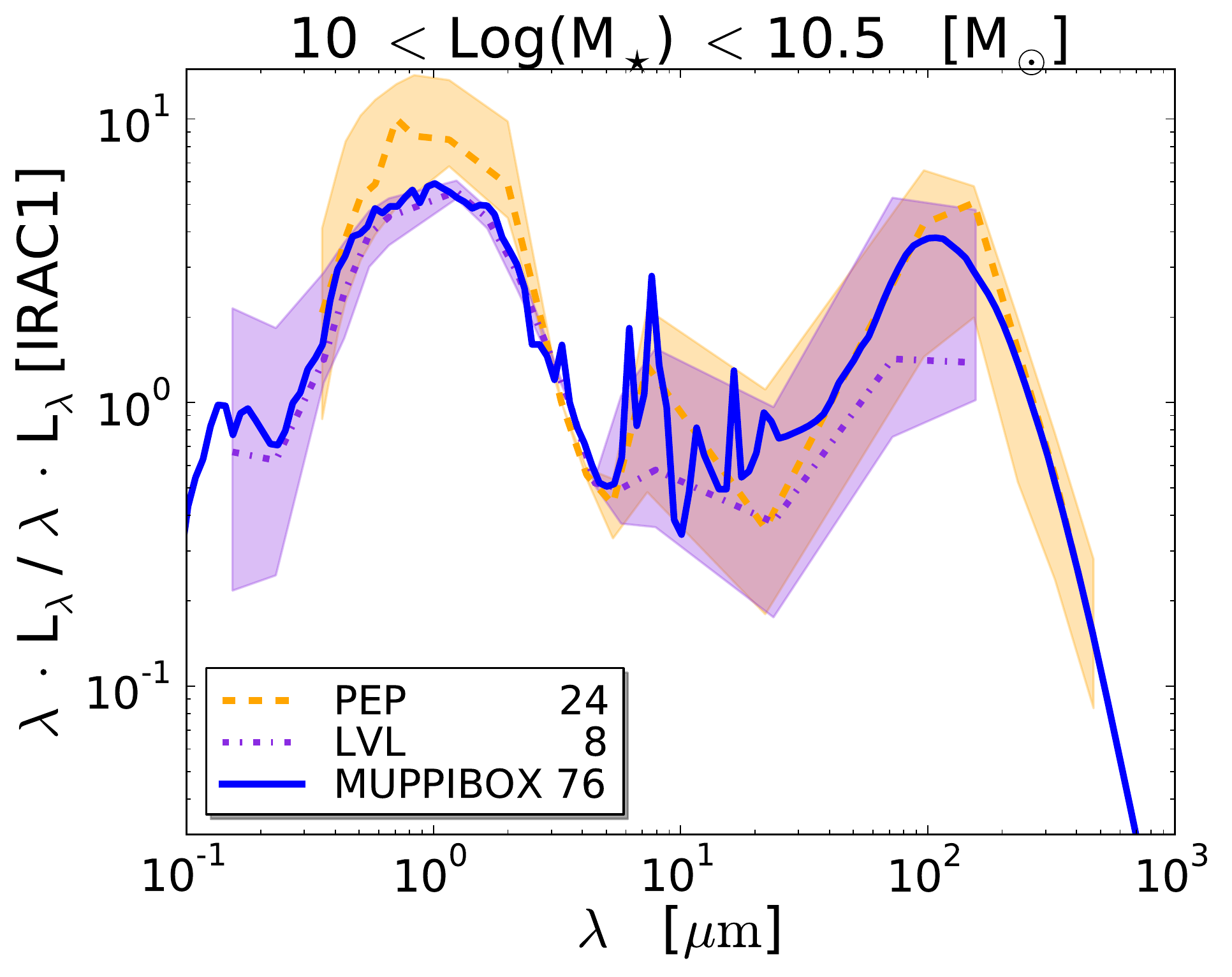}
}
\caption
{Median SEDs ($\lambda L_\lambda$) of galaxies with
  $M_\star>2\cdot10^9$ M$_\odot$ (upper left panel) and in three bins
  of stellar mass as indicated above each panel, normalized to unity at the 3.6
  $\mu$m IRAC1 band. Blue continuous line: simulated galaxy SEDs,
  obtained with {\gra} using best-fit parameters $t\rm{_{esc}}=3$
  Myr, and $M\rm{_{MC}}=10^{6}$ M$_{\odot}$. They are compared with
  the median SED of LVL (dot-dashed purple line) and PEP (dashed orange
  line) samples. Coloured areas give their 1$\sigma$ uncertainty,
  obtained using the 16th and 84th percentiles.
  Every plot reports the number of galaxies in the MUPPIBOX, PEP and LVL samples.}
\label{fig:calibration_best_fit}
\end{figure*}

As explained in Section~\ref{section:g3d}, radiative transfer in
  a dusty medium, performed by {\gra}, requires to specify the escape
  time of stars from MCs, $t_{\rm esc}$, and the MC optical depth
  through their putative mass $M_{\rm MC}$ (this enters in the
  computation only through MC mass surface density, and $R\rm_{MC}$ is
  fixed to 15 pc). It has been shown several times
  \citep{Silva_1998,Granato_2000,Granato_2015} that GRASIL is able to
  reproduce normal star-forming galaxies at low redshift by assuming
  $t\rm{_{esc}}$ in the range of $2-5$ Myr and $R_{\rm MC}\sim15$ pc
  for a $M\rm{_{MC}}=10^{6}$ M$_{\odot}$ cloud. We thus adopt
  $t\rm{_{esc}}=3$ Myr and $M\rm{_{MC}}=10^{6}$ M$_{\odot}$. In order
  to assess the impact of variations of these parameters on the
  resulting SEDs, we proceed as follows. We compute galaxy SEDs using
  $t_{\rm esc}=2$, $3$, $4$ and $5$ Myr, and $M_{\rm MC}=10^5$, $10^6$
  and $10^7$ M$_\odot$. We divide the sample into three bins of
  stellar mass: $9.3-9.5$, $9.5-10$ and $10-10.5$ in ${\rm Log}
  M_\star/{\rm M}_\odot$, excluding the few most massive galaxies that
  are affected by lack of quenching. For each mass bin and for each
  choice of the two parameters we compute average SEDs, and compare
  them with the average SED of LVL and PEP samples, subject to the
  same mass limit. Details of these analyses are reported in
  Appendix~\ref{appendix:calibration}. This analysis shows that
  results are rather robust for variations of $M_{\rm MC}$, while
  $t_{\rm esc}$ has a sizable effect in the MIR and, to a lesser
  extent, in the FUV. The adopted values give very good results in
  many mass bins, confirming their validity.
  We have also verified that normalizing the SEDs to their
  total IR luminosity, from 8 to 1000 $\mu$m, gives very similar results.

Having set {\gra} parameters, we can now address the agreement
of predicted galaxy SEDs with observed ones.
Figure~\ref{fig:calibration_best_fit} shows the average SED of
galaxies in the simulated sample, normalized to unity at $3.6 \mu$m,
the Spitzer/IRAC1 band, compared with the median LVL and PEP SEDs.
The number of objects is reported in each panel.
Coloured areas give the scatter in the observed SEDs, quantified by
the 16th and 84th percentiles. The four panels give the average SED
for the whole sample (upper-left panel) and for three bins of stellar
mass, as indicated above each panel. The LVL and PEP samples are
selected in a different way, based on optical or FIR observations in
the two cases and their SEDs show some differences, though they are
generally compatible within their scatter. The agreement of simulated
and observed SEDs is very good at all stellar masses. In the
  upper-left panel, the median SED of simulated galaxies lies always
  between the two observational medians. Even the PAH region is found
  within the observed range, while some underestimate is appreciable
  in the FUV, though the median curve is still within the 1-$\sigma$
  observed range. Considering the average SEDs in bins of stellar
  mass, tension is visible in the smallest stellar mass bin; this is
  the stellar mass range for which discrepancies with observations were noticed in
  Section~\ref{section:global}. Here the average SED is underestimated
  in the range from 30 to 100 $\mu$m and in the UV, where it lies on
  the 16th percentile. The agreement is much better in the two higher
  mass bins: some underestimation in the UV bands is still present in
  the middle stellar mass bin, while the galaxies in the most massive
  stellar mass bin agree very well with observation, with only some
  overestimation limited to $\sim20-30\ \mu$m, a range that is very
  sensitive to the exact value of the $t_{\rm esc}$ parameter (see
  Appendix~\ref{appendix:calibration}).

\begin{figure}
\centering{
\includegraphics[angle=0,width=\linewidth]{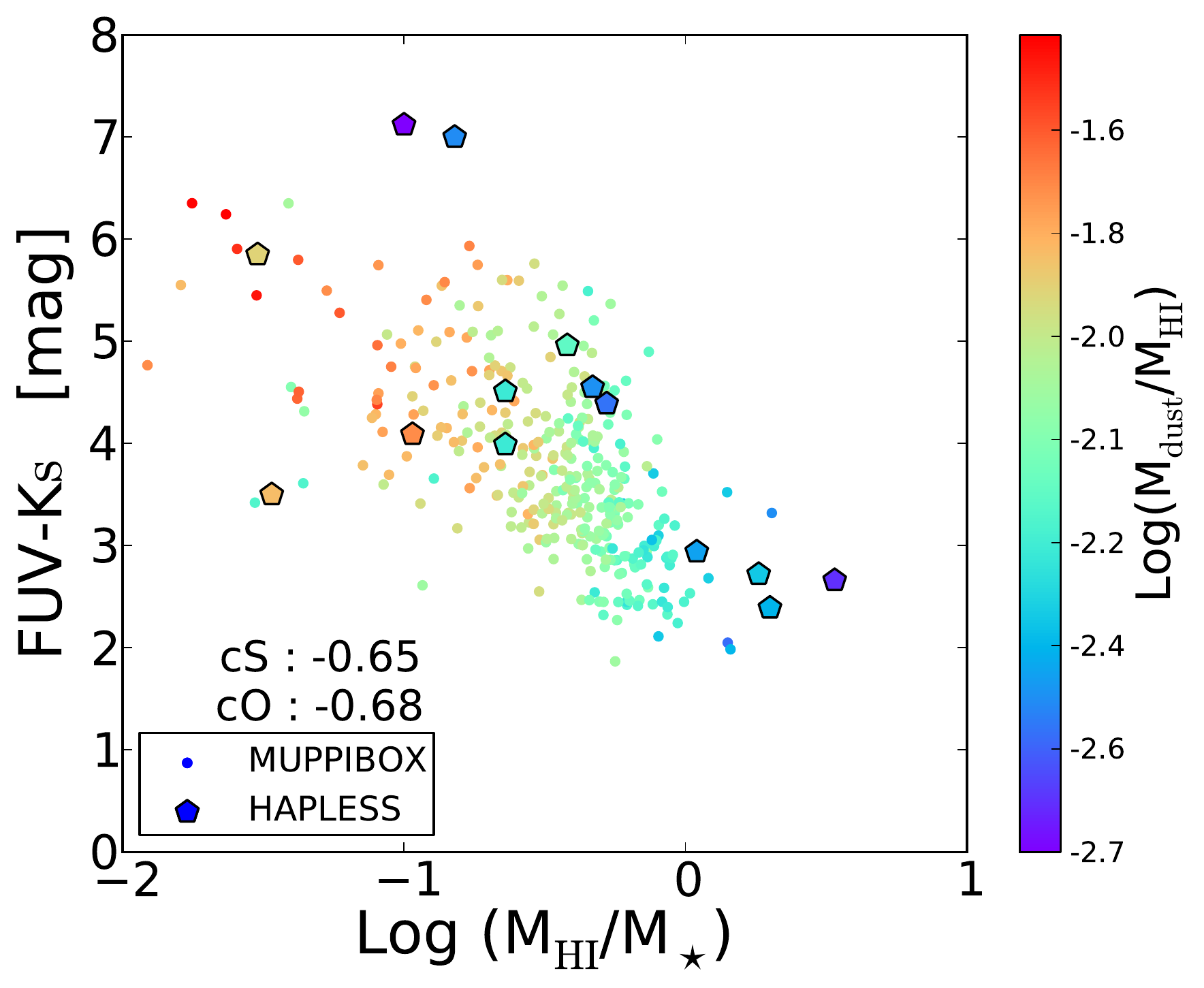}
}
\caption
{$FUV-K_s$ colour versus gas fraction $M_{HI}/M_\star$; points are
  colour-coded with $M_{\rm dust}/M_{\rm HI}$. cS and cO show the Spearman
  coefficients for MUPPIBOX and HAPLESS samples, respectively.}
\label{fig:FUVcolor}
\end{figure}

To understand the origin of the underestimation of UV light, we address the
$FUV-K_s$ colour and correlate it with galaxy gas richness $M_{HI}/M_\star$ and
metallicity to understand whether FUV reddening is caused by unproper
SED synthesis or simply by the higher metallicities of galaxies of low mass.
Because the LVL sample does not provide gas or dust masses, we use the
HAPLESS sample; we restrict ourselves to $HI$ gas mass because HAPLESS
does not give molecular masses, and we use $M_{\rm dust}/M_{\rm HI}$ as a proxy for
metallicity. 
Figure~\ref{fig:FUVcolor} shows that observed and simulated
galaxies lie on the same broad relation of $FUV-K_s$ versus gas
richness $M_{HI}/M_\star$, with galaxies with lower metallicities being in
general more gas-rich and less attenuated. The figure also reports the
Spearman correlation coefficient for the two relations, for MUPPIBOX
(cS) and HAPLESS (cO) galaxies, that turn out to be very similar. This
test shows that the same $FUV-K_s$ colours are obtained for galaxies with the
same gas richness, so the difference in the average SEDs
is induced by physical differences in input galaxy properties.

\begin{figure*}
\centering{
\includegraphics[angle=0,width=0.33\linewidth]{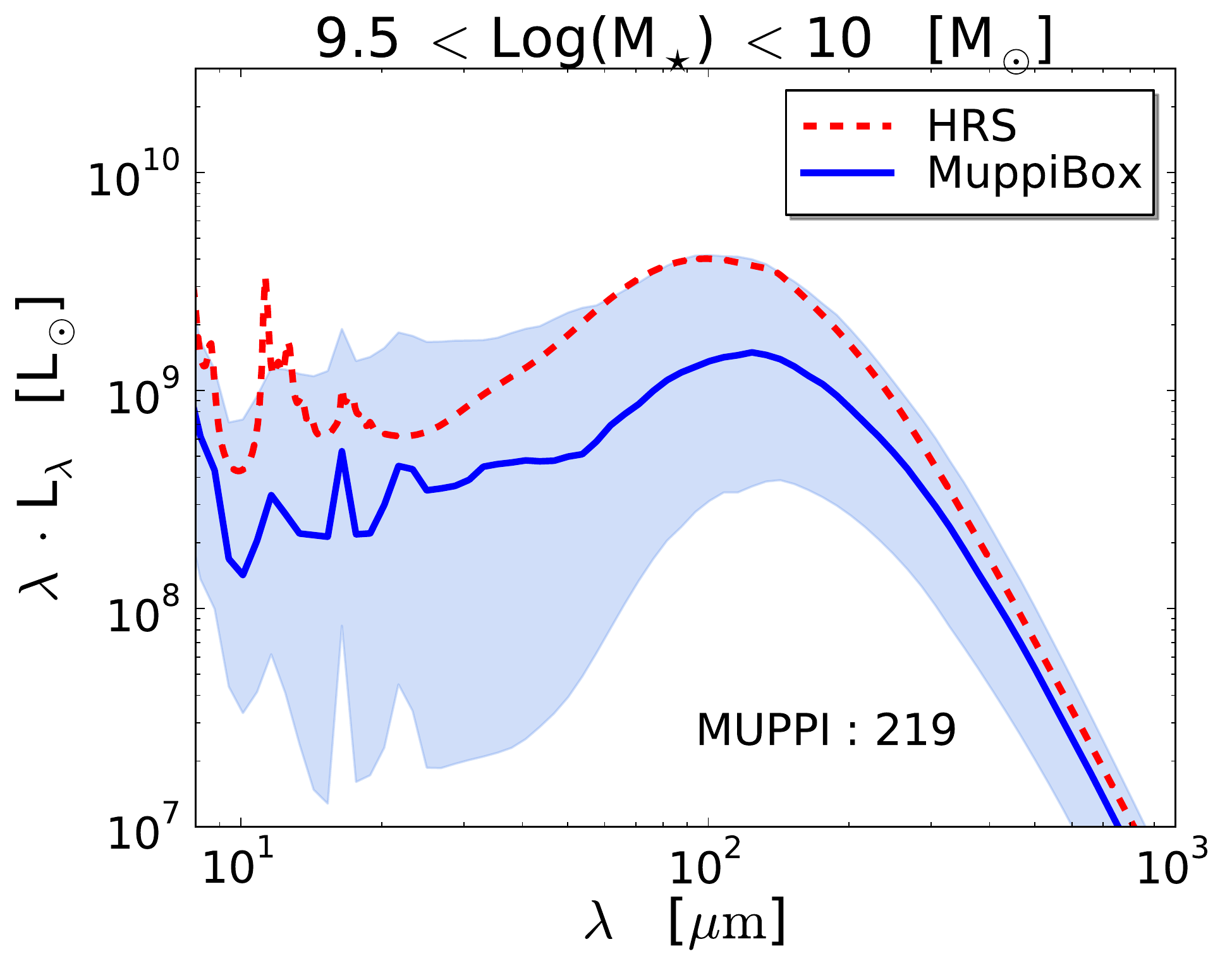}
\includegraphics[angle=0,width=0.33\linewidth]{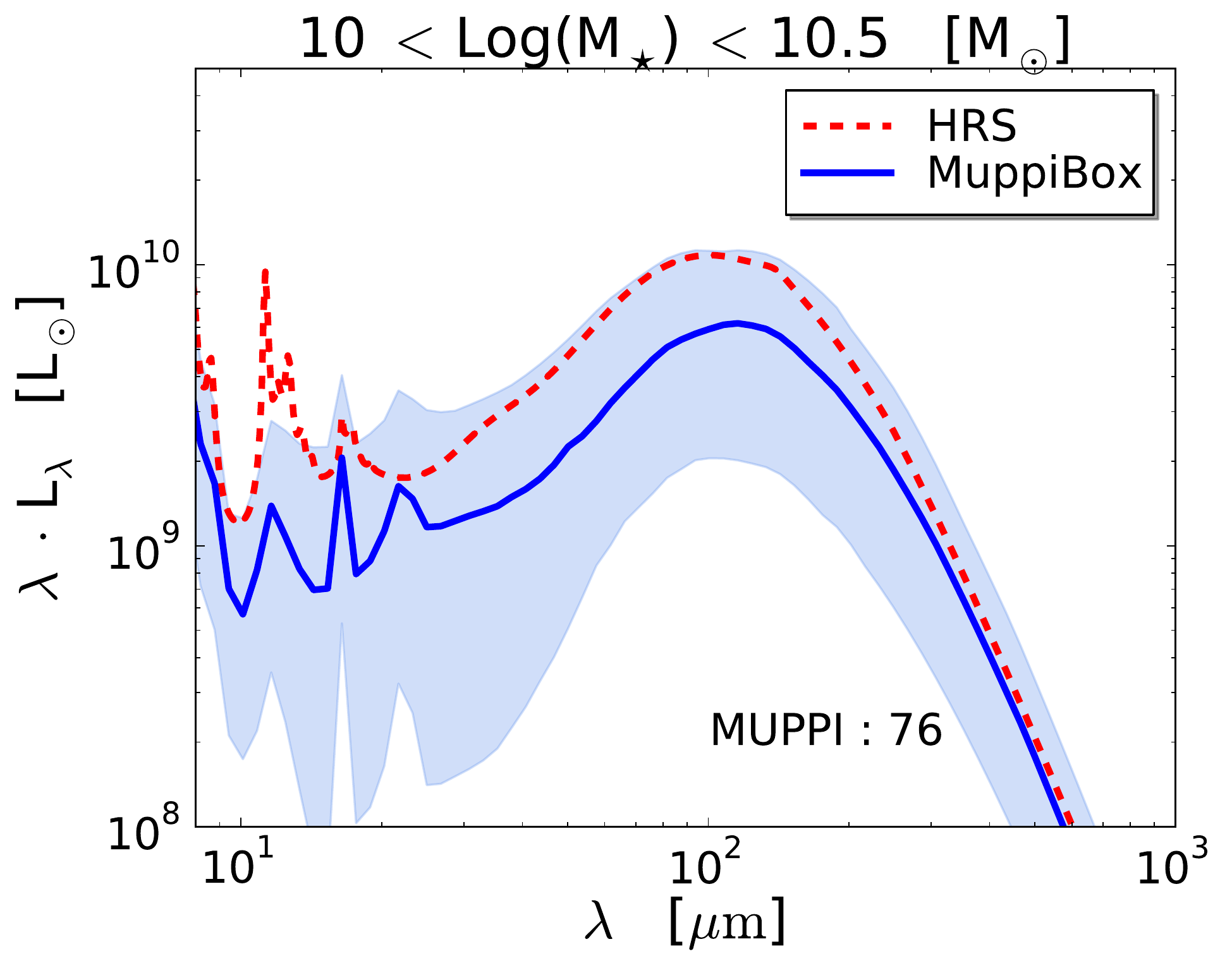}
\includegraphics[angle=0,width=0.33\linewidth]{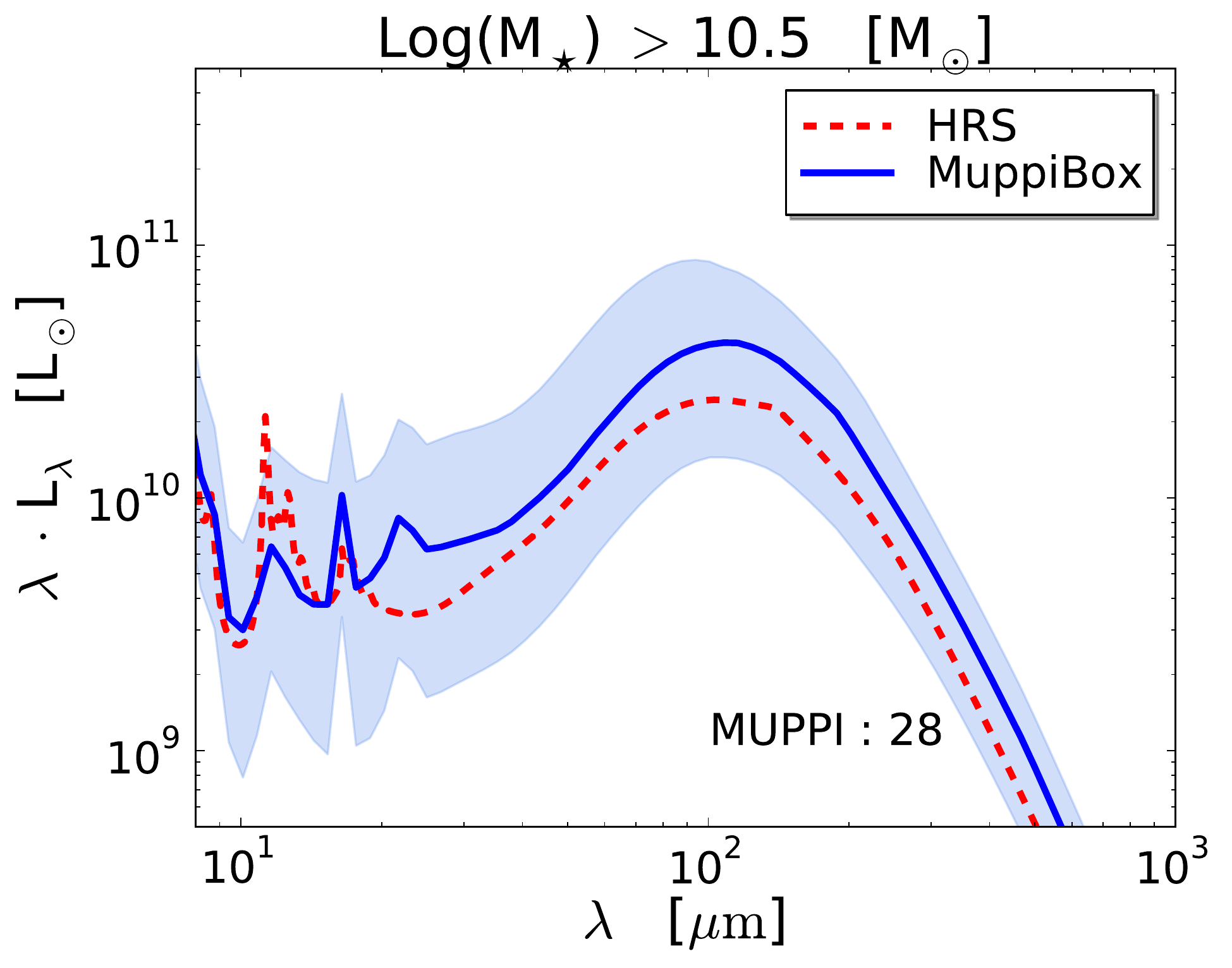}
}
\centering{
\includegraphics[angle=0,width=0.33\linewidth]{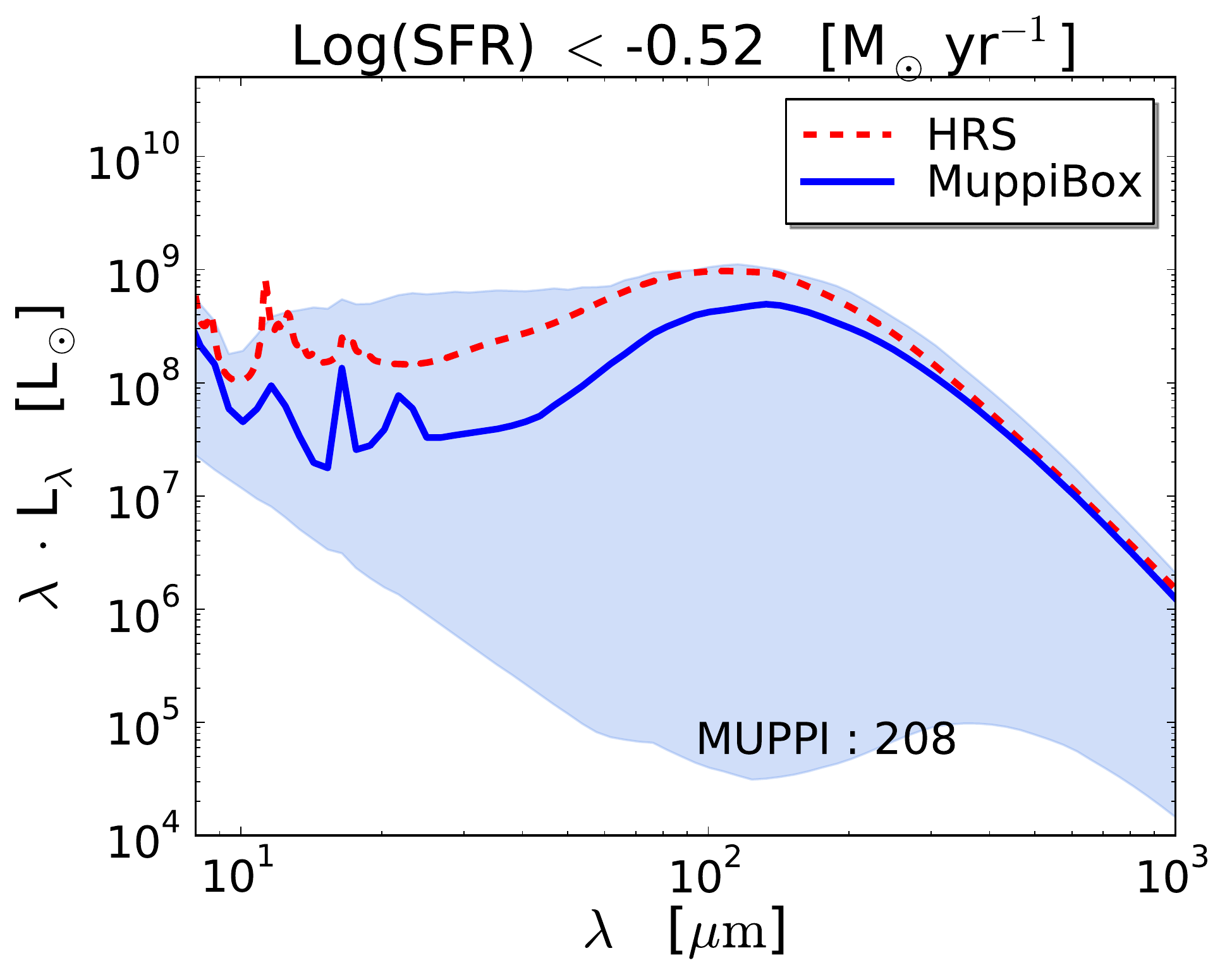}
\includegraphics[angle=0,width=0.33\linewidth]{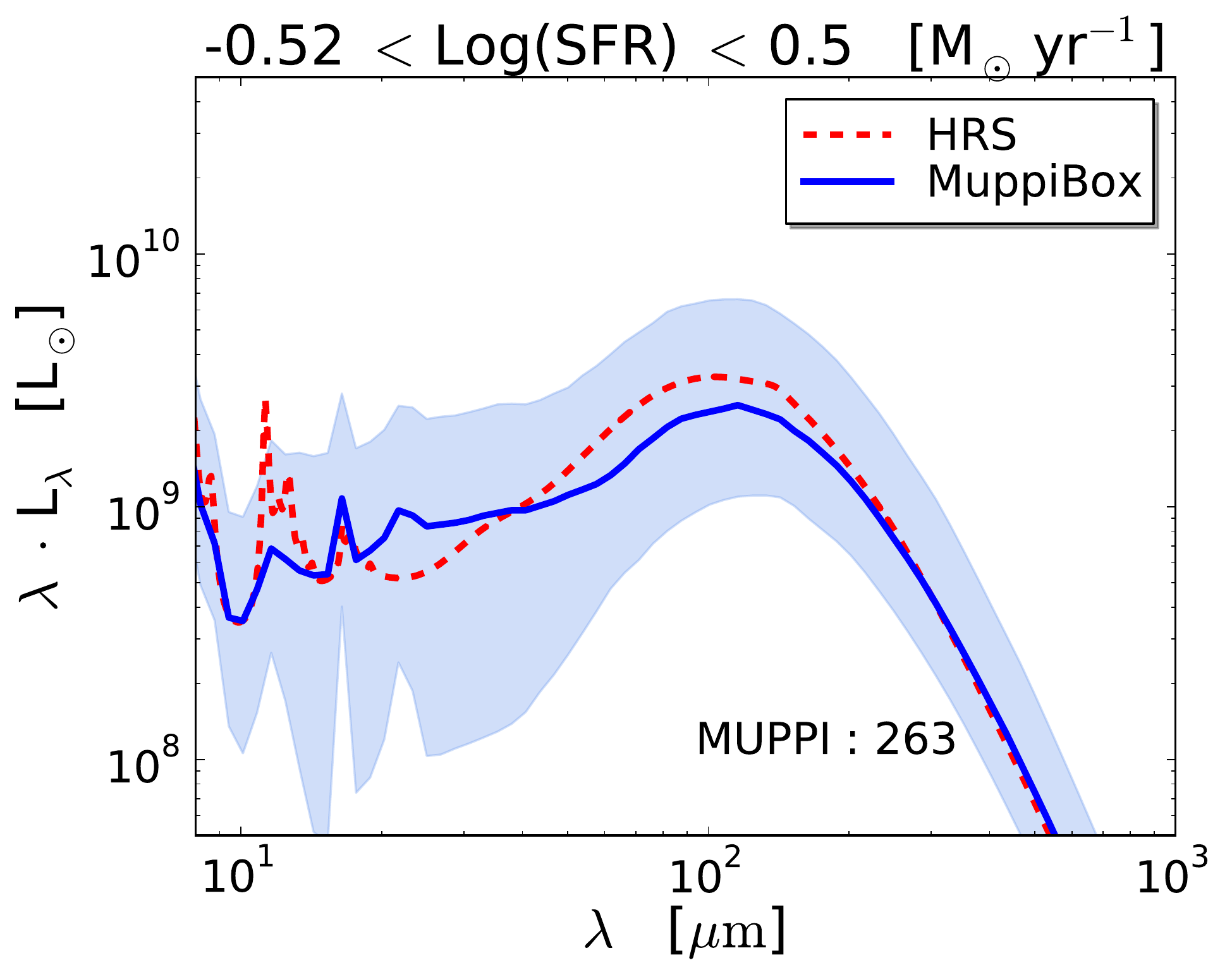}
\includegraphics[angle=0,width=0.33\linewidth]{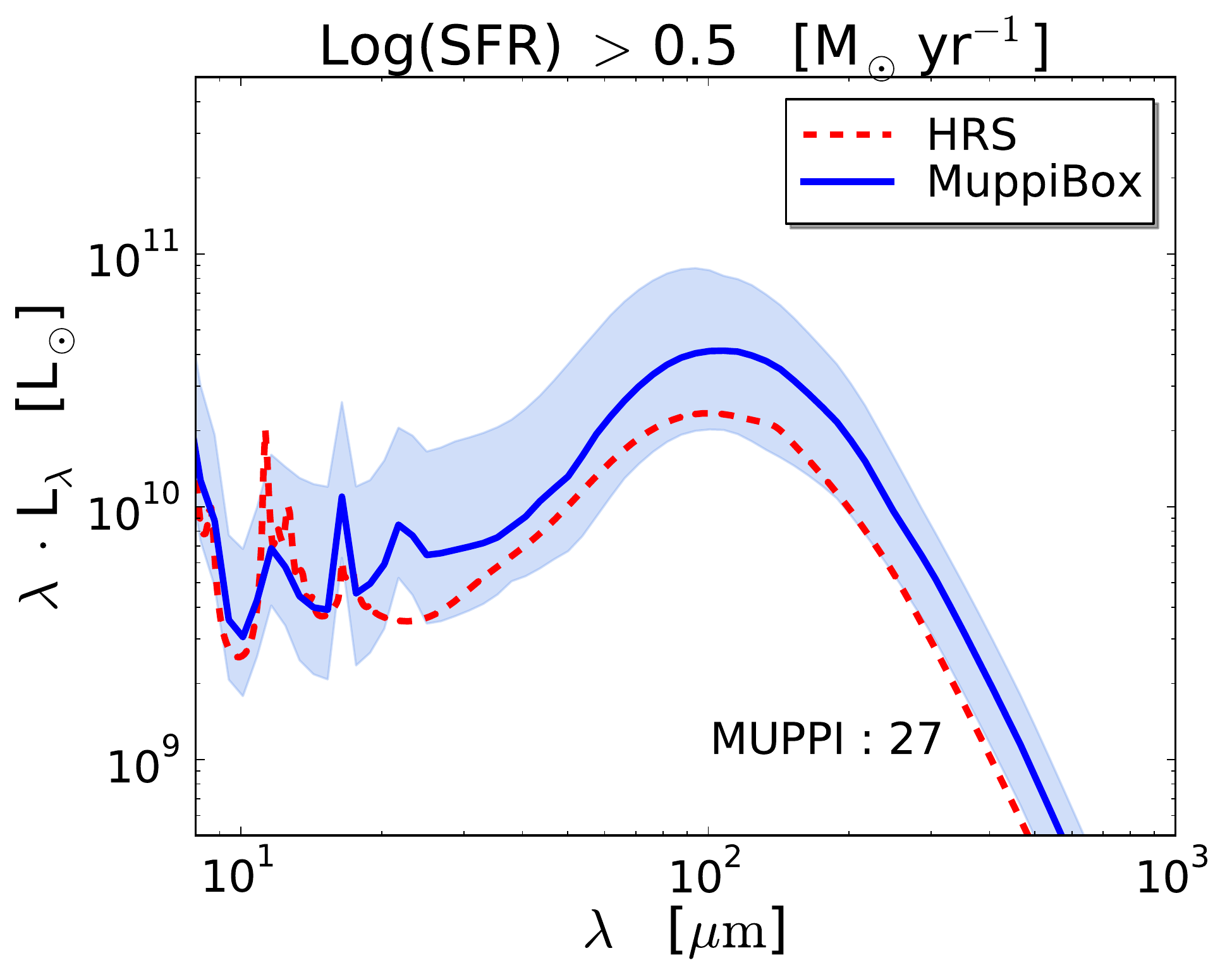}
}
\caption
{Comparison between median SEDs of simulated galaxy sample (blue
  continuous lines) and HRS (red dashed lines) for galaxies in various
  bins of stellar mass (first row) or SFR (second row). 
  The blue filled region marks the $1\sigma$
  uncertainty of the simulated sample, defined by the 16th and 84th
  percentiles. Above each panel we report the selection criterion used
  to define the subsample, while the number of simulated galaxies
  (MUPPI [number]) is reported in each figure.}
\label{fig:hrs_sed}
\end{figure*}

We further test our SEDs in the range from $8$ to $1000\ \mu$m by
comparing them to the highly complete HRS sample, this time checking
SED normalizations and not only their shapes. For this comparison we
use the results of \cite{Ciesla_2014}, who computed average SEDs for
subsamples binned by stellar mass, SFR, dust mass, metallicity and
morphological type. We show here results for bins in stellar mass and
SFR; we recall that the stellar mass limit of HRS is $\sim10^9$
M$_\odot$, a factor of $\sim 2$ lower than our resolution limit, so
the comparison with SEDs binned in stellar mass will be limited to
$M_\star>2\cdot10^9$ M$_\odot$, while the lower SFR bins will be
affected by this mild mismatch in limiting mass.
Figure~\ref{fig:hrs_sed} shows the average SEDs for three bins of
stellar mass (upper panels) and of SFR (lower panels); sample
definition is given above each panel. Here the blue continuous line gives
the median SED of the simulated sample; this time we report the
scatter of simulated galaxies, computed using the 16th and 84th
percentiles. HRS SEDs are given by red dashed lines;
\cite{Ciesla_2014} do not give scatter for their SEDs.
The comparison in bins of stellar mass (upper row of panels in the
figure) confirms the trend of Figure~\ref{fig:relations} of an
underestimate of SFR (and thus of IR luminosity) for small galaxies:
the normalization at the FIR peak is down by a factor of three for the
smallest galaxies (from $9.5$ to $10$ in ${\rm Log} M_\star/{\rm
  M}_\odot$), of two for the intermediate mass bin (from $10$ to
$10.5$), while a modest overestimate is present for the largest
galaxies. SED shapes are very similar to the observed ones,
peak positions are closely reproduced with the exception of the
smallest galaxies where dust is colder; we will show below that this
is a consequence of the lower level of star formation. The
scatter of model galaxies, that was not shown in
Figure~\ref{fig:calibration_best_fit} for sake of clarity, 
is somewhat higher than what found in observations, 
especially at smaller masses.

When galaxies are divided in bins of SFR (second row in
Figure~\ref{fig:hrs_sed}) differences in normalization are much lower,
while the change in SED shape is broadly reproduced by {\gra}. This shows
that SFR is the main driver of IR light. The galaxies with lowest SFR
are those where the agreement is worst, but this is the bin that is
most populated by the smallest galaxies and most affected by the
different stellar mass limit. We also find that scatter is dominated
by galaxies with low SFR.

\subsection{Correlation of IR luminosity with galaxy physical properties}
\label{section:IRluminosity}

\begin{figure}
\centering{
\includegraphics[angle=0,width=\linewidth]{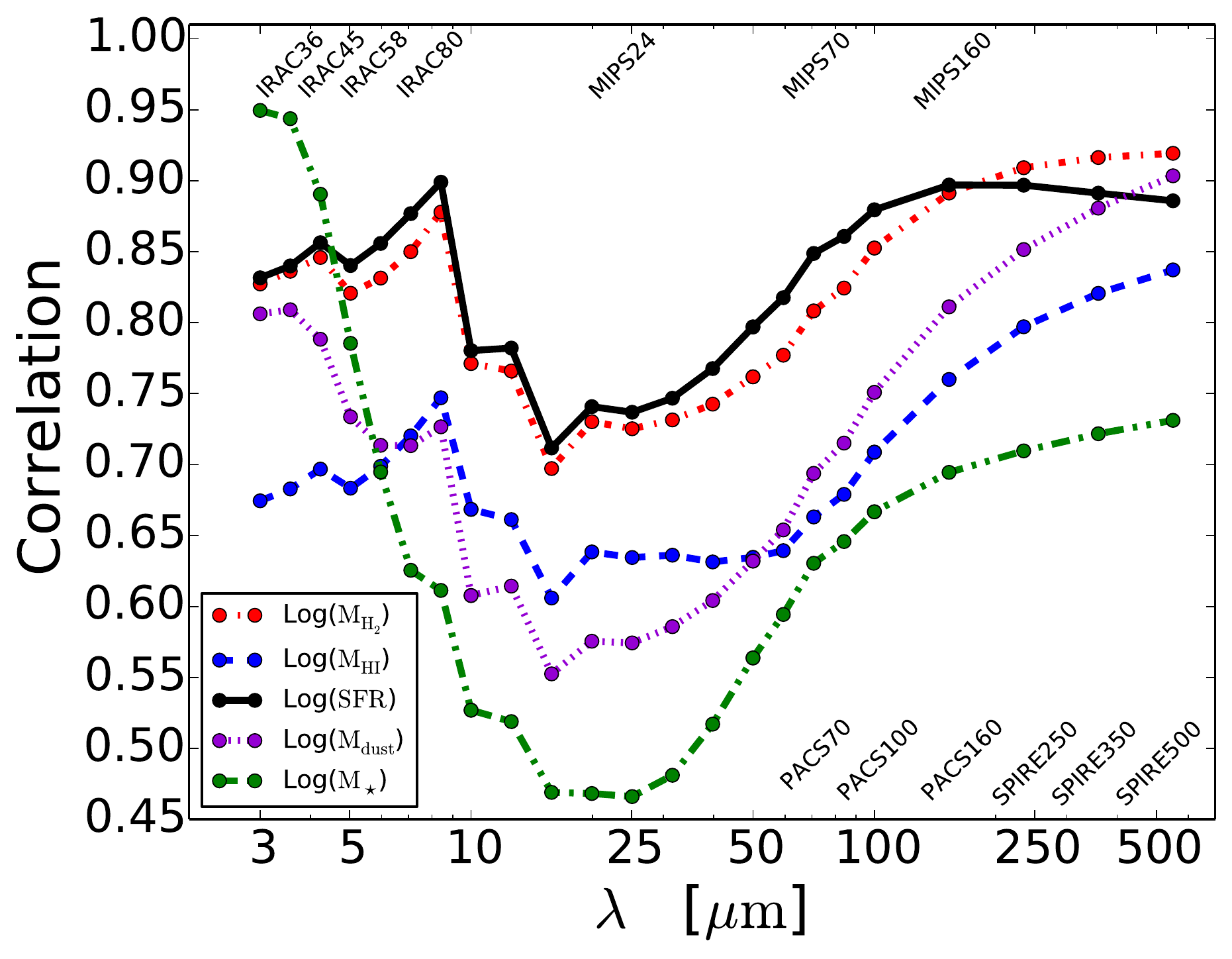}
}
\caption
{Spearman Correlation coefficients of IR luminosity in \emph{Spitzer}
  (IRAC and MIPS) and \emph{Herschel} (PACS and SPIRE) bands with
  $M_{\rm H_2}$ (dot-dashed red line), $M_{\rm HI}$ (dashed blue line), 
  SFR (continuous black line), $M_{\rm dust}$ (dotted violet line)
  and $M_{\rm \star}$ (dot-dot dashed green line).}
\label{fig:correlations}
\end{figure}

After having quantified the level of agreement of simulated SEDs
  with observations, we 
consider here the correlation of IR/sub-mm
luminosity with global properties of MUPPIBOX galaxies. This is done
to understand to which physical quantity IR light is most sensitive,
and which bands best trace that given quantity. 
We consider here bands from 3.6 to 500 $\mu$m,
covering Spitzer and Herschel ones plus a fine sampling of the MIR region from 8 to 24 $\mu$m.
For each simulated galaxy we compute its luminosity
and correlate it with atomic gas mass $M_{HI}$, molecular
gas mass $M_{H_2}$, SFR, dust mass $M_{\rm dust}$ and stellar mass $M_\star$.
 
These quantities
are of course all strongly correlated among themselves, so they will
all correlate with IR luminosity. For each correlation we compute
Spearman and Pearson correlation coefficients; since they show the same
behaviour, we restrict ourselves to the Spearman one.
Figure~\ref{fig:correlations} shows these correlation coefficients. IR
light best correlates with SFR up and beyond the $\sim100\ \mu$m peak,
with $M_{H_2}$ giving just slightly lower coefficients before taking
over at $\sim160\ \mu$m. The best correlation with SFR is obtained at
8.0 $\mu$m. \cite{Crocker_2013} showed that only about 50 per cent of
the emission at 8$\mu$m from a galaxy is dust-heated by stellar
populations 10 Myr or younger, and about $2/3$ by stellar populations
100 Myr or younger, so the correlation is possibly enhanced by our
choice of averaging SFR over 100 Myr. At longer wavelengths luminosity
has a better correlation with molecular and, eventually, dust mass,
though the correlation with SFR remains very significant. Rest-frame
24 $\mu$m, deep in the PAH region, is found to be the worst band to
trace SFR. Correlation with atomic gas mass is always weaker, though
coefficients stay above a value of 0.65; the same is true for
correlation with dust mass, with the exception of sub-mm wavelengths
of 500 $\mu$m, where it becomes slightly more significant than that with SFR.

The fact that luminosities beyond the IR peak, at $\sim$ 160 $\mu$m,
show very good correlation with SFR was noticed by \cite{Groves_2015}.
Conversely, the increase of the correlation coefficient with gas mass at
long wavelengths suggests that this sub-mm emission, dominated by cold
dust, is not driven exclusively by the most massive stars. For
instance, \cite{Calzetti_2007} found that stochastically heated dust
can trace both young and more evolved stellar populations.

\begin{figure}
\centering{
\includegraphics[angle=0,width=\linewidth]{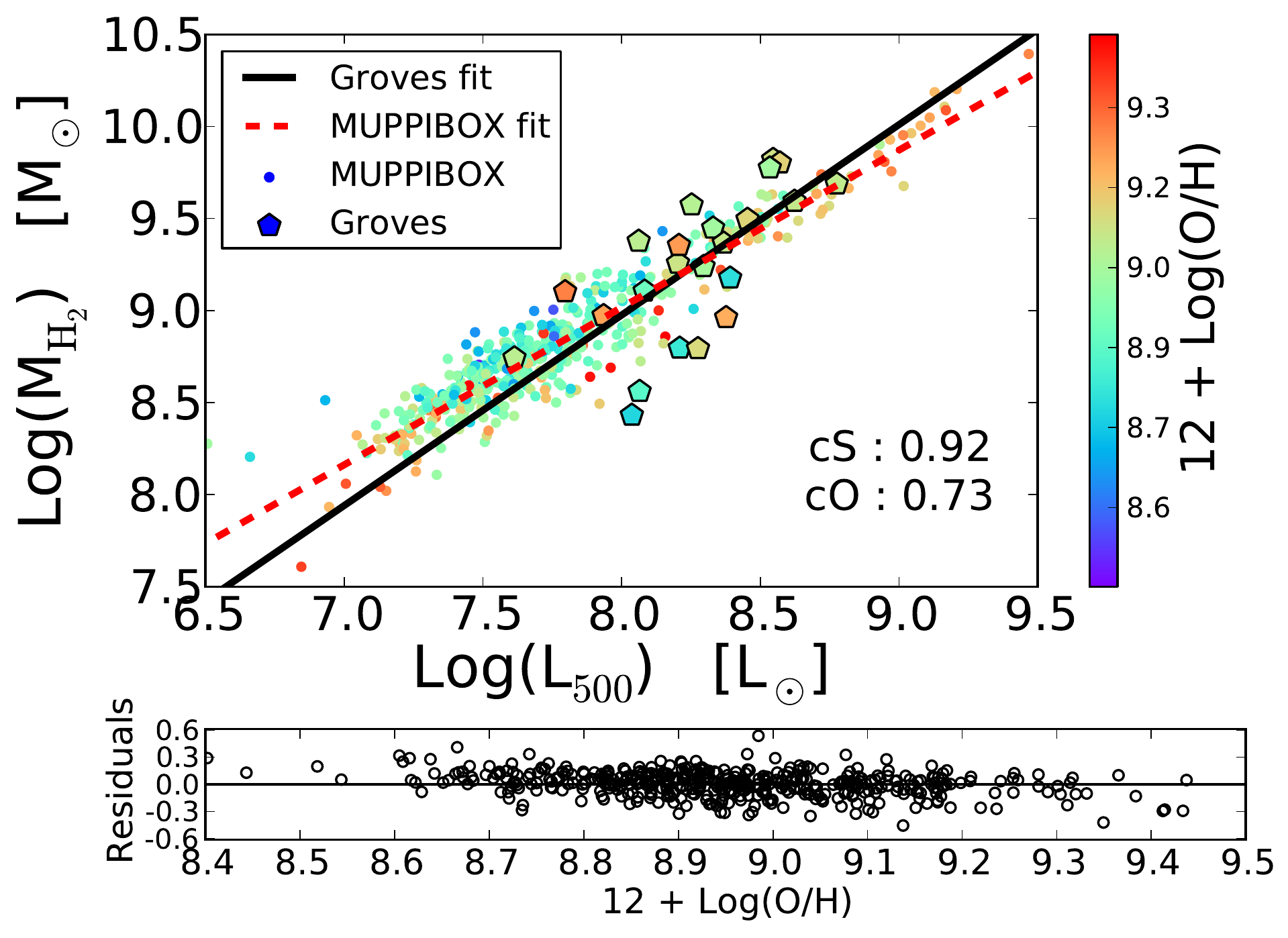}\\
\includegraphics[angle=0,width=\linewidth]{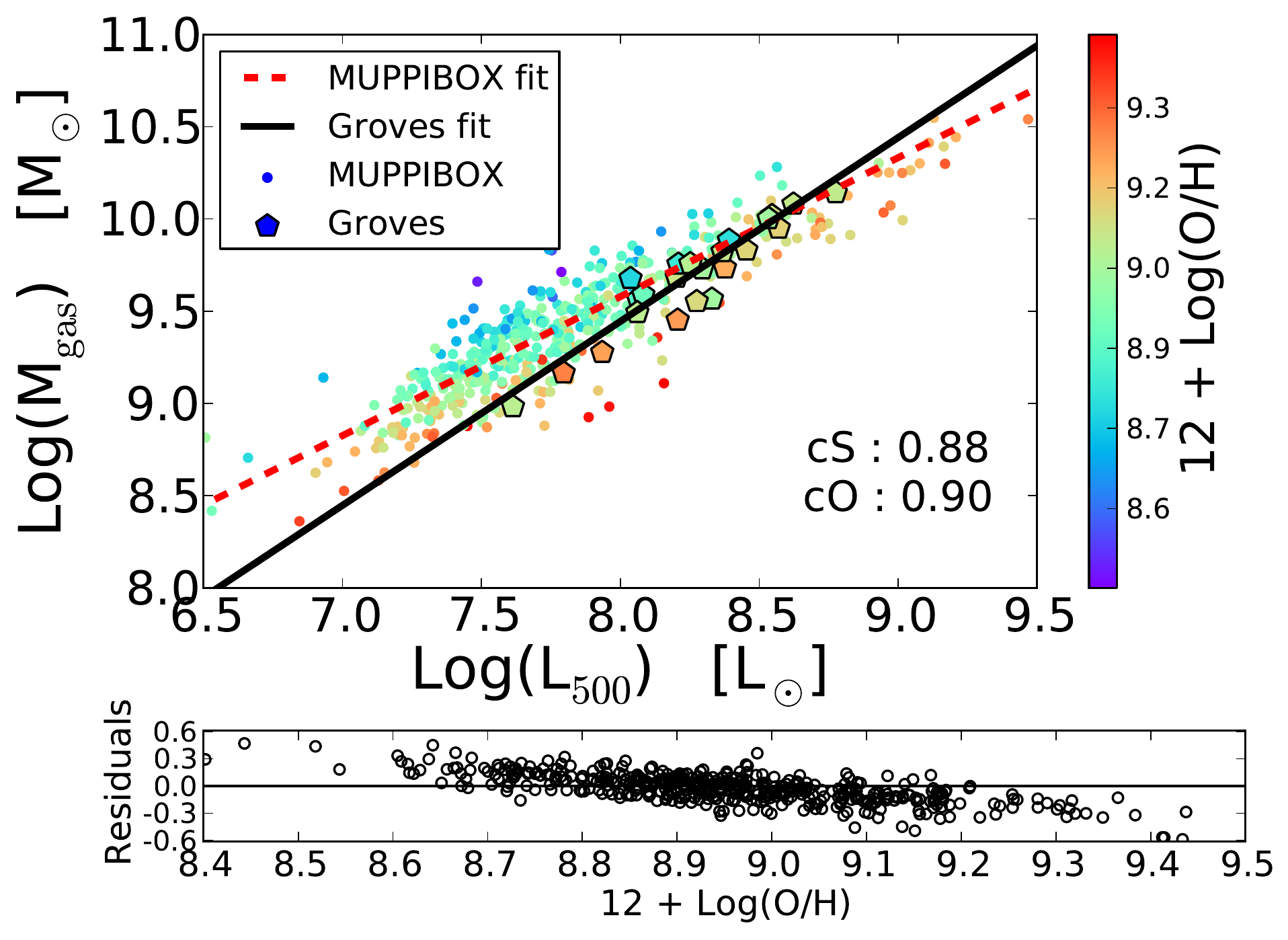}\\
\includegraphics[angle=0,width=\linewidth]{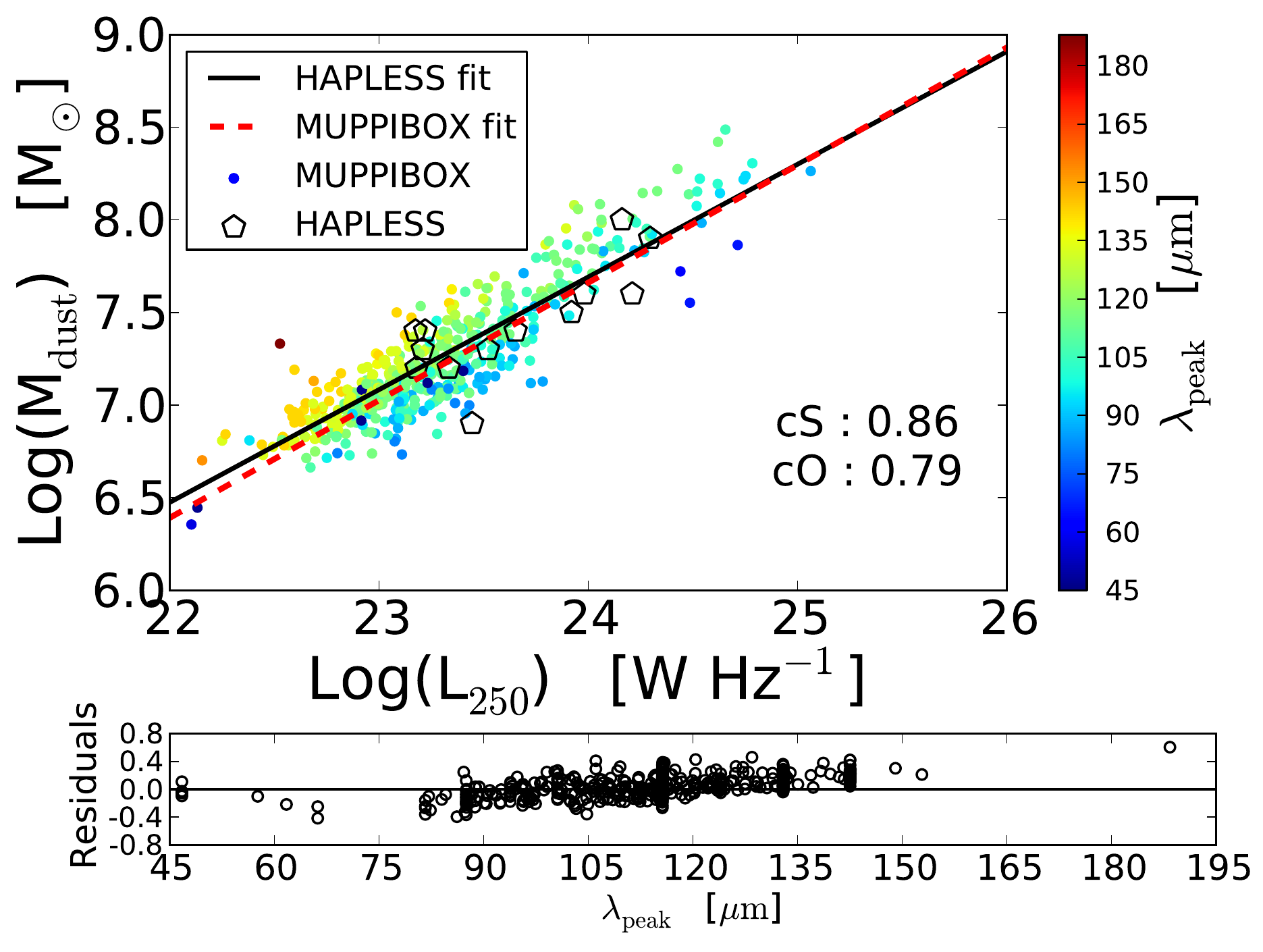}}
\caption
{Correlation between IR luminosity and galaxy physical properties, for
  simulated galaxies and for galaxies from the Groves and HAPLESS
  samples. In the first two panels, data are colour-coded with gas
  metallicity, according to the colour bar on the right, while in the
  lower panel colour coding follows $\lambda_{\rm peak}$. Black
  continuous and red dashed lines give linear fits of the relations in
  the log--log plane, each panel reports the Spearman correlation
  coefficients for the simulation (cS) and for the observations (cO).
  Residuals of simulated galaxies are shown below each panel against
  gas metallicity (first two panels) and $\lambda_{\rm peak}$ (bottom panel).
  Top panel: $L_{500}$ luminosity against molecular mass $M_{H_2}$. Middle
  panel: $L_{500}$ against total gas mass $M\rm{_{gas}}= M_{HI} +
  M_{H_2}$. In these two cases data are from the Groves sample. Bottom
  panel: $L_{250}$ against dust mass $M\rm{_{dust}}$, data taken from
  the HAPLESS sample.}
\label{fig:LversusM}
\end{figure}

To strengthen our interpretation of these results, we show in
figure~\ref{fig:LversusM} the correlations of IR luminosity with $H_2$
gas, total gas and dust masses. Data are taken from the Groves and
HAPLESS samples; we use IR luminosities at 500 $\mu$m and 250 $\mu$m
respectively, to be consistent with the data available for the two
surveys. The upper and middle panels show the correlation of $L_{500}$
with $H_2$ gas mass and total gas mass; correlation coefficients are
reported in the panels. In each panel we show linear fits (in the
log-log space) of samples, performed with the Theil-Sen algorithm that
is robust to outliers. Points are colour-coded with gas metallicity,
data are taken from the Groves sample. The correlation with molecular
gas is consistent, in normalization, slope and scatter, with the
sparse observational evidence: the slope of simulated and Groves
  samples are, respectively, $0.85\pm0.03$ and $1.03\pm0.53$. The two
lowest points in the Groves sample are galaxies at low metallicities;
apart from these outliers, the scatter does not seem to correlate with
metallicity in either samples. Indeed, the residuals of simulated
galaxies do not show a correlation with gas metallicity. We tested
that, using dust emissivity index $\beta= 1.6$ in place of 2, the
correlation between $M_{H_2}$ and $L_{500}$ is shallower, in worse
agreement with the Groves data.

Conversely, correlation with total gas mass of MUPPIBOX galaxies shows
systematic differences from the data; the tension is marginal, as the
measured slopes are $0.75\pm0.03$ and $0.98\pm0.18$ for the simulated
and observed samples respectively. Galaxies in the Groves sample
appear consistent with the lower envelope of the simulated relation.
The figure shows a clear correlation of residuals of simulated
galaxies with gas metallicity. Moreover, galaxies in the Groves sample
appear to have systematically higher metallicities. We noticed from
figure~\ref{fig:relations} that simulated galaxies tend to have higher
gas metallicities and lower $HI$ gas content than observed ones, and
this is especially true for the small galaxies that dominate the
sample in number. Groves galaxies have metallicities typically larger
than $12+{\rm Log}(O/H)\sim8.9$, that is at $+1\sigma$ of the
\cite{Tremonti_2004} relation. Therefore the difference in metallicity
cannot be due to the bias in metal content of simulated galaxies but
is an observational bias, presumably due to the fact that low
metallicity galaxies tend to be $CO$-blind and so are not present in
this sample. If we limit ourselves to galaxies in the same metallicity
range $>9.0$, the slope changes to $0.78\pm0.02$ and the two slopes
become consistent at 1 $\sigma$.

Besides, our simulations predict that lower
metallicity galaxies will be, at fixed total gas mass, less luminous
at 500 $\mu$m, though this effect is less evident when molecular gas
is used. In light of the results of Figure~\ref{fig:correlations},
where gas and dust masses were found to correlate better with $L_{500}$
than SFR, we conclude that this behaviour is due to the reprocessing
of starlight by the diffuse component, whose dust content scales with
metallicity.

The lower panel of Figure~\ref{fig:LversusM} shows the correlation of
$L_{250}$ with dust mass. Data are taken from the HAPLESS sample, where
dust masses were estimated from SED fitting. 
The simulated and observed datasets show very good
agreement in the degree of correlation between the two quantities, the slopes being
$0.63\pm0.03$ and $0.60\pm0.19$ respectively. Again,
the agreement worsens when a dust emissivity index $\beta$ = 1.6 is
used. \cite{Clark_2015} noticed that residuals on this relation
correlate with the temperature of the cold dust component, obtained by
fitting the SED with two modified black bodies of higher and lower
temperature. {\gra} computes dust temperatures in a self-consistent
way, but obtains a whole range of values. We prefer to use the
position of the $\sim100\ \mu$m peak, $\lambda_{\rm peak}$, of the IR
emission as a proxy of the average temperature of the cold dust
component, larger values denoting lower temperatures. Simulated
galaxies are thus colour-coded with $\lambda_{\rm peak}$. A clear
dependency of residuals on the position of the peak is visible, with
colder dust having lower 250 $\mu$m luminosity at fixed dust mass;
this is the expected behaviour of a (modified) black body. Moreover,
we did not find any significant dependency of residuals on gas
metallicity.

\subsection{On the position of the FIR peak}
\label{section:peak}

\begin{figure*}
\centering{
\includegraphics[angle=0,width=0.5\linewidth]{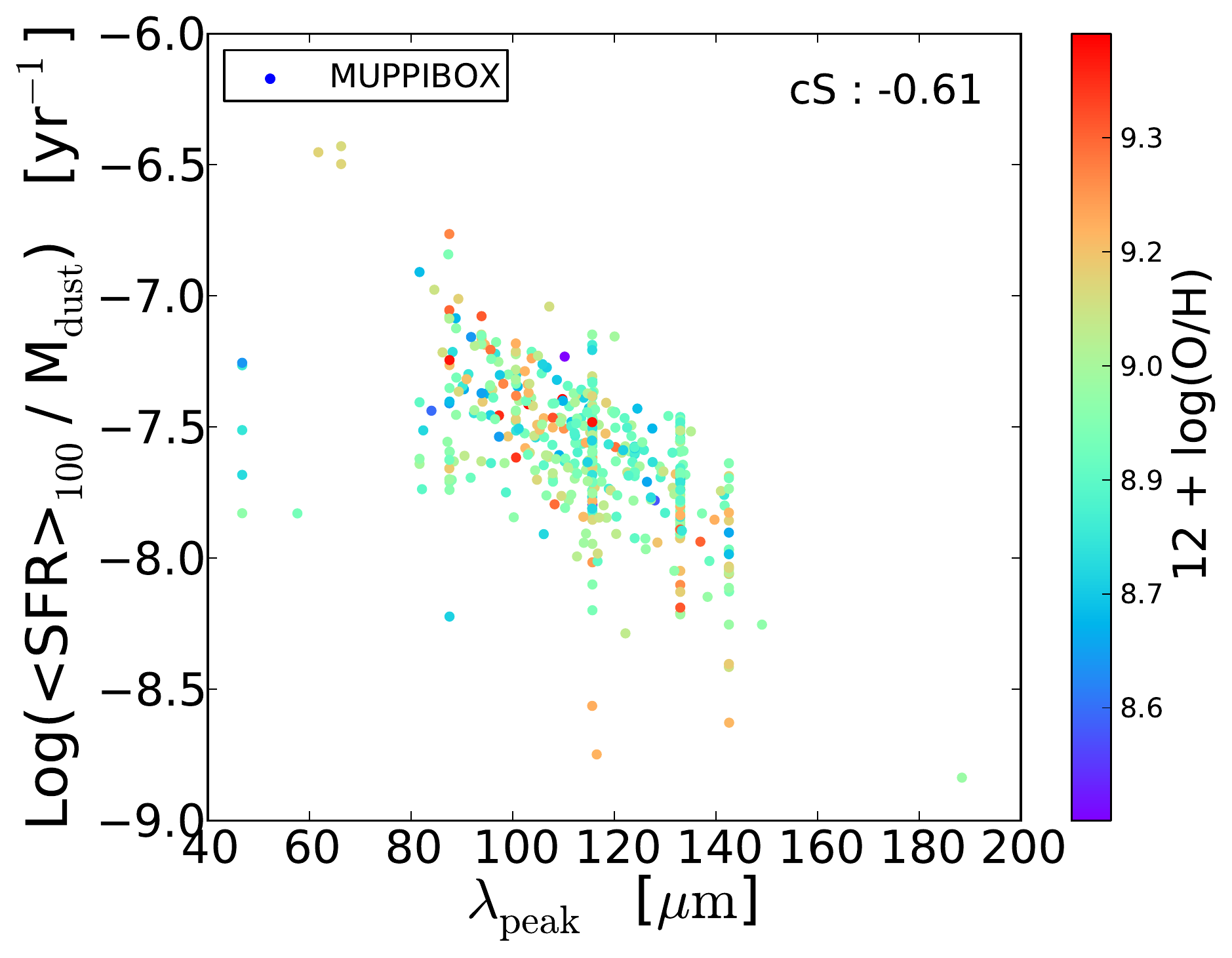}
\includegraphics[angle=0,width=0.49\linewidth]{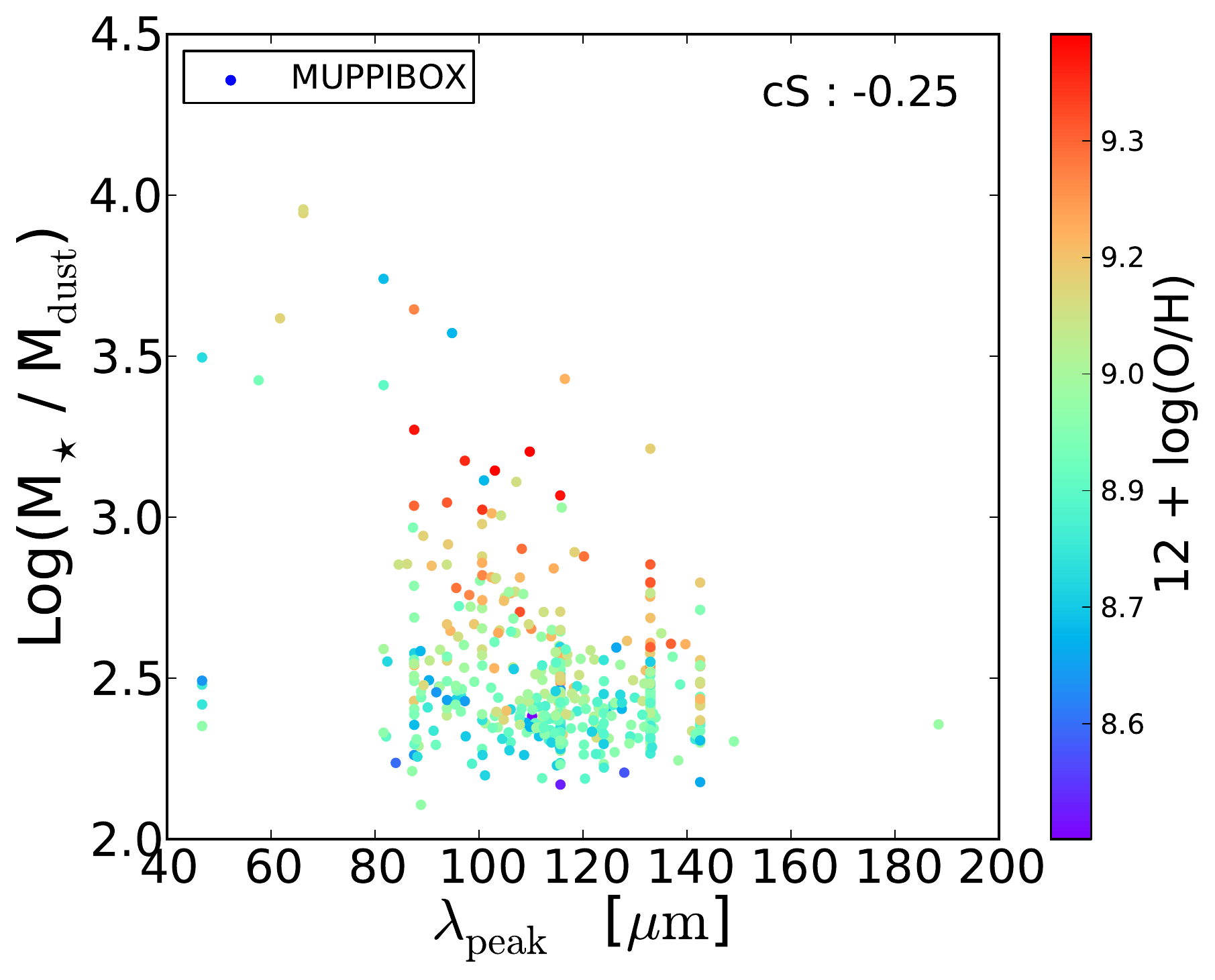}
}
\caption
{Correlation of $\lambda_{\rm peak}$ with SFR per unit dust mass (left
  panel) and stellar mass per unit dust mass (right panel). Points
  denote MUPPIBOX galaxies, colour coding is based on gas metallicity,
  as given by the colour bar. Each panel reports the correlation
  coefficient of the points.}
\label{fig:colddustT}
\end{figure*}

An issue addressed by \cite{Clark_2015} is the source of heating of
cold dust. The general idea is that the colder component of dust is
prevalently heated by old stellar populations
\citep[e.g.][]{Draine_2007,Kennicutt_2009}, even though the exact
geometry of the galaxy will influence the details. The results of
Figure~\ref{fig:correlations} are also in line with the idea that
colder components, dominating the longest wavelengths, are less well
correlated with SFR, as if young stars were not any more the main
source of energy. To verify this hypothesis, \cite{Clark_2015}
considered the correlation of cold gas temperature, obtained by
fitting the SED with two modified black bodies for cold and warm dust,
with proxies of light coming from young stars or from the bulk of
stellar mass. For the first quantity they chose the ratio of SFR,
measured from GALEX FUV and WISE 22 $\mu$m fluxes, and the estimated
dust mass, while for the second quantity they chose the ratio between
the $K_s$-band luminosity, a proxy of stellar mass, and dust mass.
They showed that the correlation of cold dust temperature with
SFR/$M_{\rm dust}$ is higher than that with $L_{K_s}/M_{\rm dust}$.
We repeat their analysis, using again the position of the
FIR peak, $\lambda_{\rm peak}$, as a proxy of cold dust temperature,
and the quantities SFR$/M_{\rm dust}$ and $M_\star/M_{\rm dust}$ to
mark the relative role of young and old stars. We recall again that
SFRs are averaged over 100 Myr. Figure~\ref{fig:colddustT} shows the
resulting correlations. In this case we do not report the HAPLESS
data, because the quantities they use are inconsistent with ours.
Nonetheless, we recover their result: $\lambda_{\rm peak}$ correlates
well with the SFR per unit dust mass, while the correlation with the
ratio of dust and stellar mass is visible only in galaxies with the
highest $M_\star/M_{\rm dust}$ ratio. This is confirmed by the
correlation coefficient reported in the panels. We conclude that in
our simulated galaxies, while the role of SFR is less prominent in
setting the temperature of the colder dust components and then the IR
luminosity in the sub-mm range, young stars (with ages $<100$ Myr) are
still the largest contributors to the radiation field responsible for
dust heating \citep[see also][]{Silva_1998}.

\section{Summary and Conclusions}
\label{section:conclusions}

We presented predictions of panchromatic SEDs of simulated galaxies at
$z=0$, obtained with a modified version of the TreePM-SPH code {\gtre}
that implements chemical evolution, metal cooling and the novel
sub-resolution model MUPPI \citep{Murante_2015} for feedback and star
formation, but contains no AGN feedback. Simulations were
post-processed using the {\gra} code \citep{Dominguez_2014}, a
radiative transfer code that is based on {\sc grasil} formalism
\citep{Silva_1998,Silva_1999,Granato_2000}. In MUPPI the molecular
fraction of a gas particle is computed on the basis of its pressure,
as in Equation~\ref{eq:br}, based on the phenomenological relation of
\cite{Blitz06}. In {\gra} stars younger than a time $t_{\rm esc}$ are
modelled as embedded in MCs with a given optical depth, while radiative
transfer is computed taking into account the atomic gas, or cirrus
component. The two codes share the rationale with which they divide
the computation into resolved and sub-resolution elements.
Post-processing by {\gra} was then performed using the molecular
fraction produced by the simulation. However, molecular gas was
re-distributed around star particles younger than $t_{\rm esc}$,
produced by a stochastic star formation algorithm, to ensure the
coincidence of the two components in a way that is consistent with
model assumptions.

In this paper we addressed the Universe at $z=0$, were high-quality
data are available. We first checked that the galaxies produced by our
code in a cosmological volume of 25 Mpc ($H_0=72$ km/s/Mpc) present
properties such as stellar masses, SFRs (averaged over
100 Myr), gas metallicities, atomic and molecular gas masses, dust
masses, $B/T$ ratios in broadly agreement with observations. 
Consistently with other simulation projects
\citep[e.g.][]{Vogelsberger_2014,Schaye_2015,Genel_2014,Lagos_2015,Bahe_2016},
galaxy properties are realistic, though at tension with observations
both at masses below $10^{10}$ M$_\odot$, where galaxies are more
passive, metallic and less gas rich than observed ones \citep[a
  well-known problem of galaxy formation in
  $\Lambda$CDM,][]{Fontanot_2009a,Weinmann_2012}, and at masses above
$10^{11}$ M$_\odot$, where the lack of AGN feedback produces blue
massive galaxies.

Our main results are the following.

(1) The SEDs of simulated galaxies resemble the observed ones,
from FUV to sub-mm wavelengths. We checked this by comparing average
SEDs, in bins of stellar mass or SFR, with data from LVL
\citep{Cook_2014,Cook_2014a}, PEP \citep{Gruppioni_2013} and HRS
\cite{Boselli_2010,Ciesla_2014} samples. The comparison with LVL and
PEP, that are subjected to different selection biases, was performed by
normalizing SEDs to the same 3.6 $\mu$m flux, while the comparison
with HRS was done considering also the normalization. SED comparison
confirms that the escape time of stars from their MCs should be
  relatively short, $t_{\rm esc}=3$ Myr. 
Simulated and observed SEDs show some disagreement only
for the smallest galaxies, where tension with observations is higher,
and their variation of shape follow the same trends with stellar mass
and SFR. An excess of attenuation in the FUV is interpreted as the
result of excessive gas metallicities in simulated galaxies.

(2) We quantified, using the Spearman coefficient, the correlation of
IR luminosities in Spitzer and Herschel bands, from 3.6 to 500 $\mu$m,
with stellar mass, SFR, molecular gas mass, atomic gas mass and
dust mass. 
The correlation with stellar mass drops for $\lambda>5\ \mu$m,
and SFR is
the quantity that best correlates with IR luminosity up to 160 $\mu$m,
while at longer wavelengths tighter correlations are shown by
molecular gas and, eventually, dust masses. This is a sign that the cold dust
components receive non-negligible contributions by old stars. The best
correlation of IR luminosity with SFR is found at 8 $\mu$m, while at
24 $\mu$m PAH emission causes a decrease of the correlation coefficient.
Dust and atomic gas masses show the worst correlations, though dust
mass overtakes SFR at 500 $\mu$m.

(3) We used Groves and HAPLESS samples to further investigate the
correlation of IR luminosity and gas or dust masses. We found very
good agreement with data, with the exception of atomic masses.
However, we found the difference to be due to a bias of Groves galaxy
in favour of high-metallicity ones. This shows that, despite the
mild tension with observations, simulated samples can be used to understand
the physical and selection properties of observed galaxy samples.

(4) We confirm the findings of \cite{Clark_2015} that the scatter in
the relation between $L_{250}$ and $M_{\rm dust}$ is due to varying
temperature of cold dust, using the position of the IR peak
$\lambda_{\rm peak}$ as a proxy for the temperature of cold dust. We
further showed that $\lambda_{\rm peak}$ is determined by SFR more
than stellar mass. Together with the correlation trends discussed
above, this implies that although the colder dust components that dominate
sub-mm emission are heated non exclusively by young stars
($<100$ Myr), their contribution remains the most prominent one.
This is partially in contrast with the canonical picture in which cold
dust is heated by evolved stellar populations
\citep[e.g.][]{Draine_2007,Kennicutt_2009}.

These results demonstrate the ability of our hydrodynamical
  simulation code to produce galaxies that, at $z=0$, resemble the
  observed galaxies in our local Universe not only in their physical,
  structural parameters but also in their observational
  manifestations. The point in common between our hydro code and the
  GRASIL3D radiative transfer solver is a description of physics at
  sub-kpc scales through sub-grid models, MUPPI on the hydro side and
  the treatment of MCs and dust on the GRASIL3D side. Our tests show
  that this combination works well: tensions of simulated galaxy SEDs
  with respect to observations can be understood as the result of
  tensions in their structural parameters, while a good degree of
  predictivity is shown in our detailed comparison of simulations and
  data of local galaxies. This work can be seen as a validation test
  to use these simulations as a tool to gain insight into observed
  galaxies, and to extend the predictions to high redshift. This will
  allow on the one hand to better constrain simulation results by
  performing comparisons in terms of observables, instead of physical
  quantities, like stellar masses or SFRs, recovered through
  techniques that require a number of assumptions that should better be done
  on the theory side. On the other hand, it will allow us to forecast
  the results of future surveys, and to understand at a deeper level
  which observables are best suited for constraining the history of
  galaxies.

\section*{Acknowledgements}

We thank D. Cook, C. Gruppioni and X.-J. Jiang for providing their
data, and Volker Springel who provided us with the non-public version
of the {\gtre} code. We acknowledge useful discussions with L. Silva,
D. Calzetti, Raffaella Schneider, F. Fontanot, P. Barai, G. De Lucia, C. Mongardi and A.
Zoldan.

This work is supported  by the PRIN-INAF 2012
grant ``The Universe in a Box: Multi-scale Simulations of Cosmic
Structures''. The simulations were carried out at the 
``Centro Interuniversitario del Nord-Est
per il Calcolo Elettronico'' (CINECA, Bologna), with CPU time
assigned under University-of-Trieste/CINECA and ISCRA grants; the analysis was
performed on the facility PICO at CINECA under the ISCRA grant IscrC\_GALPP.

We thank  the MICINN and MINECO (Spain) for  grants  AYA2009-12792-C03-03 
and AYA2012-31101 from the PNAyA. 
Aura Obreja was financially supported through a FPI contract from MINECO (Spain).

Parts of this research were conducted by the Australian Research Council Centre of 
Excellence for All-sky Astrophysics (CAASTRO), through project number CE110001020. 
This work was supported by the Flagship Allocation Scheme
of the NCI National Facility at the ANU.

This research has made use of
IPython\footnote{http://www.ipython.org/} \citep{IPython},
SciPy\footnote{http://www.scipy.org/} \citep{SciPy},
NumPy\footnote{http://www.numpy.org/} \citep{NumPy}, and
MatPlotLib\footnote{http://www.matplotlib.org/} \citep{MatPlotLib}.
 
\appendix

\section{Tests on {\gra}}
\label{section:test}

This Appendix is devoted to study in more detail the effects on the
galaxy SED of (i) the grid size adopted by {\gra} to perform the
radiative transfer calculation, (ii) the aperture size to identify
galaxies, and (iii) the impact of the dust emissivity index on the FIR
emission.

To perform these tests we use two galaxies, a normal one (ID12318) and
a recent merger (ID3); both stellar masses are close to the median
value of MUPPIBOX sample. We also use the massive late-type galaxy
(GA1), extracted from a zoom-in simulation and fully described in
\cite{Murante_2015} and \cite{Goz_2015} (their lower-resolution case);
its mass resolution and force softening are comparable to those of
MUPPIBOX. In Table~\ref{table:GA1}, we list the main characteristics
of the three selected galaxies. All the quantities marked by
(R$\rm{_{gal}}$) are evaluated within the galaxy radius, taken as
$1/10$ of the virial radius of the host halo, while by (R$\rm{_{P}}$)
within the Petrosian radius defined in
Section~\ref{section:petrosian}. Gas mass includes, as in the main
text, multi-phase gas particles and single-phase ones with temperature
lower than 10$^{5}$ K.

\begin{table*}
\begin{center}
\begin{tabular}{| c | c | c | c | c | c | c | c | c | c | c |}
\hline \hline
Galaxy   &R$\rm{_{(gal)}}$   &R$\rm{_{(P)}}$   &B/T$\rm{_{(R_{gal})}}$   &B/T$\rm{_{(R_{P})}}$ &M$_{\star \rm{(R_{gal})}}$  
&M$_{\star \rm{(R_{P})}}$   &M$\rm{_{gas (R_{gal})}}$   &M$\rm{_{gas (R_{P})}}$  &M$\rm{_{dust (R_{gal})}}$ &M$\rm{_{dust (R_{P})}}$\\

\hline

GA1          &30.37            &12.41          &0.22                   &0.23               &1.35 $\cdot$ 10$^{11}$ 
&1.15 $\cdot$ 10$^{11}$ &2.44 $\cdot$ 10$^{11}$  &1.84 $\cdot$ 10$^{10}$  &2.69 $\cdot$ 10$^{8}$ &1.89 $\cdot$ 10$^{8}$\\

ID12318      &9.39             &6.52           &0.49                   &0.53                &4.31 $\cdot$ 10$^{9}$
&3.76 $\cdot$ 10$^{9}$  &1.65 $\cdot$ 10$^{9}$   &1.14 $\cdot$ 10$^{9}$   &1.48 $\cdot$ 10$^{7}$ &1.03 $\cdot$ 10$^{7}$\\

ID3          &17               &4.43           &0.34                   &0.59               &5.1 $\cdot$ 10$^{9}$
&1.85 $\cdot$ 10$^{9}$  &2.45 $\cdot$ 10$^{9}$   &7.12 $\cdot$ 10$^{8}$   &2.01 $\cdot$ 10$^{7}$ &4.95 $\cdot$ 10$^{6}$\\

 \hline \hline
\end{tabular}
\end{center}
\caption{Basic characteristics of selected galaxies. 
Column 1: Simulation name; 
Column 2: Galaxy radius (kpc), set to 1/10 of the virial radius; 
Column 3: Petrosian radius (kpc), defined in Eq.~\ref{eq:petrosian}; 
Column 4: B/T ratio inside the galaxy radius;
Column 5: B/T ratio inside the Petrosian radius;
Column 6: Total galaxy stellar mass (M$_{\odot}$), inside the galaxy radius;
Column 7: Total galaxy stellar mass  (M$_{\odot}$), inside the Petrosian radius;
Column 8: Total galaxy gas mass (M$_{\odot}$), inside the galaxy radius;
Column 9: Total galaxy gas mass (M$_{\odot}$), inside the Petrosian radius;
Column 10: Total galaxy dust mass (M$_{\odot}$), evaluated by {\gra}, inside the galaxy radius;
Column 11: Total galaxy dust mass (M$_{\odot}$), evaluated by {\gra}, inside the Petrosian radius.}
\label{table:GA1}
\end{table*}

\subsection{Stability with grid size}
\label{appendix:resolution}

Force softening sets the space resolution of the simulation. {\gra}
uses a Cartesian grid to compute the density of star and gas
components, whose size is clearly set by softening to within a factor
of order unity.

\begin{figure}
\centering{
\includegraphics[angle=0,width=\linewidth]{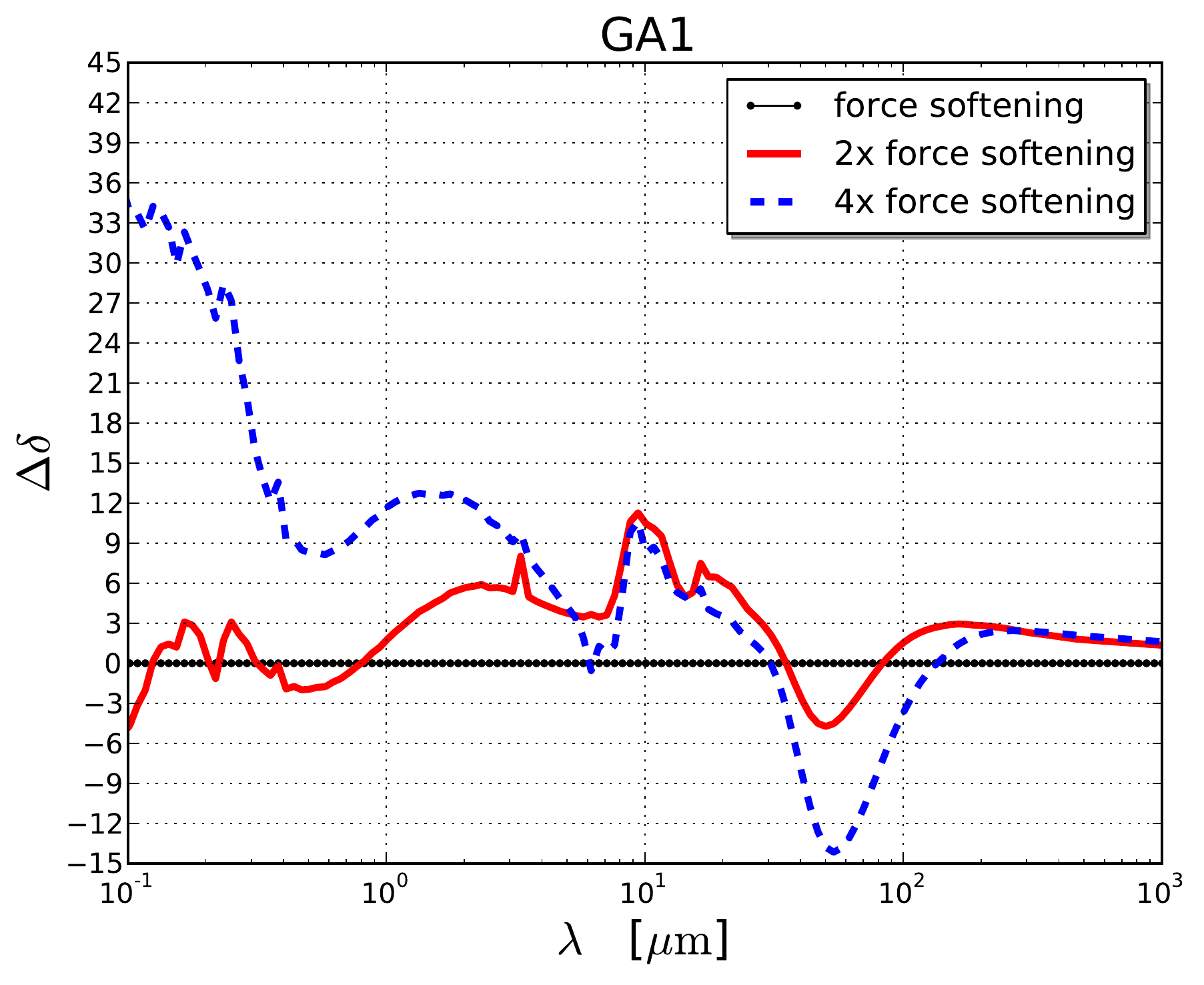}
}
\caption
{The variation on the SED, defined as in Eq~\ref{eq:sed_resolution}, 
in function of the wavelength for the GA1 galaxy. The grid size is set to
2 times (red line) and 4 times (blue dashed line) the force softening
of the simulation.}
\label{fig:resolution}
\end{figure}

We first tested on the GA1 galaxy the effect of setting the grid size
to 1, 2 or 4 times the force softening.
The variation, $\Delta \delta$, of the SED is defined as:
\begin{equation}
 \Delta \delta(\lambda) \equiv \left(\frac{L(\lambda) - L_{\rm ref}(\lambda)}{L_{\rm ref}(\lambda)} \cdot 100 \right) \%
\end{equation}
\label{eq:sed_resolution}
\noindent
where $L_{\rm ref}(\lambda)$ is the reference SED calculated with the grid
size set by the force softening.
Fig.~\ref{fig:resolution} shows how $\Delta \delta$ varies with the wavelength.
Comparing the cases of 1 and 2
force softenings, we found variations by less than $\sim3$ per cent in the
FUV-NIR ($\leq$ 1 $\mu$m) and FIR ($\geq$ 100 $\mu$m) parts of the
spectrum, while in the PAH region ($\simeq$ 10 $\mu$m) variations are
limited to at most $\sim10$ per cent. Using the coarser grid of 4
softening lengths, we found variations of $\sim30$ per cent in the FUV part
and $\sim10$ per cent in the optical-MIR, while the FIR showed variations by
less than $\sim3$ per cent.

On ID12318 galaxy and ID3 merger we found that the SED is mostly
unaffected when using grids of 1 or 2 softening lengths, with the
exception of variations of at most 10 per cent in the PAH region. On
the contrary, grid sizes of 4 times the force softening, showed
significant variations of the SED up to $\simeq20$ per cent in the
FUV, $\simeq50$ in the PAH region and $\sim30$ per cent at the FIR
peak. We extended this test to the average SED of the whole MUPPIBOX
sample, showing negligible variations when the grid is set to 1 or 2
softening lengths. 

We concluded that results are very stable as long as the grid size is
$\le2$ softening lengths. Computational times are very sensitive to
the value of the grid, so we used the softening length to compute SEDs
for our main results, but used 2 softening lengths to perform the
robustness test of {\gra} parameters (Section~\ref{section:calibrating}).

\subsection{Choice of galaxy aperture}
\label{section:petrosian}

A convenient way to define the radius of a galaxy, observed in the
presence of a sky background, has been proposed by
\citep{Petrosian_1976}, who estimated it as the radius at which the
surface brightness (at that radius) equals a fraction $\eta$ of the
average surface brightness (within that radius). A typical value for
the parameter $\eta$ \citep[e.g.][]{Shimasaku_2001} is $\eta = 0.2$.
To define the radius of a galaxy with stellar mass surface density
profile $\Sigma_\star(r)$ (in the reference frame of the galaxy,
aligned with its inertia tensor) we implemented a similar definition.
The Petrosian radius $R_P$ is such that:

\begin{equation}
\Sigma_\star(R_{P}) = \eta \frac{\int_{0}^{R_{P}}  \Sigma_\star(r) 2 \pi r dr}{\pi R_{P}^{2}} 
\end{equation}
\label{eq:petrosian}

\noindent
with $\eta=0.2$. Table~\ref{table:GA1} reports the values of Petrosian
radii of the three galaxies selected for our tests, together with
properties both within $R\rm{_{gal}}$ and $R\rm{_{P}}$.

We post-processed using {\gra} these galaxies selecting their
galaxy particles within $R_{\rm P}$. We then 
checked how galaxy SEDs change when the galaxy particles
are selected within $R_{\rm gal} = 0.1 R_{\rm vir}$ or $R_{\rm P}$.
The Petrosian aperture translates in a lower luminosity at all
wavelengths, so its effect is quantified by the change in stellar and
gas masses. More massive galaxies, $M_\star \gtrsim
10^{11}\ M_{\odot}$ tend to have extended stellar halos, so their
luminosities have a more marked dependence on aperture (this is found
in observed galaxies as well, as shown for instance by
\citealt{Bernardi_2013}). In fact, the stellar mass of GA1 decreases by 17
per cent from $R_{\rm gal}$ to $R_{\rm P}$, while the lower mass galaxy
ID12318 changes by 14 per cent. For the peculiar ID3 object, the
Petrosian aperture leads to the exclusion of the merger and the
stellar mass of the resulting object decreases by nearly a factor of 3.

We decided to adopt the more straightforward $R_{\rm gal}$ definition,
since none of the conclusions given in this paper changes if $R_{\rm P}$ is
instead used.

\subsection{Modified black-body emission}
\label{section:black-body}

Dust emission at FIR to sub-mm wavelengths is often modelled as a
modified black-body emission of the form:

\begin{equation}
 I_{\lambda} \propto B_{\lambda}(T_{\rm{dust}}) \lambda^{-\beta}
\end{equation}

where $B_{\lambda}(T_{\rm{dust}})$ is a black-body at temperature
$T_{\rm{dust}}$ and $\beta$ is the dust emissivity index. This
approach is based on laboratory experiments; these however show
temperature-dependent spectral slope variations \citep{Boudet_2005}.
The most recent results also show that the emissivity index varies with
wavelengths \citep{Coupeaud_2011} and suggest $1.5 < \beta < 2.5$ (see
discussion in \citealt{Jones_2014}). {\gra} employs the
\emph{canonical} approach by \cite{Draine_1984}, which yields a
power-law decline $\beta = 2$, while recent Planck observations of MW
dust in the diffuse ISM \citep{Planck_2014} indicate a dust emissivity
index $\beta = 1.62 \pm 0.10$.
We tested on the GA1 galaxy that, adopting the lowest value $\beta$ =
1.6, the SED is basically unaffected below the peak at $\lambda_{\rm
  peak}\simeq 100\ \mu$m, while above the luminosity is enhanced by
about 60 per cent at $\simeq$ 500 $\mu$m. The SEDs of ID12318 and ID3
galaxies were much less affected by this change. As we mentioned in
the main text, the correlations of IR luminosity with $H_2$ and dust
masses show a worse agreement with data when $\beta=1.6$ is used.

\section{Dependency of SEDs on {\gra} parameters}
\label{appendix:calibration}

\begin{figure*}
\centering{
\includegraphics[angle=0,width=0.33\linewidth]{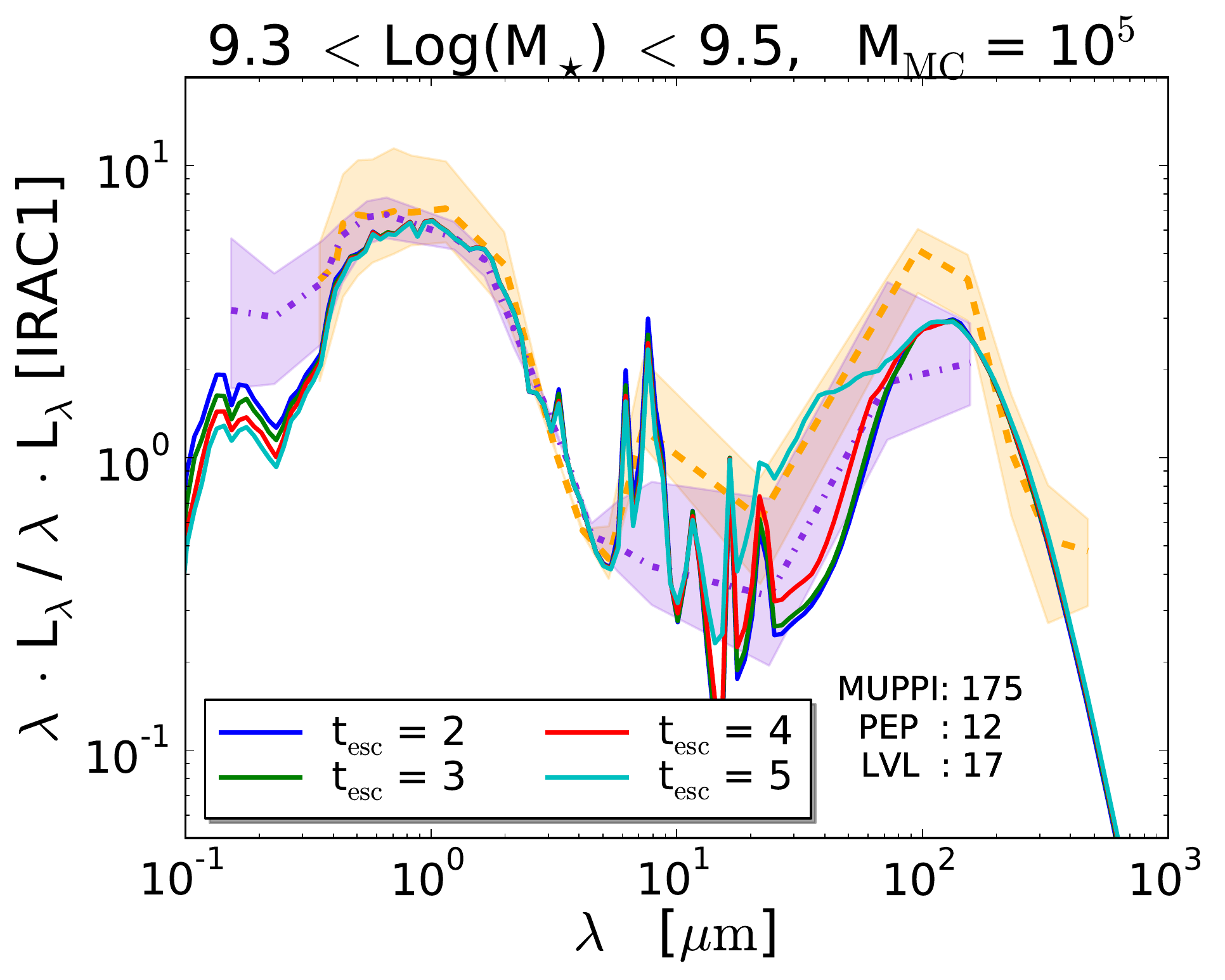}
\includegraphics[angle=0,width=0.33\linewidth]{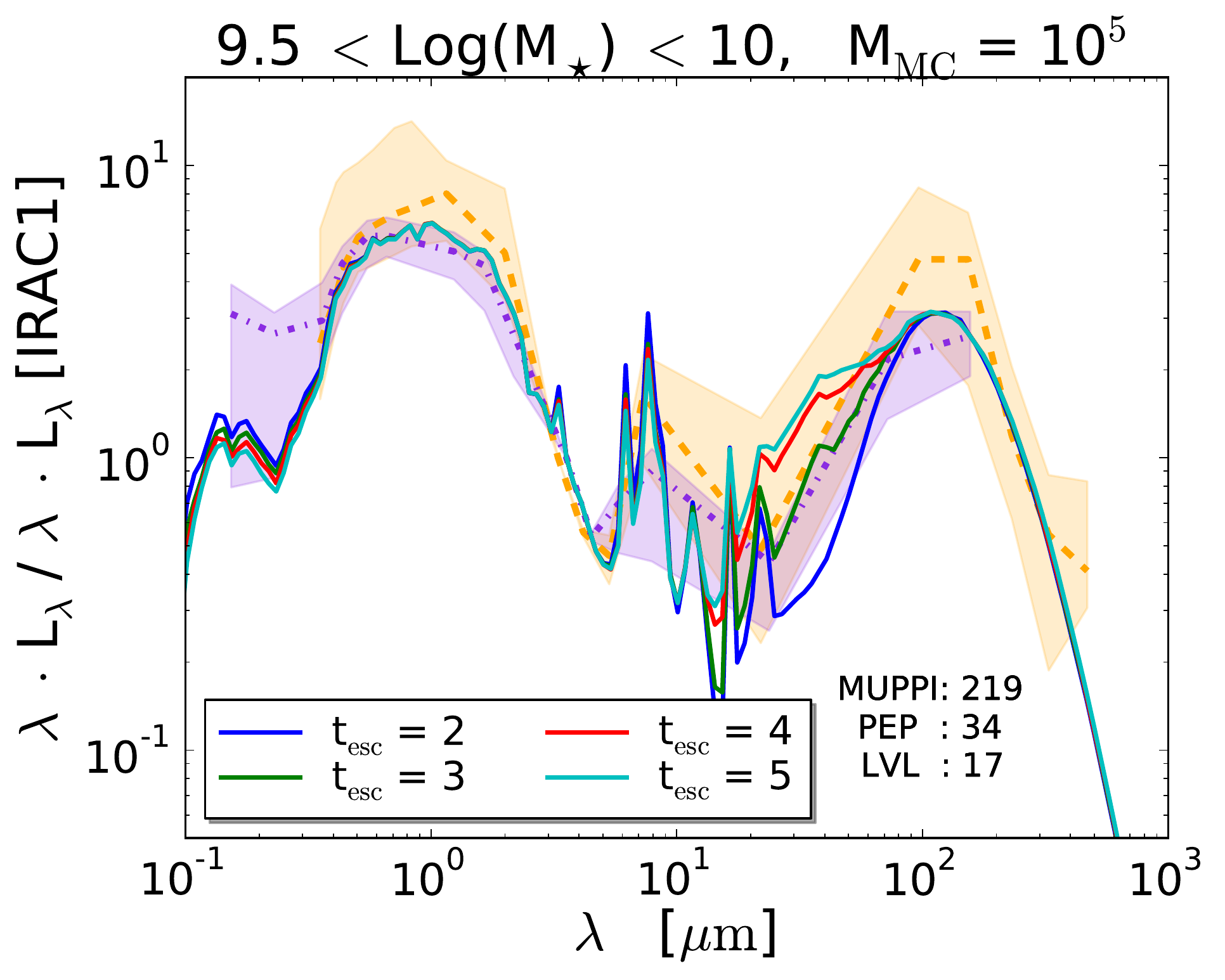}
\includegraphics[angle=0,width=0.33\linewidth]{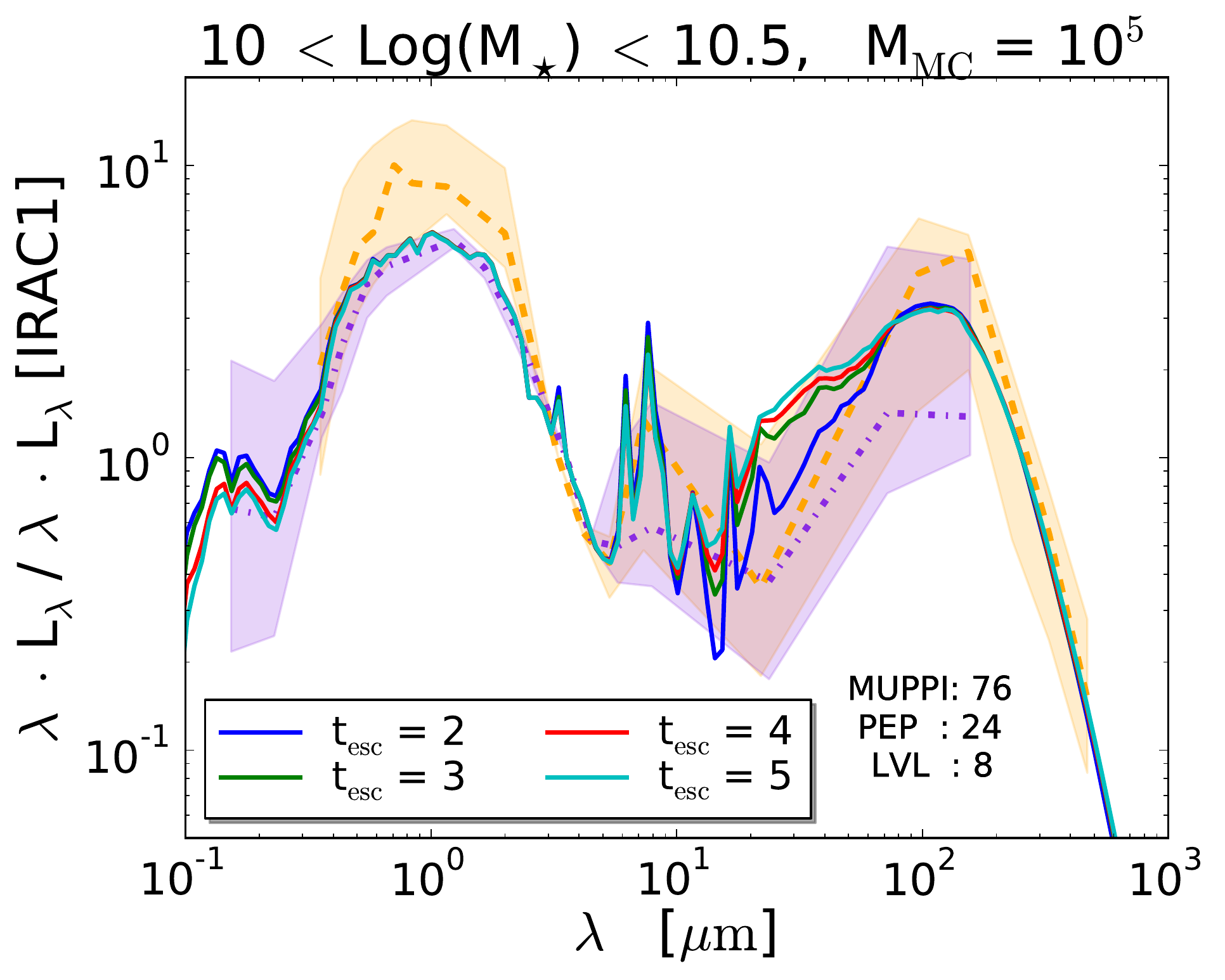}\\

\includegraphics[angle=0,width=0.33\linewidth]{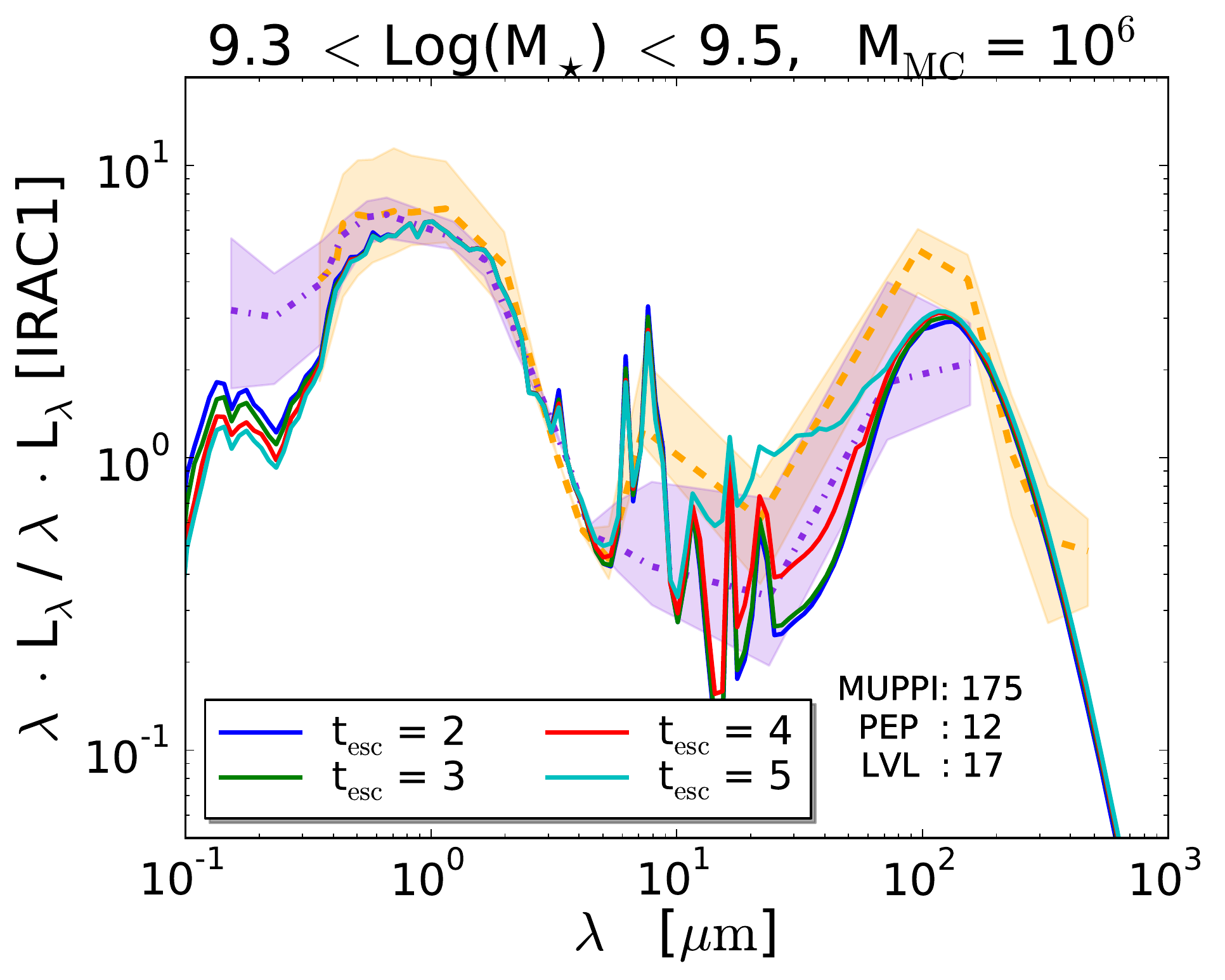}
\includegraphics[angle=0,width=0.33\linewidth]{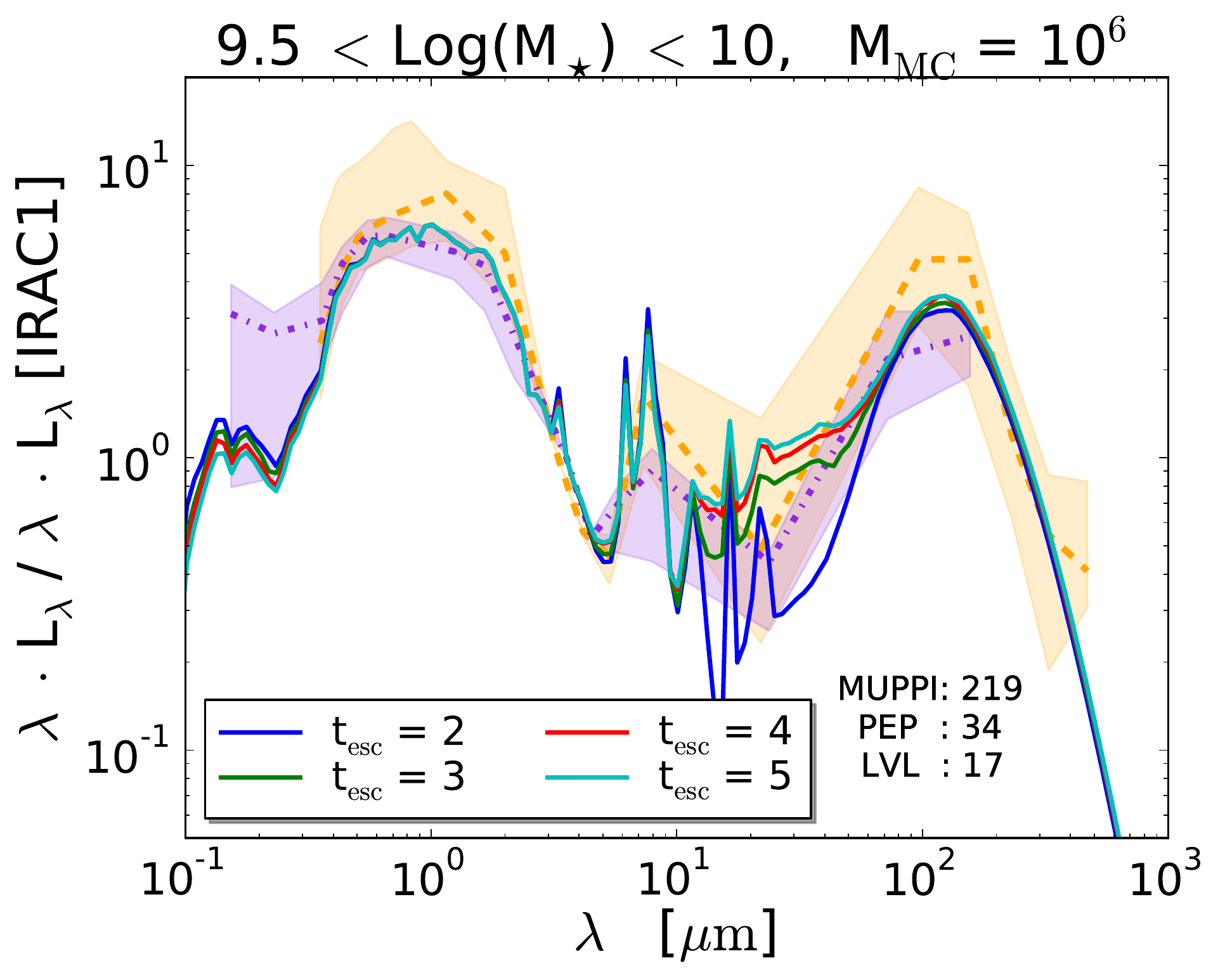}
\includegraphics[angle=0,width=0.33\linewidth]{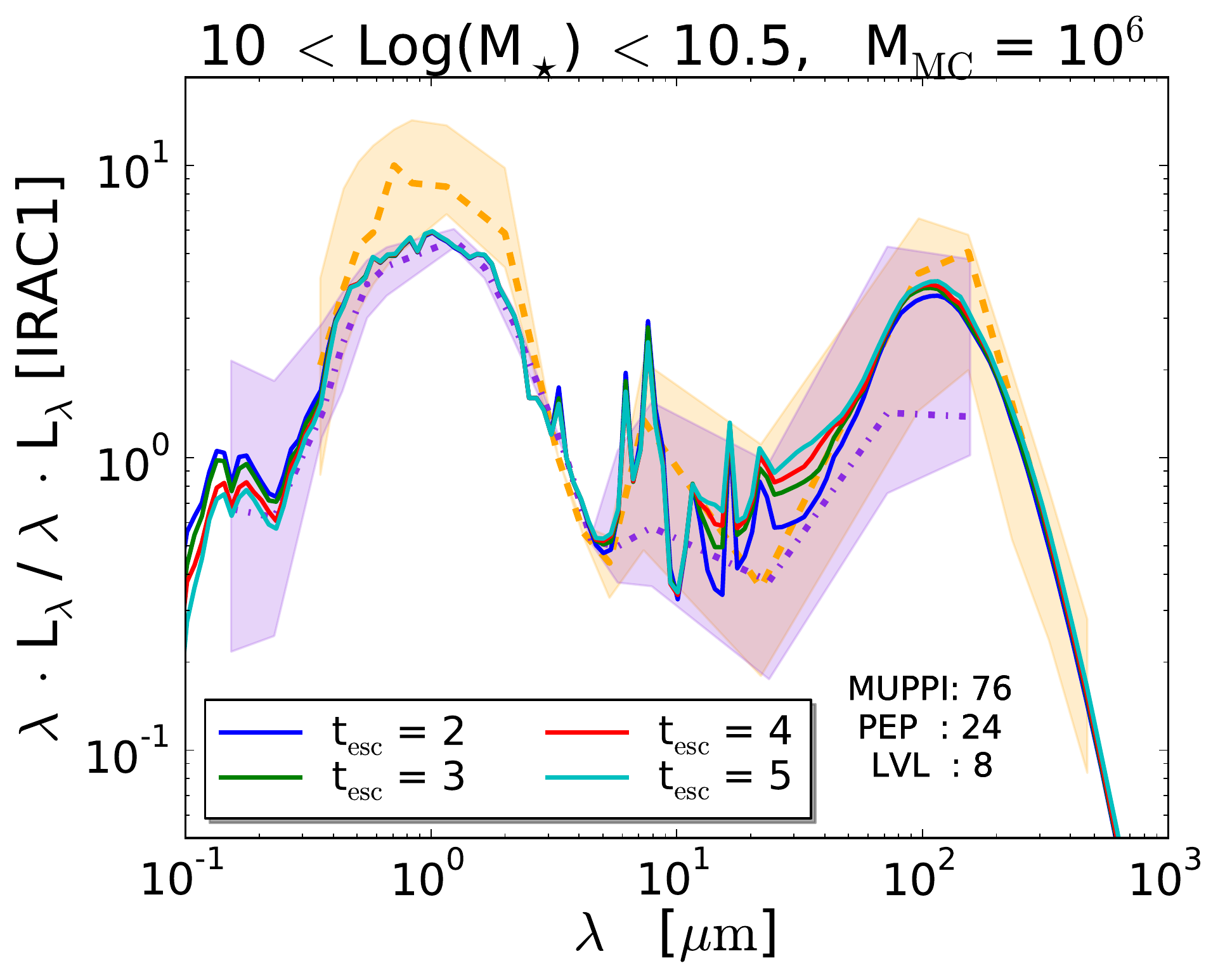}\\

\includegraphics[angle=0,width=0.33\linewidth]{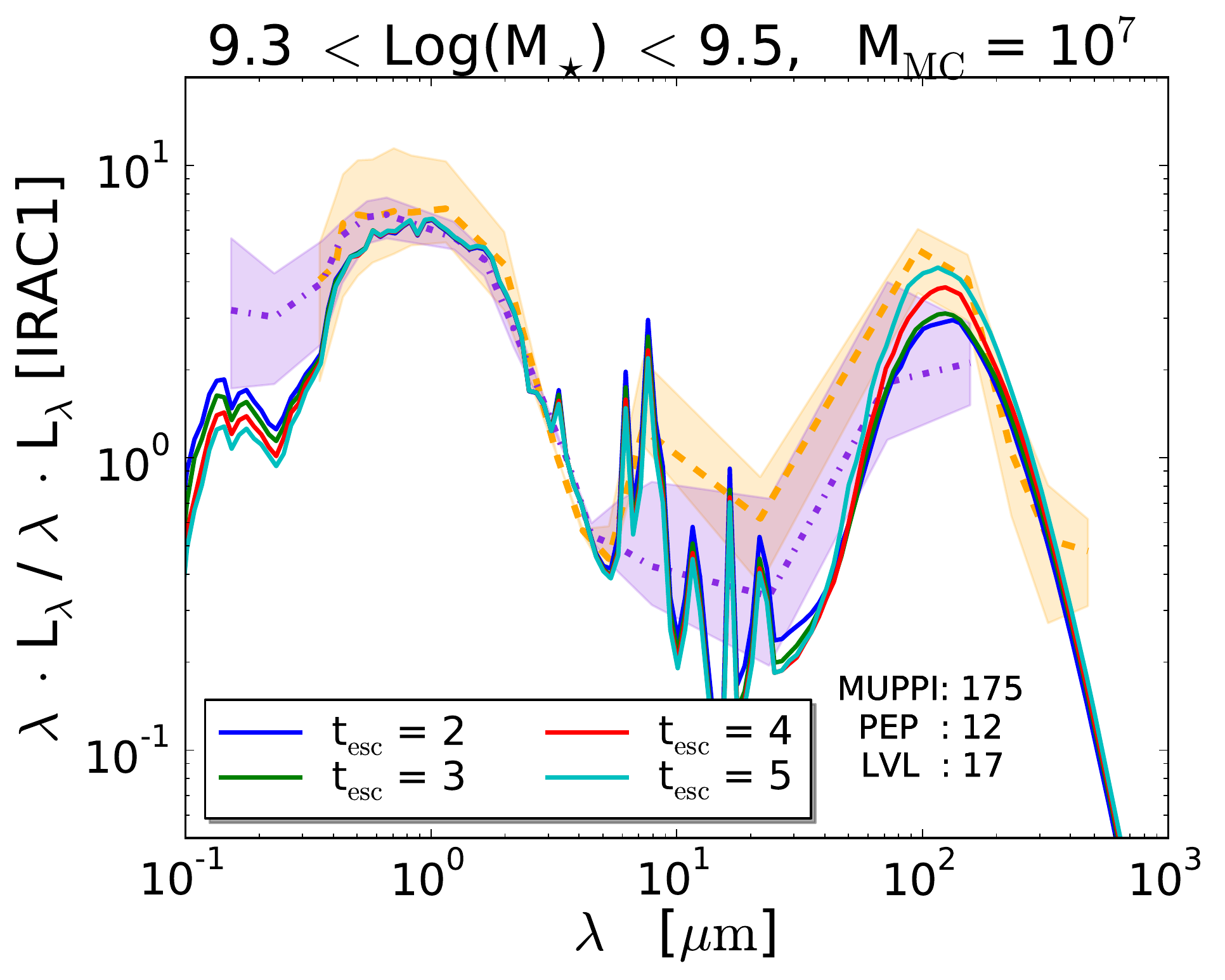}
\includegraphics[angle=0,width=0.33\linewidth]{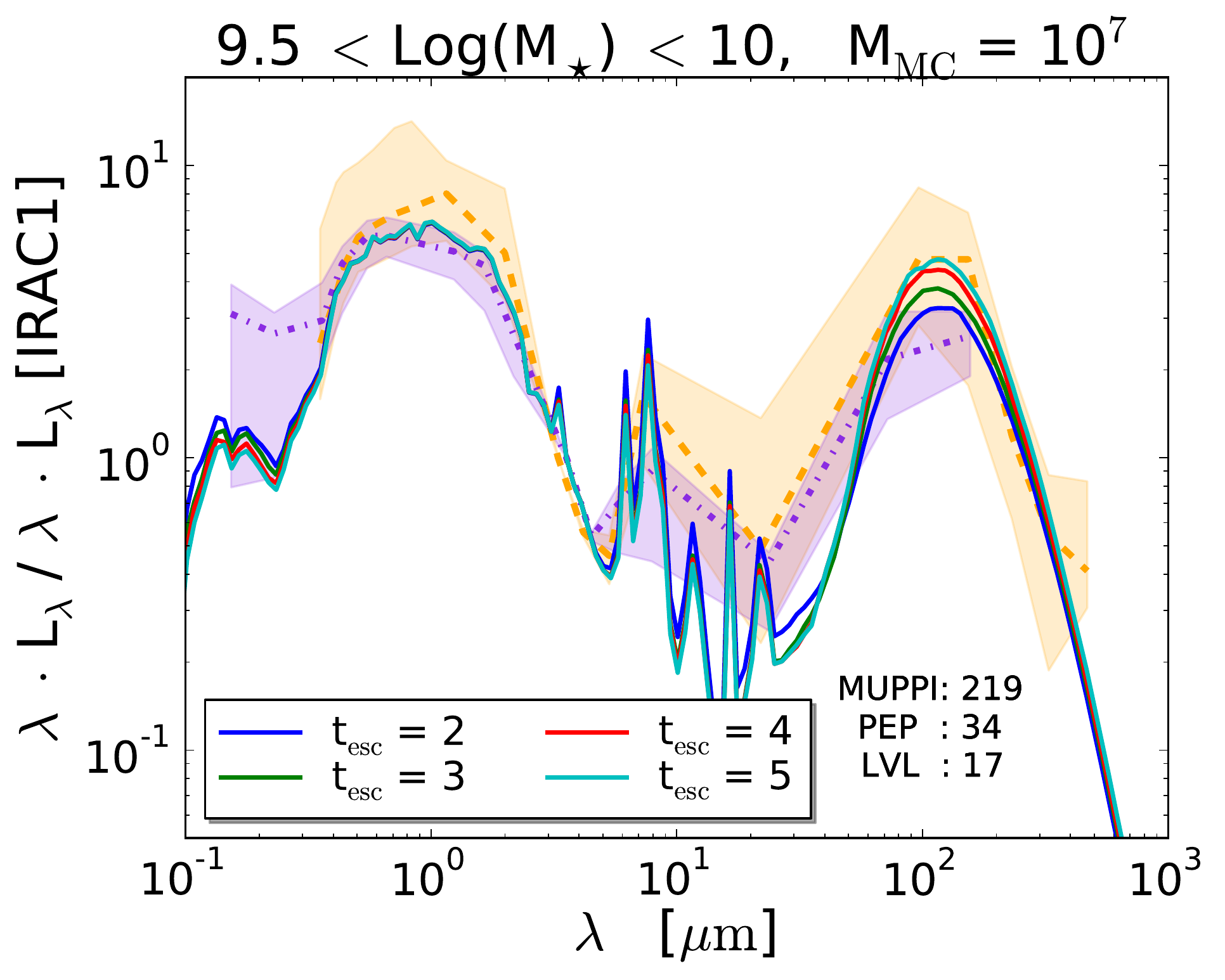}
\includegraphics[angle=0,width=0.33\linewidth]{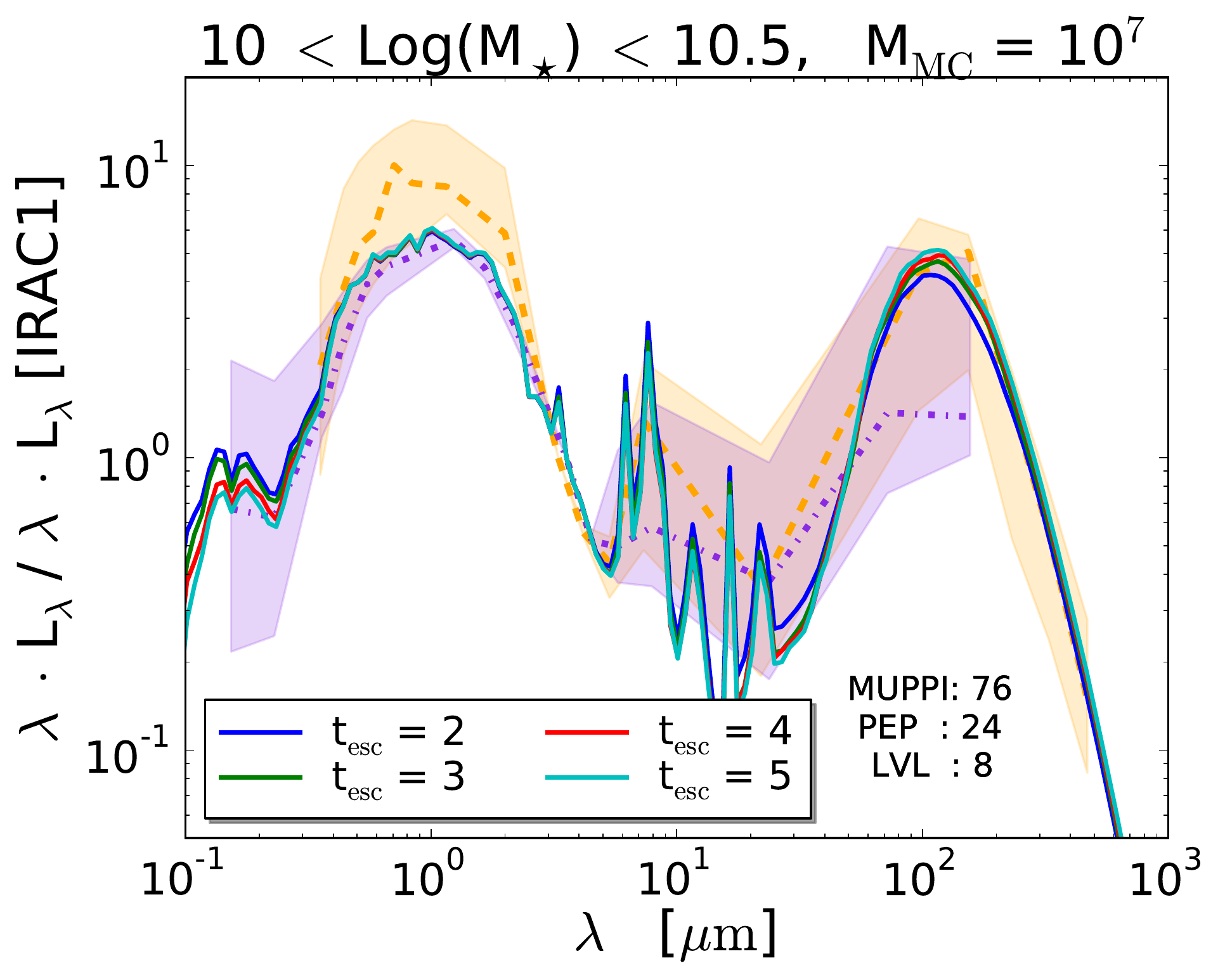}
}
\caption
{Dependency of SEDs on {\gra} parameters. In the column on the left the galaxies are selected in the mass range 9.3 $<$ Log(M$_\star$) $<$ 9.5 M$_{\odot}$,
 on the middle in the mass range 9.5 $<$ Log(M$_\star$) $<$ 10 M$_{\odot}$ and on the right in the mass range 10 $<$ Log(M$_\star$) $<$ 10.5 M$_{\odot}$
, for M$\rm{_{MC}}$ = 10$^{5}$ M$_{\odot}$ (top row), M$\rm{_{MC}}$ = 10$^{6}$ M$_{\odot}$ (middle row) and M$\rm{_{MC}}$ = 10$^{7}$ M$_{\odot}$ (bottom row).
In each plot all the SEDs are normalized to the IRAC1 band (3.6 $\mu$m), continuous colour lines show the median values for different
$\rm{t_{esc}}$, while orange and violet dot-dashed lines represent the median value for PEP and LVL samples respectively, and finally
the corresponding filled regions give the 1$\sigma$ uncertainty. Every plot reports the number of galaxies in the MUPPIBOX,
PEP and LVL samples.}
\label{fig:calibration}
\end{figure*}

\begin{figure*}
\centering{
\includegraphics[angle=0,width=0.33\linewidth]{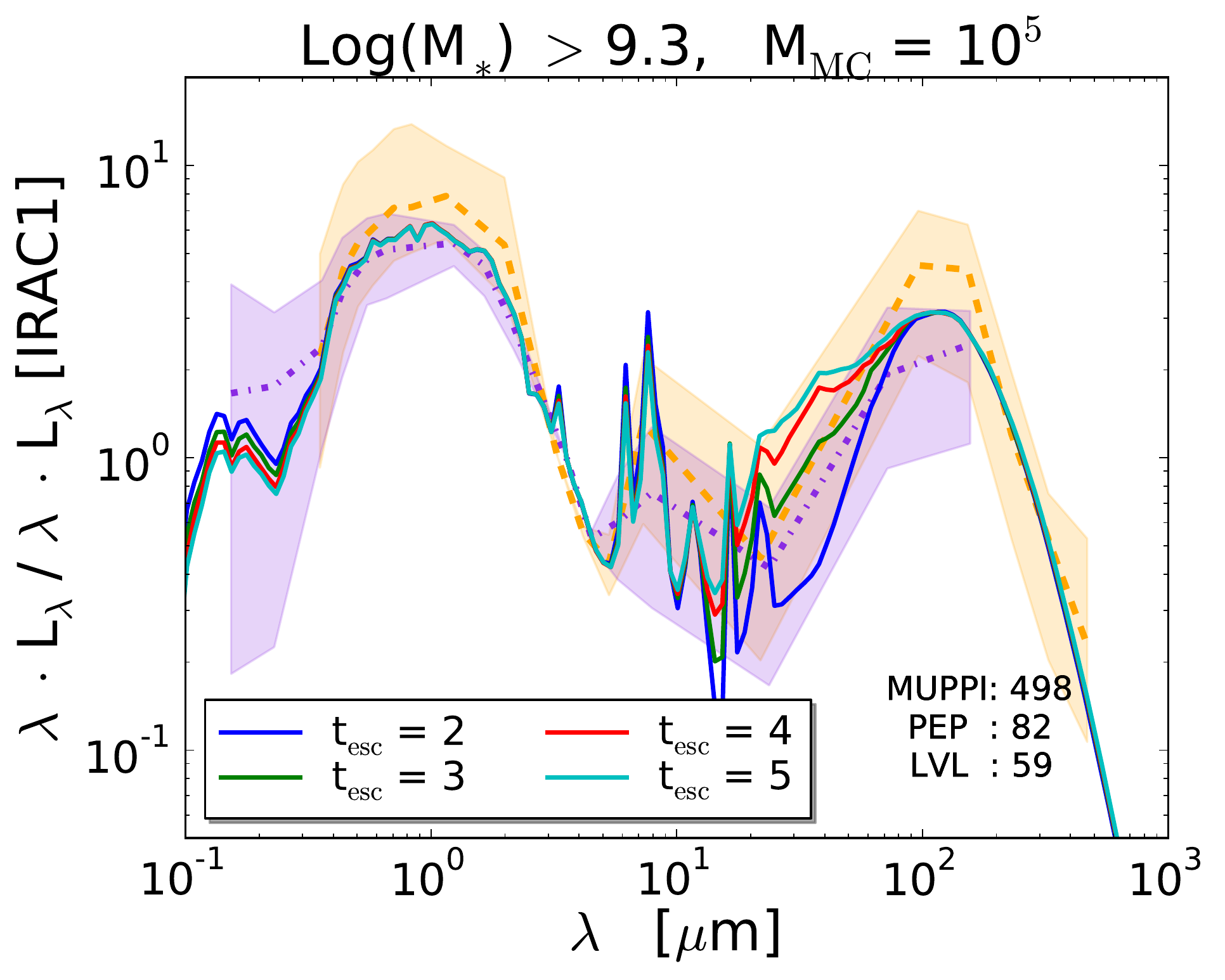}
\includegraphics[angle=0,width=0.33\linewidth]{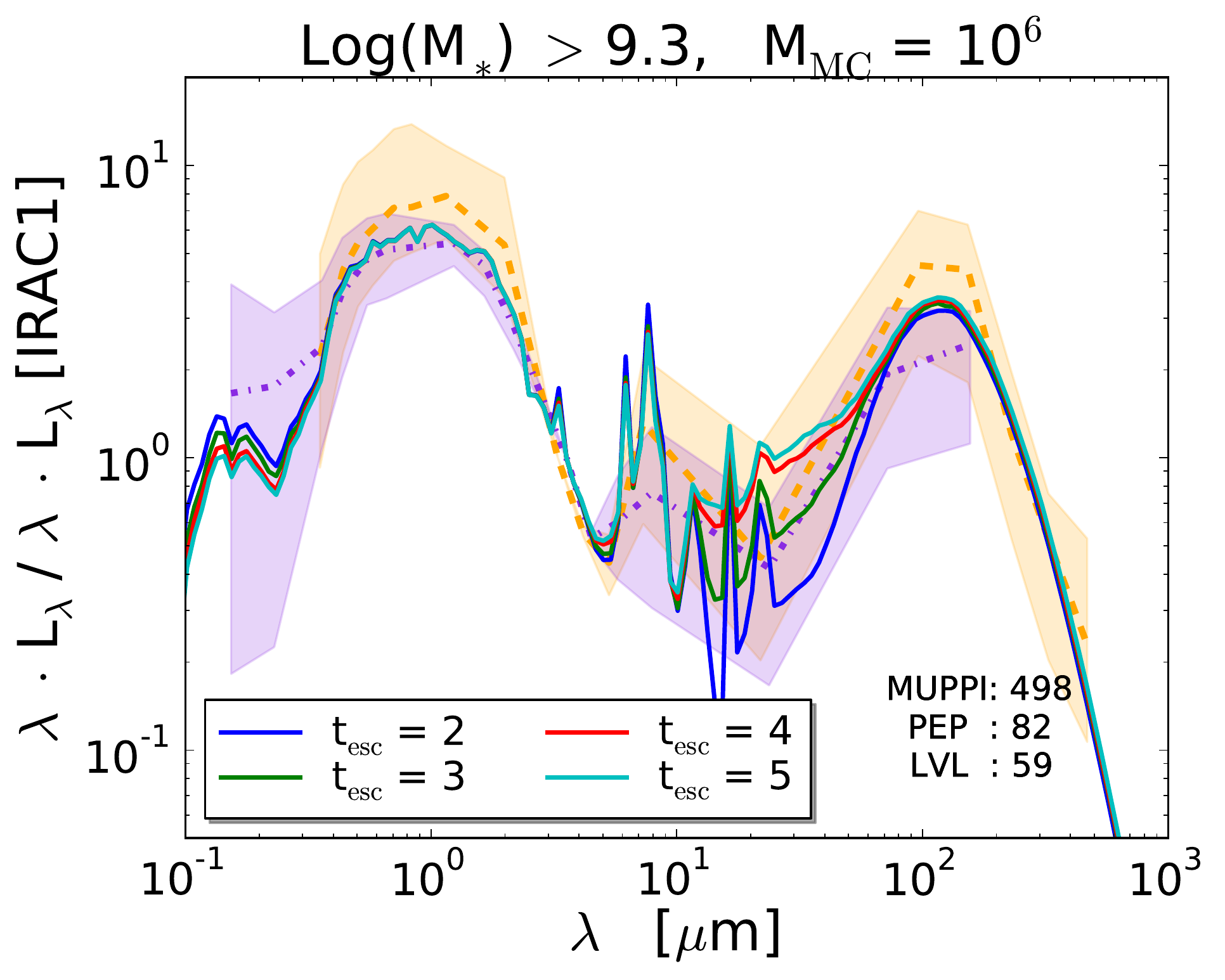}
\includegraphics[angle=0,width=0.33\linewidth]{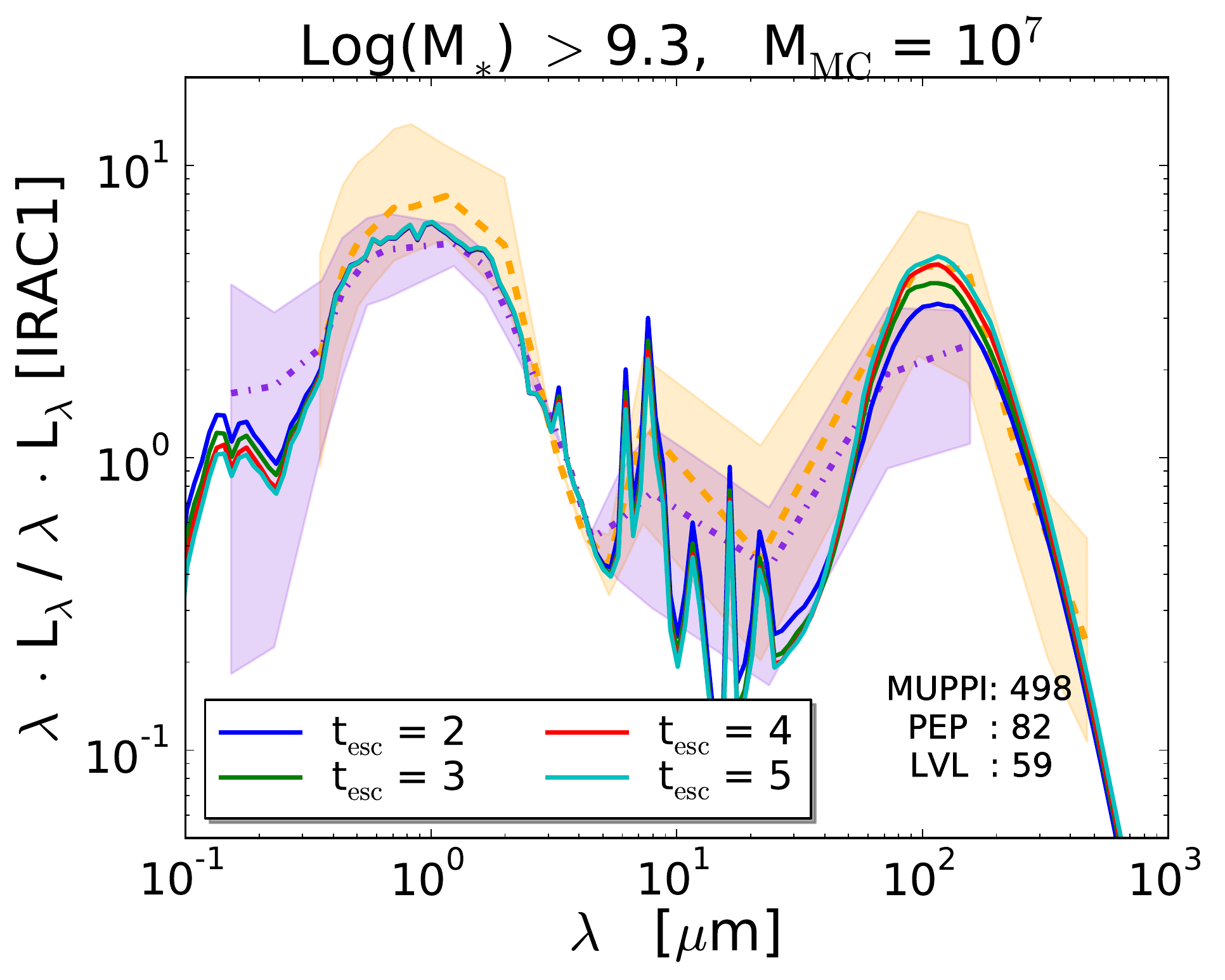}
}
\caption
{Dependency of SEDs on {\gra} parameters. In all plots only galaxies with Log(M$_\star$) $>$ 9.3 M$_{\odot}$ and
$M\rm_{MC}$ = 10$^{5}$ M$_{\odot}$ (left), $M\rm_{MC}$ = 10$^{6}$ M$_{\odot}$ (middle), $M\rm_{MC}$ = 10$^{7}$ M$_{\odot}$ (right)
are taken into account.
In each plot all the SEDs are normalized to the IRAC1 band (3.6 $\mu$m), continuous colour lines show the median values for different
$\rm{t_{esc}}$, while orange and violet dot-dashed lines represent the median value for PEP and LVL samples respectively, and finally
the corresponding filled regions give the 1$\sigma$ uncertainty. Every plot reports the number of galaxies in the MUPPIBOX,
PEP and LVL samples.}
\label{fig:calibration_THR_MASS}
\end{figure*}

We divide LVL, PEP and our MUPPIBOX samples in different stellar mass
bins (columns in Fig.~\ref{fig:calibration}), and for each of them we
post-process with {\gra} the simulated galaxies, at fixed $M\rm_{MC}$
(rows in Fig.~\ref{fig:calibration}), employing different values for
$t\rm_{esc}$. 
We used grid values equal to twice the softening for this test
(see Appendix ~\ref{appendix:resolution}) in order to save computational time.
In these plots, masses are in M$_{\odot}$, $t\rm_{esc}$
in Myr and the resulting SEDs are normalized to the IRAC1 band (3.6
$\mu$m). In every plot orange and violet dot-dashed lines represent the
median value for PEP and LVL samples respectively, while the
corresponding filled regions show the 1$\sigma$ uncertainty
obtained using the 16th and 84th percentiles and, finally,
the continuous colour lines show the median values of 
MUPPIBOX galaxies for different escape times $t\rm_{esc}$.

We explore reasonable ranges for $t\rm{_{esc}}$ (2,3,4,5 Myr), and
M$\rm{_{MC}}$ (10$^{5}$-10$^{6}$-10$^{7}$ M$_{\odot}$), as adopted in
previous works
\citep[e.g.][]{Silva_1998,Granato_2000,Dominguez_2014,Obreja_2014,Granato_2015}.
As expected, with $t\rm{_{esc}}$ increasing from 2 to 5 Myr at fixed
$M\rm_{MC}$, the UV emission decreases, but the MC cloud emission
increases and, at the same time, the cirrus emission in the PAH region
decreases. Hence the net result is that the emission increases from
the PAH region up to $\sim$ 100 $\mu$m, leaving the peak unaffected.
On the other hand, as $M\rm_{MC}$ increases from $10^{5}$ up to
$10^{7}$ M$_{\odot}$ at fixed $t\rm{_{esc}}$, the cirrus emission is
unaffected, but the bulk of the MC emission moves to longer
wavelengths (lower dust temperature) due to the increased optical
depth. The net result is a decreased emission in the PAH region but a
higher IR-peak.

Now we discuss the results for every mass bin, represented by the
columns in Fig.~\ref{fig:calibration}.

\begin{itemize}[leftmargin=0.05in]
 \item Left column - 9.3 $<$ Log(M$_\star$) $<$ 9.5 M$_{\odot}$:

    From \emph{FUV} up to mid-IR ($\sim$ 10 $\mu$m) there is not substantial variation 
    among different values of $M\rm_{MC}$ which, on the contrary, mostly affect the IR-peak. 
    The PAH emission is considerably enhanced by the highest value $t\rm{_{esc}}$ = 5 Myr, 
    even if for $M\rm_{MC}$ = 10$^{7}$ M$_{\odot}$ the bulk of the MC emission
    is considerably shifted at lower frequency due to the increased optical depth, 
    depleting for all the available $t\rm{_{esc}}$ the PAH emission.
 
 \item Middle column - 9.5 $<$ Log(M$_\star$) $<$ 10 M$_{\odot}$:

  The global trend reflects what has been already discussed for the previous mass bin. 
  The SEDs with values $t\rm{_{esc}}$ = 3-4 Myr
  are quite well in agreement with PEP and LVL samples in the PAH region, 
  except for the the highest value $M\rm_{MC}$ = 10$^{7}$ M$_{\odot}$.

 \item Right column  - 10 $<$ Log(M$_\star$) $<$ 10.5 M$_{\odot}$:
 
 The SEDs with lower values $t\rm{_{esc}}$ = 2-3 Myr are better in 
 agreement with PEP and LVL samples in the PAH region.
\end{itemize}

In all the explored mass bins the highest $M\rm_{MC}$ = 10$^{7}$
M$_{\odot}$ appears to overestimate the optical depth enhancing the
IR-emission and at the same time to decrease the PAH-emission.
$M\rm_{MC}$ = 10$^{6}$ M$_{\odot}$ fits better in all the mass bins
the PEP and LVL's median values. Furthermore the best values for the
escape time are low-intermediate ones, i.e. $t\rm{_{esc}}$ = 2-3 Myr.

In Fig.~\ref{fig:calibration_THR_MASS} we show the results for all the
galaxies (LogM$_\star$ $>$ 9.3 M$_{\odot}$). $t\rm{_{esc}}$ = 3 Myr
fits better the median PEP and LVL's values in the PAH region. There
is not substantial difference between 10$^{5}$ and 10$^{6}$
M$_{\odot}$ for the $M\rm_{MC}$.

This test shows that the dependency of SEDs on the value of
  $M_{\rm MC}$ is modest, while $t_{\rm esc}$ influences the FUV bands
  and, especially, the MIR region dominated by the PAH lines, down to
  $\sim30\ \mu$m. The values adopted in the paper of $t_{\rm esc} = 3$
  Myr is a very good compromise value.

\bibliographystyle{mn2e}
\bibliography{master}

\label{lastpage}

\end{document}